\documentclass[pra,twocolumn]{revtex4-1}
\usepackage{graphicx}
\usepackage{epsfig,color}
\usepackage{amsmath,amssymb}
\usepackage[colorlinks=true,citecolor=blue]{hyperref}
\usepackage{bbold}
\usepackage[normalem]{ulem}
\usepackage{tabstackengine}
\usepackage{makecell}
\usepackage{upgreek}
\usepackage{colortbl}
\usepackage[normalem]{ulem}
\usepackage[english]{babel}

\newcommand{\ket}[1]{|{#1}\rangle}
\newcommand{\bra}[1]{\langle{#1}|}

\newcommand{\Up}{{\uparrow}}
\newcommand{\Dn}{{\downarrow}}
\newcommand{\Id}{\mathbb{1}}
\newcommand\bwt         {\begin{widetext}}
\newcommand\ewt         {\end{widetext}}
\def\e{{\rm e}}
\def\rd{{\rm d}}

\begin{document}

\title{$\mathbb{Z}_N$ lattice gauge theory in a ladder geometry}

\date{\today}

\author{Jens Nyhegn}
\author{Chia-Min Chung}
\author{Michele Burrello}

\affiliation{Niels Bohr International Academy and Center for Quantum Devices, Niels Bohr Institute, University of Copenhagen, Universitetsparken 5,
2100 Copenhagen, Denmark}

\begin{abstract}
Under the perspective of realizing analog quantum simulations of lattice gauge theories, ladder geometries offer an intriguing playground, relevant for ultracold atom experiments. Here, we investigate Hamiltonian lattice gauge theories defined in two-leg ladders. We consider a model that includes both gauge boson and Higgs matter degrees of freedom with local $\mathbb{Z}_N$ gauge symmetries. We study its phase diagram based on both an effective low-energy field theory and density matrix renormalization group simulations. For $N\ge 5$, an extended gapless Coulomb phase emerges, which is separated by a Berezinskii-Kosterlitz-Thouless phase transition from the surrounding gapped phase. 
Besides the traditional confined and Higgs regimes, we also observe a novel quadrupolar region, originated by the ladder geometry.

\end{abstract}

\maketitle

\section{Introduction}

Gauge theories are both the backbone of the standard model of particle physics and the key to understand a wide variety of condensed matter systems \cite{Fradkin}. Their pervasive importance, however, is flanked by the extreme difficulty in obtaining exact solutions for such strongly correlated models: many non-perturbative phenomena of quantum chromodynamics and other gauge theories remain open challenges at the core of intense research efforts \cite{Brambilla2014}. These difficulties prompted a long-standing endeavor in the simulation of gauge theories, generally based on the framework of lattice gauge theories (LGT) \cite{Wilson} and Monte Carlo techniques, which achieved many accurate results, including, for example, the definition of both the hadrons and light
mesons spectra \cite{Aoki2013}.

In addition to these traditional simulations, novel strategies to analyze gauge theories are being explored in the last years, based on the knowledge acquired in the field of classical and quantum simulations of many-body quantum systems (see, for example, the reviews \cite{Zohar2015,Montangero2016,Banuls2019,Banuls2020}). These efforts develop on several directions and include new effective theoretical approaches to LGTs, the experimental realizations of the building blocks for their quantum simulation (for instance in trapped ion \cite{Martinez2016} and ultracold atom systems \cite{Dai2017,Aidelsburger2019,Gorg2019,Mil2019,Yang2020}), and tensor network calculations based on the Kogut and Susskind Hamiltonian formulation of LGTs \cite{KogutSusskind}.

The general development of these novel approaches relies on implementing progressive steps of increasing complexity on several levels. On one side, Abelian LGTs have been the first basic platform to test these techniques, before considering non-Abelian models. On the other, one-dimensional (1D) quantum systems offer the easiest playground to test tensor-network simulations before addressing higher dimensions.

Concerning the simulation of Abelian gauge theories and quantum electrodynamics (QED), for most numerical and experimental quantum simulations, it is useful to restrict the number of degrees of freedom in the considered many-body systems. Two main possibilities have been considered: (i) to maintain a continuous U(1) gauge symmetry by truncating the maximal value of the electric field flux propagating in each link; this is consistent with a quantum link model approach \cite{zoller2013,rico2014,kuehn2014} or with a truncation of the choices of the gauge boson states corresponding to different representations of the U(1) group \cite{burrello15,burrello15b,orus17}; (ii) to consider $\mathbb{Z}_N$ gauge theories \cite{zohar2013,notarnicola2015,cobanera2016,zohar2017,zohar2017b,ercolessi2018,magnifico2019,notarnicola2020,cirac2020,emonts2020} which reproduce the U(1) physics in the large $N$ limit and rapidly converge to the exact compact QED observables in one dimension \cite{kuehn2014} (see also \cite{Haase2020}).

Concerning the dimensionality of the systems, very recently there have been first attempts to investigate two-dimensional models with Abelian gauge symmetries. Truncated $U(1)$ models have been studied based on both theoretical \cite{santos2020,celi2019} and finite-size tensor-network investigations \cite{montangero2019}; the phase diagram of the  pure $\mathbb{Z}_3$ LGT, instead, has been studied with both infinite-size tensor networks \cite{cirac2020} and finite tensor networks combined with variational Monte Carlo procedures \cite{emonts2020}.

In this work, we address $\mathbb{Z}_N$ LGTs in the geometry of ladder systems.  This geometry offers an interesting compromise between one and two dimensions: on one side, it is the simplest geometry featuring plaquette interactions, thus enabling a full investigation of the Kogut and Susskind Hamiltonian; on the other, its quasi 1D nature allows us to develop an effective quantum field theory based on bosonization \cite{giamarchi}, which guides us in the exploration of the phase diagram of the model. Furthermore, ladder geometries  have been very recently adopted for small-size digital quantum simulations with superconducting qubits of a truncated non-Abelian SU(2) lattice gauge model \cite{klco2020}.

Our aim is to investigate the $\mathbb{Z}_N$ models independently on their continuum and U(1) limit. Models with discrete Abelian gauge symmetries, indeed, have recently been a focus of attention on their own; for instance, several $\mathbb{Z}_N$ symmetric models have been discussed in the context of 2D topological order \cite{brennen2007,orus2012,burrello13,Zarei2020}. Furthermore, the last generation of experimental platforms for quantum simulators based on Rydberg atoms displayed the emergence of phases with discrete $\mathbb{Z}_N$ symmetries \cite{keesling2019}.

In general, these Abelian LGTs constitute simplified models in which the gauge bosons mediating the interactions among the matter particles behave like photons and do not directly interact with themselves. Despite this simplification, Abelian models are known for displaying phenomena, such as confinement, which are common to more complex non-Abelian theories such as, for example, quantum chromodynamics, which is characterized by a non-Abelian SU(3) gauge symmetry \cite{Brambilla2014}. In particular, several mechanisms proposed to explain the confinement forces in SU($N$) symmetric models are tied to their Abelian $\mathbb{Z}_N$ center symmetries (see, for instance, Ref. \cite{greensite2003}). Therefore, the study of confinement in $\mathbb{Z}_N$ models can shed light also on more advanced non-Abelian theories. In this respect, we will focus on several properties related to the confinement and the screening of the electric $\mathbb{Z}_N$ charges in the ground states of our model. In particular, we choose to study a $\mathbb{Z}_N$ symmetric model with electric charges represented by bosonic Higgs matter degrees of freedom and we will identify the emerging thermodynamic phases and regimes based on the behavior of static and dynamical charges of the system.

In the following, we will investigate the phase diagram and main features of the $\mathbb{Z}_N$ Abelian gauge theories with Higgs matter. In Sec. \ref{sec:model} we introduce the model in the ladder geometry. In Sec. \ref{sec:pure} we study the pure lattice gauge theory limit and we show that the electric field term in this quasi-1D system always dominates over the magnetic energy, thus leading the model into a confined phase for every $N$. In Sec. \ref{sec:full} we introduce the Higgs matter and we set up a low-energy field theoretical description of the model based on bosonization. Sec. \ref{sec:gen} is devoted to the renormalization group (RG) analysis and density matrix renormalization group (DMRG) simulation of the model, focusing on its thermodynamic phases and their properties in terms of the main observables.  We show that for $N=2,3,4$ the phase diagram is, in general, trivial and displays only a single gapped phase interpolating between the confined and Higgs regimes. For $N\ge 5$, instead, an extended gapless phase appears which we interpret as a Coulomb phase.
In Sec. \ref{sec:2D} we discuss the possible extension of the model to a larger number of legs and in Sec. \ref{sec:concl} we present our conclusions. The Appendices are devoted to several details of the analysis of the model and its renormalization group study.

\section{The gauge theory in the ladder geometry} \label{sec:model}

\begin{figure}[t]
\includegraphics[width=\columnwidth]{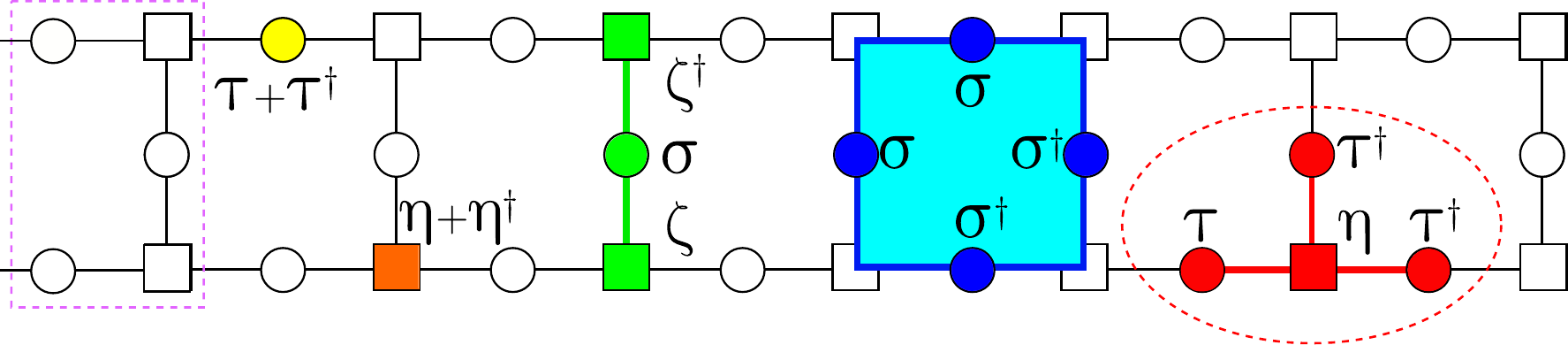}
\caption{Schematic representation of the terms in the gauge-invariant Hamiltonian \eqref{Hinv}. Squares and circles refer to Higgs matter and gauge boson degrees of freedom lying, respectively, on the vertices and links of the ladder. The yellow circle depicts the electric field energy, the orange square the matter mass, the green link represents a rung tunneling term and the blue plaquettes displays the magnetic plaquette interaction. In red, we depict one of the gauge constraints (dashed ellipse). The dashed purple rectangle depicts the first unit cell on the left (\textit{rough} boundary); the boundary conditions on the right, instead, are of the \textit{smooth} kind. } \label{fig:ladder}
\end{figure}

The quantum simulation of ladder systems has been broadly investigated in connection with the introduction of artificial gauge fluxes; in ultracold atom setups, ladders can be realized either in real two-dimensional systems \cite{atala2014} or using inner degrees of freedom to implement a synthetic dimension \cite{fallani2015,spielman2015}.
On the theoretical side, their study is closely related to the coupled wire construction of interacting topological phases of matter (see \cite{meng2020} and references therein).

In connection with lattice gauge theories, ladder models offer the simplest scenario to study the plaquette interactions which are at the basis for the appearance of deconfined and topological phases in two space dimensions.
In the following, we consider an Abelian $\mathbb{Z}_N$ theory, in which the connection and plaquette operators are unitary \cite{horn1979}. Our model is composed by bosonic $\mathbb{Z}_N$ gauge field degrees of freedom living on the edges of the ladder, which represent $N$ possible values of the electric field, and ``frozen'' Higgs matter degrees of freedom lying on the ladder vertices, which represent $N$ different charge states (see Fig. \ref{fig:ladder}). Both the edges and vertices, therefore, are characterized by an $N$-dimensional local Hilbert space and we introduce two pairs of clock operators acting on the gauge and matter degrees of freedom respectively. The first pair is given by $\tau \sim \e^{i\frac{2\pi}{N}E}$ and $\sigma \sim \e^{iA}$, which affect the gauge degrees of freedom and are respectively related to the electric field operator $E$ and the magnetic connection $A$.
They obey the algebra of $\mathbb{Z}_N$ clock operators:
\begin{equation} \label{clockpr}
\sigma^N=\tau^N=\Id\,, \quad \sigma^\dag=\sigma^{-1}\,, \quad \tau^\dag=\tau^{-1}\,, \quad \sigma\tau=\e^{i\frac{2\pi}{N}}\tau\sigma\,.
\end{equation}
In the following, we will label with indices $\sigma_{r,\Up}$ and $\sigma_{r,\Dn}$ the clock operators on the links in the upper and lower leg of the ladder at position $r$. $\sigma_{r,0}$ refers instead to the clock operators along the rung $r$. The same indices apply to the $\tau$ operators.

 The second pair of clock operators is given by $\zeta_{r,y}$ and $\eta_{r,y}$ which, instead, act on the Higgs matter degrees of freedom lying in the upper ($y=\Up$) or lower ($y=\Dn$) legs. $\zeta$, in particular, represents the phase of a Higgs field, whose radial mode is frozen to unity (London limit). Therefore, $\zeta$ and $\zeta^\dag$ respectively annihilate and create a $\mathbb{Z}_N$ electric charge, whereas $\eta=\e^{i\frac{2\pi q}{N}}$ is linked to the charge operator $q= 0,\ldots,N-1$ defined modulo $N$. They obey the same onsite algebraic relations as the previous clock operators:
	\begin{equation}\label{clockpr2}
\zeta^N=\eta^N=\Id\,, \quad \zeta^\dag=\zeta^{-1}\,, \quad \eta^\dag=\eta^{-1}\,, \quad \zeta\eta=\e^{i\frac{2\pi}{N}}\eta\zeta\,.
\end{equation}
Based on these definitions, the Kogut-Susskind Hamiltonian \cite{KogutSusskind,fradkin1979} in the ladder geometry is:
\begin{multline} \label{Hinv}
H = -\frac{1}{g} \sum_{r=1}^{L-1} \left(\sigma_{r,0}\sigma_{r+1,\Up}\sigma_{r+1,0}^\dag\sigma_{r+1,\Dn}^\dag + {\rm H.c.}\right) \\ 
-g \sum_{s=\Up,\Dn,0}\sum_{r=1}^{L} \left(\tau_{r,s} +\tau^\dag_{r,s}\right) 
-\frac{1}{\lambda}\sum_{s=\Up,\Dn}\sum_{r=1}^{L} \left(\eta_{r,s} +\eta^\dag_{r,s}\right)\\
-\lambda\left[\sum_{s=\Up,\Dn}\sum_{r=1}^{L-1}\zeta^\dag_{r,s}\sigma^\dag_{r+1,s}\zeta_{r+1,s}+ \sum_{r=1}^{L}\zeta^\dag_{r,\Up}\sigma_{r,0}\zeta_{r,\Dn} + {\rm H.c.}\right].
\end{multline}
The terms appearing in this Hamiltonian are represented in Fig. \ref{fig:ladder}.
The first  corresponds to the plaquette interaction; it is a function of the $\sigma$ connection operator which define the mass of the $\mathbb{Z}_N$ magnetic fluxes in the $r^{\rm th}$ plaquette of the ladder. The second term, with coupling constant $g$, provides a dynamics to the magnetic fluxes and defines the electric field energy density on each link of the ladder. The term in $1/\lambda$ represents the mass of the electric charges of the model, whereas the last line corresponds to their tunneling mediated by the gauge degrees of freedom. In this work, we consider $\lambda$ and $g$ as free parameters, which are not directly related to the continuum U(1) theory. In this respect, we adopted the notation in Ref. \cite{Fradkin} and we label by $g$ the coupling constant for the gauge boson interactions, rather than using the standard particle physics $g^2$ notation \cite{KogutSusskind}.

The Hamiltonian \eqref{Hinv} corresponds to \textit{rough} boundary conditions on the left side and \textit{smooth} boundary conditions on the right side (see Fig. \ref{fig:ladder}). In this situation, the ladder geometry can be described in terms of $L$ unit cells, each including 2 matter sites and 3 gauge-boson links. Such Hamiltonian can be supplemented by boundary operators:
\begin{equation}\label{hambound}
H_{\text{left bound.}}= -\lambda_b \sum_{s=\Dn,\Up}\left(\sigma_{1,s}^\dag\zeta_{1,s} +\sigma_{1,s}\zeta_{1,s}^\dag\right)\,.
\end{equation}
This additional term allows for single matter charges to enter or leave the system from the left boundary, thus breaking their global charge conservation if $\lambda_b \neq 0$. 

The total Hamiltonian $H+H_{\text{left bound.}}$ is symmetric with respect to the following local gauge transformations for the bulk vertices of the ladder:
\begin{equation} \label{gop}
G_{r,\Up}=\tau_{r,\Up}\tau_{r,0}\tau^\dag_{r+1,\Up}\eta_{r,\Up} \,,\quad G_{r,\Dn}=\tau_{r,\Dn}\tau^\dag_{r,0}\tau^\dag_{r+1,\Dn}\eta_{r,\Dn}\,,
\end{equation}
with $1<r<L-1$; depending on the boundary conditions, additional boundary gauge constraints may appear. For our choice of rough-smooth boundaries we have the boundary gauge symmetries: 
\begin{equation} \label{gop2}
G_{L,\Up}=\tau_{L,\Up}\tau_{L,0}\eta_{L,\Up}\,, \quad G_{L,\Dn}=\tau_{L,\Dn}\tau^\dag_{L,0}\eta_{L,\Dn}\,.
\end{equation}
The physical Hilbert space (without static charges) is defined by the gauge constraint:
\begin{equation} \label{gconstr}
 G_{r,s}\ket{\psi_{\rm phys}} = \ket{\psi_{\rm phys}}\,, \quad\text{for each }r,s\,,
\end{equation}
 which imposes a $\mathbb{Z}_N$ Gauss law on each vertex.

The gauge constraints can be used to rewrite the Hamiltonian in specific gauge choices. In the following we will adopt either the axial gauge, in which the local gauge transformations are used to set all the gauge degrees of freedom along the two legs to the trivial state $\sigma_{r,\Up/\Dn}\ket{\psi_{\rm axial}}= \ket{\psi_{\rm axial}}$, or the unitary gauge in which all the matter sites are set to the trivial state $\zeta_{r,y}\ket{\psi_{\rm uni}}= \ket{\psi_{\rm uni}}$. The latter choice is adopted in our tensor network simulations. 

In the next sections we will focus on several properties related to the confinement and screening of the electric charges of the model. These properties are conveniently examined by introducing opposite pairs of electric charged in different positions along the ladder. The introduction of charges, however, implies the introduction of suitable electric fields as well, in order not to violate the gauge constraints. In the following we will distinguish static and dynamical charges. Static charges are necessary to study the behavior of the system in the pure LGT limit $\lambda\to 0$, which is the focus of Sec. \ref{sec:pure}. The pure LGT is confined when the interaction energy among static charges grows linearly with their distance. Dynamical charges, instead, appear naturally for any finite $\lambda$, and, typically, they are nucleated in pairs by the tunneling term in the Hamiltonian \eqref{Hinv}. These pairs of opposite electric charges are connected by electric flux lines and constitute the mesons of the theory. The study of the mesons in the ground state of the theory is addressed in Sec. \ref{sec:gen} and allows us to distinguish different thermodynamic regimes of our model and examine the screening properties of its ground states.

\section{Confinement of the pure gauge theory} \label{sec:pure}

We begin our analysis from the pure gauge limit ($\lambda\to 0$), in which the matter degrees of freedom are frozen in $\eta_{r,y}\ket{\psi}=\ket{\psi}$. This limit is conveniently studied in the axial gauge in which the Hamiltonian is expressed as a function of the rung degrees of freedom only (we drop, for convenience the index $0$):
\begin{multline} \label{pureLGT}
H_{\rm gauge} = -\frac{1}{g} \sum_{r=1}^{L-1} \left(\sigma_{r}\sigma_{r+1}^\dag + {\rm H.c.}\right) -g \sum_{r=1}^{L} \left(\tau_{r} +\tau^\dag_{r}\right)\\
 -2g \sum_{r=1}^{L}\left[\prod_{j=r}^{L} \tau_j + \prod_{j=r}^{L} \tau_j^\dag\right].
\end{multline}
Based on the axial gauge choice, the plaquette term is mapped on a ferromagnetic coupling between the gauge bosons on neighboring rungs, whereas the electric field energy of the rung is left invariant. These two terms are therefore mapped into a standard one-dimensional $\mathbb{Z}_N$ quantum clock model (see, for example, \cite{fendley2012,ortiz2012}). The electric field energy of the gauge bosons in the links, instead, corresponds to the last non-local term in $H_{\rm gauge}$, and it distinguishes this LGT from the model studied in \cite{burrello2018} in the context of topologically ordered systems (see also \cite{Vaezi2018} for a study of the $\mathbb{Z}_2$ toric code on the ladder geometry). This term can be easily derived by considering the products of all the gauge constraints $\prod_{j=r}^L G_{j,s}$ for the $s$ leg. These string operators are equivalent to the identity on the physical states and, for the pure LGT, relate the electric field operator $\tau_{r,s}$ on both legs with the string operators appearing in Eq. \eqref{pureLGT}. 

The pure LGT Hamiltonian with the boundary term \eqref{hambound} enjoys a global $\mathbb{Z}_N$ symmetry given by the t'Hooft string $\mathcal{G}=\prod_r \tau_{r,0}$. This corresponds to the $\mathbb{Z}_N$ symmetry underlying the clock model defined by the first two terms of Eq. \eqref{pureLGT} only. For our choice of boundary conditions, this symmetry directly appears as the first of the non-local terms $(r=1)$ in the second line of \eqref{pureLGT}, proportional to the coupling $g$.

The clock model (local terms of $H_{\rm gauge}$) is characterized by two gapped phases for $N=2,3,4$: an ordered phase, in which the $\mathbb{Z}_N$ $\mathcal{G}$ symmetry is broken for small $g$, and a disordered phase for large $g$. For $N\ge 5$ a critical phase appears at intermediate values of $g$ \cite{lecheminant2002,ortiz2012,milsted2014,Chen2017,Tu2019,sun2019}. This picture is modified by the addition of the non-local terms, which include all the disorder operators acting on each link. Physically these operators correspond to the introduction of quanta of magnetic flux $\pm 2\pi/N$ on each plaquette and these disorder operators always favor the disordered phase. For each finite value of $N$, $H_{\rm gauge}$ displays indeed only a single gapped (disordered) phase for each value of $g>0$. 

The only exception is given by the limit $g\to 0$ in which only the plaquette terms survive. In this limit, the ground state is a state without any magnetic vortex and, for rough-smooth boundary conditions, it presents an $N$-fold degeneracy corresponding to the breaking of the t'Hooft $\mathcal{G}$ symmetry. This symmetry can be explicitly broken by different boundary conditions, including, for example, rough boundary conditions with an additional 3-site boundary plaquette terms.

The non-local Hamiltonian $H_{\rm gauge}$ can be mapped through a bond-algebraic duality \cite{ortiz2012} to the local Hamiltonian of a quantum clock model with both transverse and longitudinal fields, generalizing the analogous $\mathbb{Z}_2$ Ising model. The Ising model in transverse and longitudinal fields displays indeed only a single gapped phase (see, for example, \cite{coldea2010,banuls2011}), and we show in Appendix \ref{app:pure} that the same is true for its $\mathbb{Z}_N$ generalization.

The analysis of the properties of the gapped phase of the pure lattice gauge theory for $g>0$ is conveniently performed by considering perturbation theory in the two limits $g\to \infty$ and $g\to 0$. In the first limit, the ground state is just the product state of rungs displaying zero electric field, thus $\tau_{r,0}\ket{\psi}=\ket{\psi}$. The plaquette operators with small amplitude $1/g$ introduce pairs of gapped local excitations which do not qualitatively modify the paramagnetic ground state. For $g\to 0$, instead, it is easy to see that the degeneracy of the ferromagnetic ground states is split by the term $g \mathcal{G}$ appearing in $H_{\rm gauge}$ and corresponding to $\tau_{1,s}$. Furthermore, we observe that the non-local symmetry $\mathcal{G}$ is mapped into an holographic symmetry in the dual model \cite{cobanera2012} and the resulting ground state is symmetric under $\mathcal{G}$ for any $g\neq 0$ (see Appendix \ref{app:pure}).

\begin{figure}[tb]
\includegraphics[width=\columnwidth]{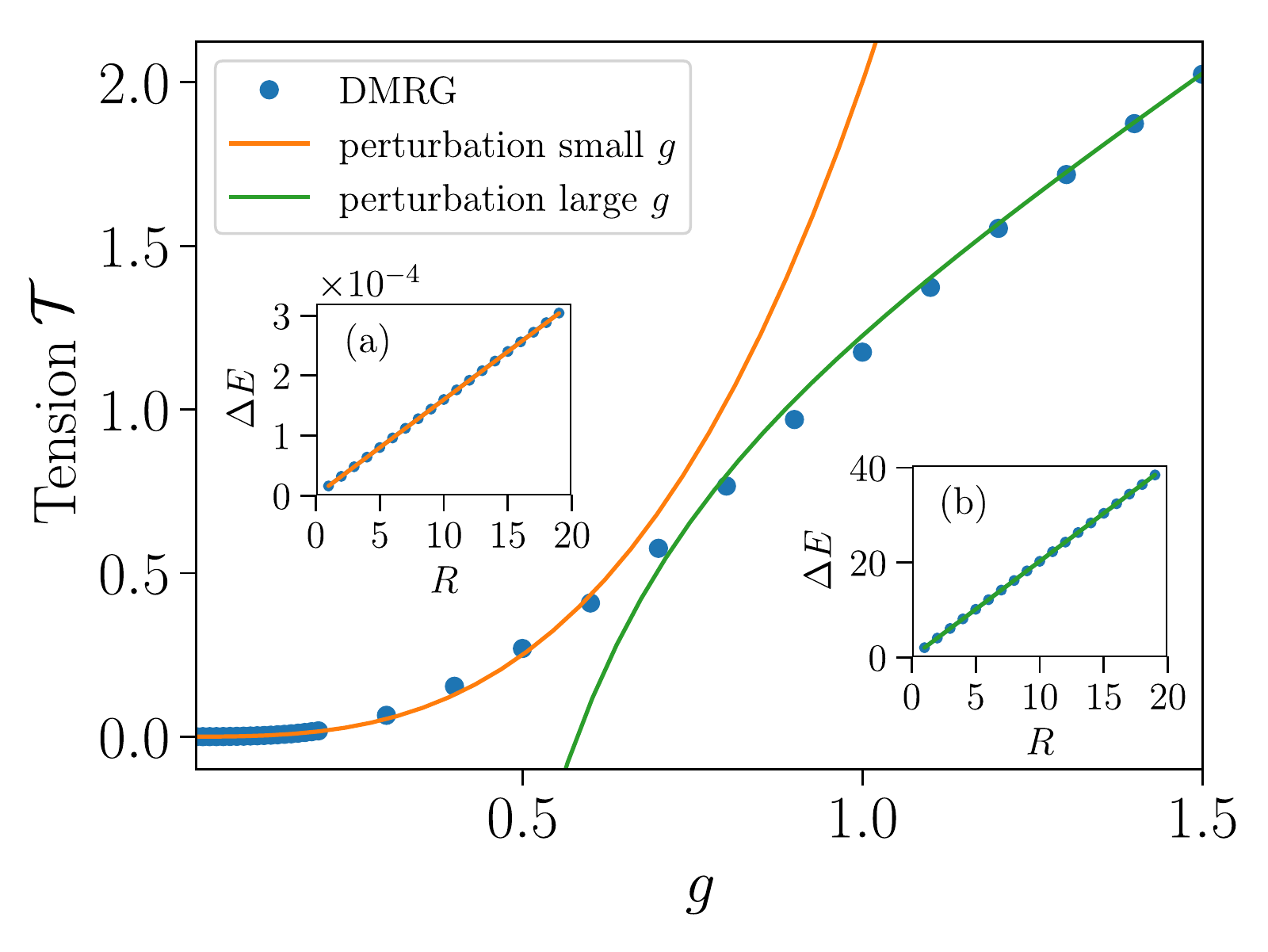}
\caption{String tension $\mathcal{T}$ of two static charges on the ladder model in the pure $\mathbb{Z}_5$ lattice gauge theory. $\mathcal{T}$ is estimated by evaluating the ground state energy difference $\Delta E$ of the system with and without static charges. $\Delta E$ behaves linearly with the separation $R$ of the static charges for each $g>0$, as shown in the insets for $g=0.02$ (a) and $g=10$ (b) (see Eqs. \eqref{deltaEg1} and \eqref{deltaEg2}).}
\label{fig:tension}
\end{figure}

This gapped phase corresponds to a confined phase for static charges; whereas only the limit $g\to 0$ results in a deconfined phase. This can be proved by the introduction of static charges in the system through a violation of the gauge constraint in arbitrary pairs of sites. In particular, we consider a system in which we introduce a pair of opposite static charges in the sites $x$ and $y$ of the lower leg. This is done by imposing that, in these sites $G_{x,\Dn}\ket{\psi}=e^{i\frac{2\pi}{N}}\ket{\psi}$ and $G_{y,\Dn}\ket{\psi}=e^{-i\frac{2\pi}{N}}\ket{\psi}$ (see Appendix \ref{app:pure} for more detail). By introducing these static charges, the energy of the ground state of the system is increased by a quantity $\Delta E$ which represents their interaction energy. We find that this interaction energy between the two charges grows linearly with the distance $R=|y-x|$, $\Delta E \approx \mathcal{T} R$, for any $g>0$. We estimated the string tension $\mathcal{T}$ as a function of $g$ through perturbation theory close to the limits $g\to 0$ and $g\to \infty$ (see Fig. \ref{fig:tension} and Appendix \ref{app:pure}), and the results of our DMRG simulations show that the string tension interpolates between the two predicted behaviors, as shown in Fig. \ref{fig:tension} for $N=5$.

We conclude that the pure $\mathbb{Z}_N$ lattice gauge theory in the ladder is always confined and, in this respect, it behaves as a 1D LGT.

\begin{figure}[tb]
\includegraphics[width=\columnwidth]{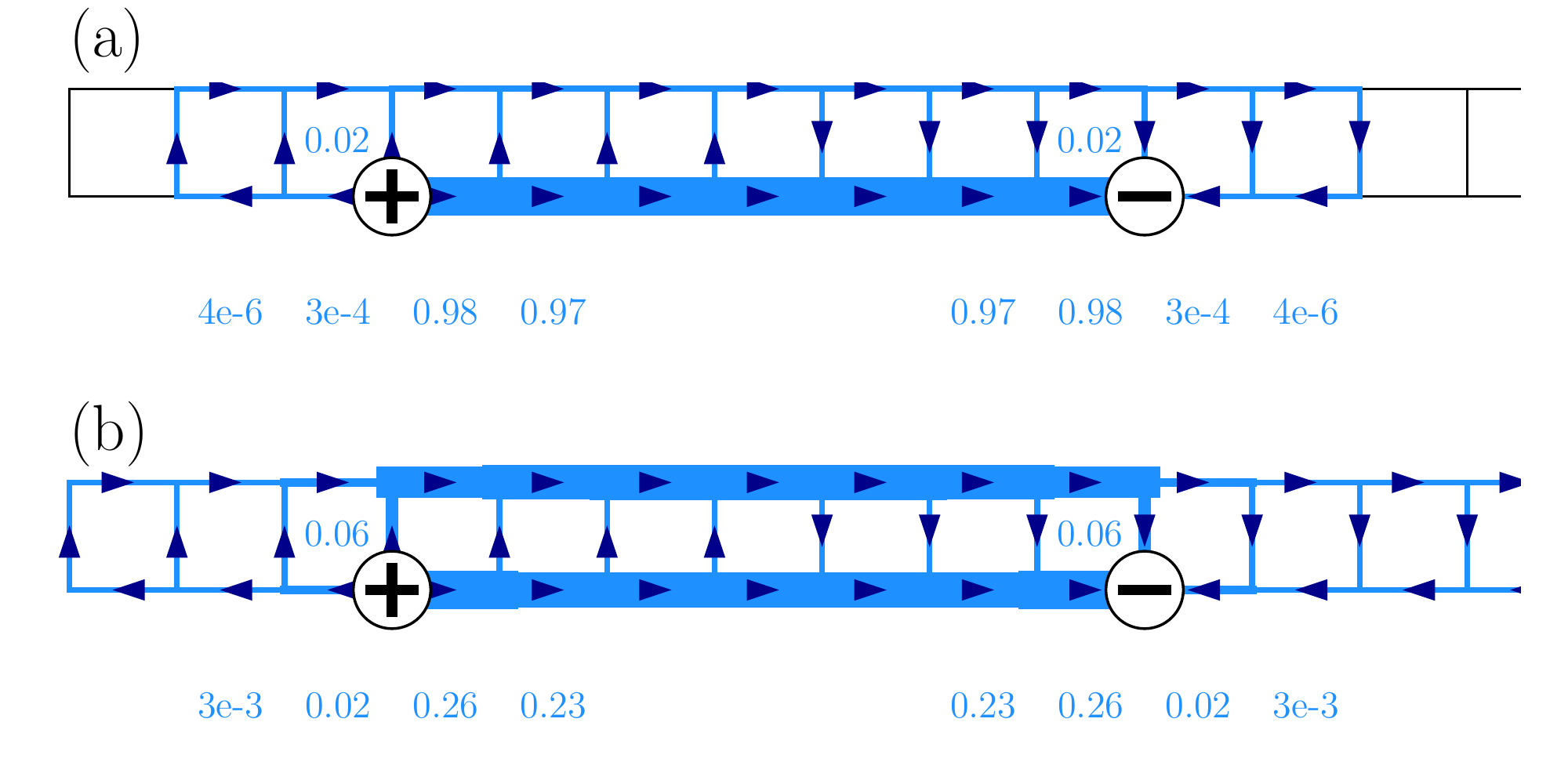}
\caption{Behavior of the electric field in the pure $\mathbb{Z}_5$ lattice gauge theory ($\lambda\to 0$) for (a) $g=1.5$ and (b) $g=0.3$ in the presence of two opposite static charges. The static charges are indicated by the circles and are introduced in the central region of a ladder of size $L=41$ with smooth boundaries. The thickness of the ladder edges indicate the expectation value of the electric field $\left\langle E\right\rangle$ in arbitrary units. Some example of their value is reported in the blue labels.
}
\label{fig:purestatic}
\end{figure}

It is interesting to inspect the behavior of the expectation values of the electric fields, $E=-i ({N}/{2\pi}) \log \tau$, in the presence of the two static charges. No phase transition characterize the pure lattice gauge theory at finite $g$; however, we can distinguish two different regimes. For large values of $g$, when the string tension is strong, the electric field propagates in a straight line from one static charge to the opposite when the two of them are on the same leg [Fig. \ref{fig:purestatic} (a)]. For larger and larger $g$, indeed, the ground state progressively becomes a product state of the $\tau$ eigenstates in all the links and the energy cost of prolonging the electric field lines from one leg to the other is excessive. For small $g$, instead, the expectation value of $\tau$ decreases in modulo. The average electric field, however, propagates in both legs in the region between the two static charges [Fig. \ref{fig:purestatic} (b)], as dictated by the strong plaquette interactions.

\section{An effective field theory description} \label{sec:full}

\subsection{The quantum clock model limit}

\begin{figure}
\includegraphics[width=\columnwidth]{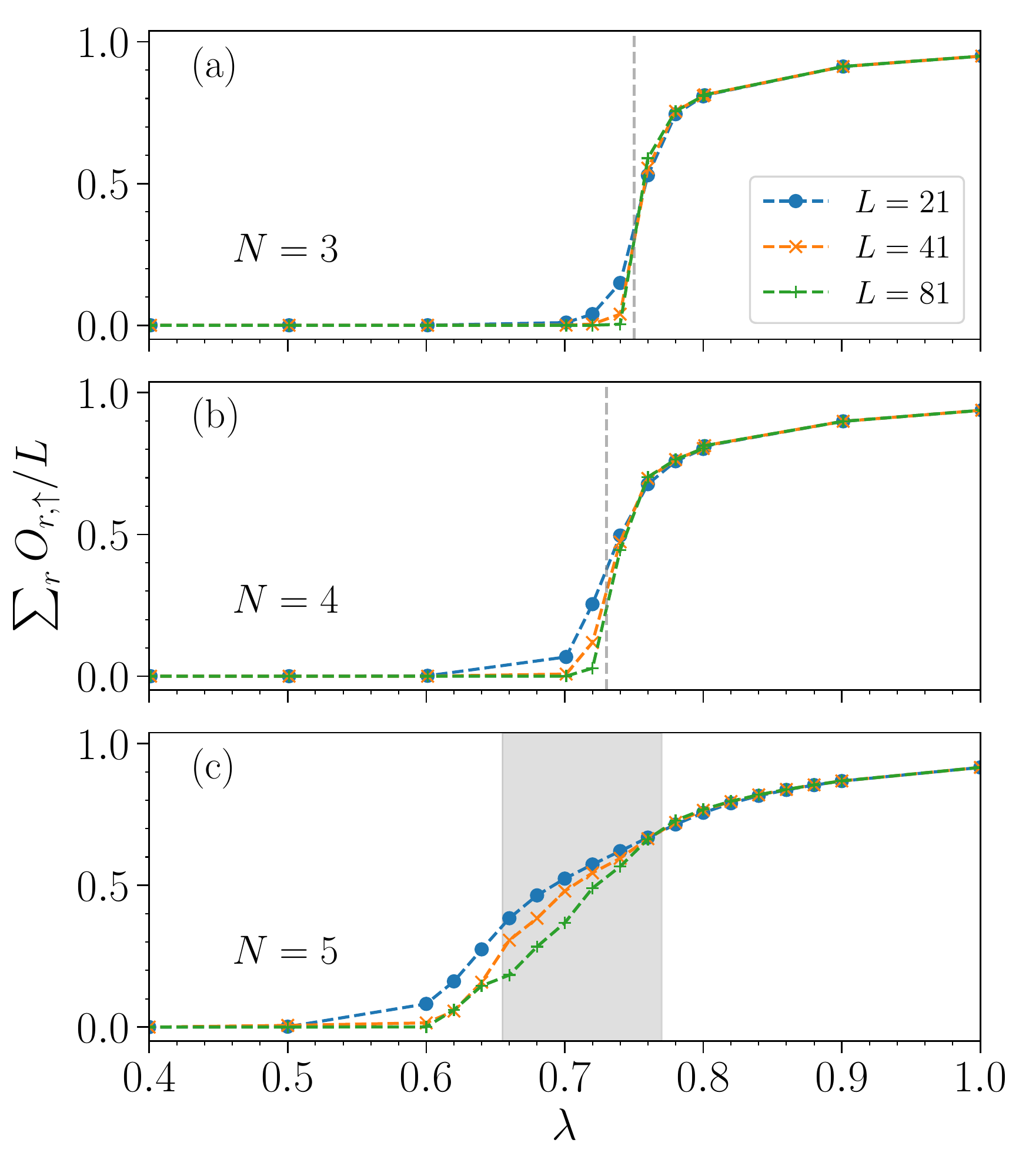}
\caption{
Averaged order parameters $\sum_r O_{r,\Up} / L$ as functions of $\lambda$ for systems in the clock limit ($g=0$) for $N=3$, $4$ and $5$.
The system has rough and smooth boundaries at the left and the right sides respectively.
For $N=3$ and $4$, the dashed lines indicate the proposed critical values of $\lambda$.
For $N=5$, the gray region represents the gapless phase determined by the fidelity susceptibility in a system of smooth-smooth boundaries (see Fig.~\ref{fig:FS}).
}
\label{fig:order}
\end{figure}

The next step in our analysis is to examine the role of the Higgs matter degrees of freedom, in the spirit of the seminal work by Fradkin and Shenker \cite{fradkin1979}. To investigate the physics of the Hamiltonian \eqref{Hinv}, we begin our study from the limit $g\to 0$. In this limit, the gauge bosons are frozen in a state without any magnetic excitation in the plaquettes of the ladder. Therefore, it is easy to rewrite the Hamiltonian in the axial gauge by imposing all the link states to be aligned in such a way that $\sigma_{r,s} \ket{\psi} = \ket{\psi}$ \cite{Fradkin}. Hence, the Hamiltonian $H$ takes the form of a $\mathbb{Z}_N$ quantum clock model on the ladder geometry:
\begin{multline} \label{Potts}
H(g=0) = -\lambda\left[\sum_{s,r}\zeta^\dag_{r,s}\zeta_{r+1,s}+ \sum_r \zeta^\dag_{r,\Up}\zeta_{r,\Dn} + {\rm H.c.}\right] \\
-\frac{1}{\lambda} \sum_{r,s} \left(\eta_{r,s} + \eta_{r,s}^\dag\right)\,.
\end{multline}
The first line corresponds to a ferromagnetic interaction between any pair of neighboring clock operators representing the Higgs matter, along both the rungs and the legs of the ladder. The second line can be interpreted as the sum of the disorder operators in all the ladder vertices.

Quantum clock models of this kind are in general characterized by a symmetry broken ordered ``ferromagnetic'' phase for large $\lambda$ and a disordered ``paramagnetic'' phase for small values of $\lambda$. We stress that, in the lattice gauge theory, the order parameter $\zeta$ is not a well-defined gauge invariant operator. The corresponding order parameter can be written in a gauge-invariant form only when considering at least an edge with ``rough'' boundary conditions by including the boundary interaction \eqref{hambound}, which explicitly breaks the global conservation of the matter charge associated to the symmetry $\prod_{r,s} \eta_{r,s}$. 
The boundary term \eqref{hambound} allows us to introduce a gauge-invariant order parameter $O_{r,s} = \left\langle \prod_{j=1}^{r} \sigma_{j,s}^\dag \zeta_{r,s}\right\rangle$ which matches $\left\langle \zeta_{r,s}\right\rangle$ in the axial gauge.

In Fig. \ref{fig:order} we illustrate the value of this order parameter in ladders with \emph{rough-smooth} boundary conditions for $N=3$, $4$ and $5$.
One can see a phase transition between the ordered and the disordered phases for $N=3$ and $4$.
For $N=5$, an intermediate gapless phase appears as indicated in the gray region in Fig.~\ref{fig:order}(c).

The appearance of a gapless phase in this model for $N>4$ is reminiscent of the study of ferromagnetic one-dimensional quantum clock models.
 In these 1D chains, it is well known that the quantum clock model displays an extended gapless phase for $N\ge 5$ separating the gapped ordered and disordered phases (see, for example, \cite{lecheminant2002,ortiz2012,milsted2014}). The phase transitions between them are of the Berezinsky-Kosterlitz-Thouless (BKT) kind \cite{sun2019}. Similar properties characterize the ladder clock model in Eq. \eqref{Potts}, such that for $N\ge 5$ a gapless phase appears in the system for intermediate values of $\lambda$. In the following, we will investigate the properties of this gapless system by describing the low-energy sector of the theory through an effective field theory inspired by bosonization, and we will numerically examine its main features through DMRG simulations.

\subsection{Bosonization of the model}

To build an effective low-energy description of the model we construct a representation of the clock operators based on vertex operators of a pair of dual bosonic massless fields, $\theta$ and $\varphi$. The following construction matches the dual sine-Gordon model description of 1D systems with $\mathbb{Z}_N$ symmetry presented in Ref. \cite{lecheminant2002} and it is inspired by standard bosonization techniques \cite{giamarchi}. A similar strategy has also been recently applied to the study of the $\mathbb{Z}_2$ lattice gauge theory on the chain \cite{moroz2020}.

Our first step is to introduce the pairs of dual bosonic massless fields $\theta_s(x)$ and $\varphi_s(x)$, with $s=0,\Up,\Dn$ that fulfill the following commutation relations:
\begin{equation} \label{comm}
\left[\theta_{s}(x),\varphi_{s'}(x')\right]= -i \frac{2\pi}{N} \Theta\left(x-x' \right)\delta_{ss'}\,,
\end{equation}
where $\Theta$ is the Heaviside step function with $\Theta(x\ge 0)=1$ and $\Theta(x<0)=0$. Based on this commutation relations, we build the following mapping between the clock operators of the ladder model expressed in the axial gauge and the vertex operators, in such a way that the algebraic properties of the clock operators \eqref{clockpr} and \eqref{clockpr2} are satisfied. For the Higgs matter operators ($s=\Up,\Dn$) the mapping reads:
\begin{equation} \label{map1}
\zeta_{j,s} \to \e^{-i\theta_s(ja)}\,,\quad \eta_{j,s} \to \e^{-i\varphi_s(ja)+i\varphi_s(ja+a)} \,.
\end{equation}
For the gauge bosons on the rungs we analogously impose the following:
\begin{equation} \label{map2}
\sigma_{j,0} \to \e^{-i\theta_0(ja)} \,,\quad \tau_{j,0} \to \e^{-i\varphi_0(ja)+i\varphi_0(ja+a)}\,.
\end{equation}
In these relations we introduced the lattice spacing $a$, which is useful to define a proper ultraviolet cutoff of the theory. In particular, we consider the bosonic fields $\varphi$ and $\theta$ to vary slowly in space with respect to the length scale set by $a$. It is easy to verify that the previous definitions fulfill \eqref{clockpr} and \eqref{clockpr2} based on Eq. \eqref{comm} (see Appendix \ref{app:bosonization}). The physical interpretation of the bosonic fields can be deduced as well from the previous equations. The fields $\varphi_{\Up,\Dn}$ represent the electric field propagating along the legs, whereas $\theta_0$ is associated to the magnetic field flux along the rungs.

Some care is required in dealing with the boundary conditions: smooth boundary conditions on the ladder, for example, impose Dirichlet boundary constraint on $\varphi_{\Up,\Dn}$, since they require that no electric flux is allowed to enter the system from outside. Rough boundary conditions, instead, impose Dirichlet boundary constraints on $\theta_{\Up,\Dn}$. Adopting the axial gauge, we obtain the following effective Hamiltonian on the continuum (see Appendix \ref{app:bosonization} for further details):
\begin{multline} \label{boson}
H=\frac{N}{4\pi} \int \rd x \, \sum_{s=0,\Up,\Dn} v\left[K_s \left(\partial_x \varphi_s\right)^2 +\frac{1}{K_s}\left(\partial_x \theta_s\right)^2 \right]\\
 -T\int \rd x\, \cos\left(\theta_\Up-\theta_\Dn-\theta_0\right) \\
-G\int \rd x\,\left[ \cos\left(\varphi_\Up+\varphi_0\right) + \cos\left(\varphi_\Dn-\varphi_0\right)\right] \\
-\sum_{s=0,\Up,\Dn} \int \rd x \,\left[P_s \cos N \theta_s + Q_s \cos N\varphi_s  \right]\,.
\end{multline}
In this Hamiltonian, the $s=\Up,\Dn$ contributions of the first line account for the tunneling term along the legs of the ladder and the onsite term for the mass of the charges. The $s=0$ contribution of the first line describes instead both the plaquette term and the electric field interaction along the rungs. These terms of the Hamiltonian in the axial gauge can indeed be mapped into a 3-component Luttinger liquid. The mapping between clock and vertex operators suggest that, in proximity to the gapless phase, $K_\Up = K_\Dn \approx 1/\lambda$ and we expect $K_0$ to be proportional to $g$ close to $g=1$. The velocity is the same for all sectors: $v=4\pi a/N$. The second line in the Hamiltonian \eqref{boson} describes the rung tunneling term; the third line corresponds instead to the electric field interaction along the leg links, which has a non-local description in the axial gauge but recovers its locality in this description. For later convenience we labeled their coupling constants as $T$ and $G$ such that:
\begin{equation}
T= \frac{2\lambda}{a}\,,\quad G= \frac{2g}{a}\,.
\end{equation}
The final terms in the Hamiltonian \eqref{boson} are aimed at restoring the $\mathbb{Z}_N$ symmetry of the model and we will refer to them as ``background interactions''. The mapping \eqref{map1} and \eqref{map2} promote indeed the clock operators from discrete operators to continuous rotors (see, for example, \cite{sachdev18a}). The background interactions have the role of breaking the system symmetries from $U(1)$ to $\mathbb{Z}_N$, consistently with the field theoretical description of clock models \cite{lecheminant2002,delfino01,Tu2019}. The values of the constants $P_s$ and $Q_s$ can be roughly estimated by comparing the energy of the kinks of these sine-Gordon interactions with the energy of the domain walls in the corresponding operators (see Appendix \ref{app:bosonization}):
\begin{align}
&P_s= \frac{N^2 \left(1- \cos 2\pi/N\right)^2}{32aK_s} \,, \label{Peq}\\
&Q_s= \frac{N^2 K_s \left(1- \cos 2\pi/N\right)^2}{32a}\,. \label{Qeq}
\end{align}

Let us finally observe that the Hamiltonian \eqref{boson} is invariant under the global $\mathbb{Z}_N$ transformation $\theta_{\Up/\Dn} \to \theta_{\Up/\Dn} + 2\pi/N$. The fields $\varphi_{\Up,\Dn}$, instead, do not enjoy such discrete global symmetry due to the leg electric field term. The $\mathbb{Z}_N$ transformation $\varphi_{\Up/\Dn} \to \varphi_{\Up/\Dn} + 2\pi/N$ corresponds to the addition of a background electric field, thus to a change of the $\uptheta$ vacuum of the theory. This is analogous to similar features in truncated 1D models with U(1) gauge symmetry (see, for example \cite{magnifico19,magnifico19b,funcke20,surace20}).

\subsection{Properties of the clock model limit}

We can obtain the main features of the system in the limit $g \to 0$ by considering the Hamiltonian \eqref{boson}. In this limit $ \theta_0=0$ everywhere ($\sigma_{r,0}=1$), consistently with the Hamiltonian \eqref{Potts}, and the electric field term disappears. We are effectively left with a two-component Luttinger liquid perturbed by the interleg tunneling and the background interactions. As customary in these cases \cite{giamarchi}, it is convenient to separate symmetric ``charge'' $\rho$ and antisymmetric ``spin'' $\sigma$ combinations of the fields (the terms ``charge'' and ``spin'' are taken from the study of one dimensional two-component fermionic systems):
\begin{align}
&\varphi_\rho = \frac{\varphi_\Up + \varphi_\Dn}{\sqrt{2}}\,,\quad \theta_\rho = \frac{\theta_\Up + \theta_\Dn}{\sqrt{2}} \,,\\
&\varphi_\sigma = \frac{\varphi_\Up - \varphi_\Dn}{\sqrt{2}}\,,\quad \theta_\sigma = \frac{\theta_\Up - \theta_\Dn}{\sqrt{2}}\,.
\end{align}
The effective bosonized Hamiltonian reads:
\begin{multline} \label{boson2}
H(g=0)=\\
\frac{N}{4\pi} \int \rd x \, \sum_{q=\rho,\sigma} v\left[K_q \left(\partial_x \varphi_q\right)^2 +\frac{1}{K_q}\left(\partial_x \theta_q\right)^2 \right]\\
 -T\int \rd x\, \cos\left(\sqrt{2}\theta_\sigma \right) \\
-2\int \rd x \,\left[P \cos\frac{N\theta_\rho}{\sqrt{2}}  \cos\frac{N\theta_\sigma}{\sqrt{2}}  + Q \cos\frac{N\varphi_\rho}{\sqrt{2}}  \cos\frac{N\varphi_\sigma}{\sqrt{2}}   \right].
\end{multline}
A first-order renormalization group analysis shows that, for $N=2$, the system at $g=0$ displays, as expected, two gapped phases separated by a critical point. For $N>2$, the competition between the rung tunneling term and the background $Q$ term is non-trivial and yields the possibility of having a phase in which only the spin sector is gapped. This can be understood by comparing the scaling dimensions of the interactions.
The scaling dimension of the rung tunneling is $D_T = K_\sigma/N$, whereas the scaling dimensions of the background interactions are $D_P = (K_\rho+K_\sigma)N/4$ and $D_Q= (K_\sigma^{-1} + K^{-1}_\rho)N/4$. In particular, the bare value of the Luttinger parameters matches $K_\sigma=K_\rho \approx 1/\lambda$ and we will label it with $K$ only. Based on the previous scaling dimension, for $K \in \left(4/N,N/\sqrt{2}\right)$, the rung tunneling $T$ term is the dominating interactions, with the $P$ interaction being irrelevant and the $Q$ interaction being suppressed by $T$. Therefore $T$ gaps the spin sector of the system. 

Moreover, a two-step renormalization group analysis (see Appendix \ref{app:mpd}) shows that additional emergent second-order terms gap also the charge sector for all values of $\lambda$ in the cases $N=3,4$; only for $N>4$ an extended gapless phase appears in the phase diagram, characterized by a gapless charge sector (see the next section and Appendix \ref{app:2nd}). 

In general, for small $\lambda$ (thus large $K$) the background interaction $Q$ dominates; the fields $\varphi$ are semiclassically pinned to one of their minima and the system is in a disordered phase of the corresponding clock model. For large $\lambda$, instead, the tunneling and background $P$ terms dominate and the system is in an ordered state. 

Let us discuss next the behavior of the system as a function of $N$. For $N=2$ and $3$, we expect a single critical point to separate these phases. This critical point falls in the Ising and Potts universality class respectively.

For $N=4$, the limit $g=0$ can be examined through the mapping of the $\mathbb{Z}_4$ quantum clock model into two separate copies of the Ising model in the same geometry \cite{ortiz2012} (see Appendix \ref{app:4} for more detail). Therefore, also in this case, the system behaves as its 1D counterpart with only two gapped phases separated by a single critical point corresponding to two copies of the Ising critical point.

We additionally observe from our numerical results that for $N=2,3,4$, the critical value of $\lambda_c$ is smaller than 1, due to the presence of the rung tunneling interaction. This is a signature that, indeed, our model interpolates between one and two dimensions. In particular, our rough numerical estimates for $N=3$ provide a value $\lambda_c \approx 0.75$ based on the expectation value of the order parameter $\left\langle O_{r,s} \right\rangle$. In the 1D clock model the critical value is $\lambda_c^{(1)}=1$, whereas in the 2D system the paramagnetic and ferromagnetic phases are separated by a first-order phase transition for $\lambda_c^{(2)} \approx 0.498$ calculated with tensor network techniques \cite{cirac2020,ciracnote}; therefore $\lambda_c^{(2)} < \lambda_c < \lambda_c^{(1)}$.

For $N=5$ (or larger), the phase diagram at $g=0$ becomes richer and the second-order renormalization group analysis confirms the existence of a gapless phase between the ordered and disordered phases [see Fig. \ref{fig:pd}(b)]. There is indeed a finite intermediate interval of $\lambda$, thus of the bare Luttinger parameter, such that the spin sector of the Hamiltonian \eqref{boson2} is gapped by the tunneling interaction, but the charge sector remains gapless. In the next section we analyze in detail the phase diagram for $N=5$ and we verify through tensor network simulations that such gapless phase exists and extends also to finite $g$.

\section{The phase diagram and the onset of the Coulomb phase} \label{sec:gen}

\begin{figure}[tb]
\includegraphics[width=\columnwidth]{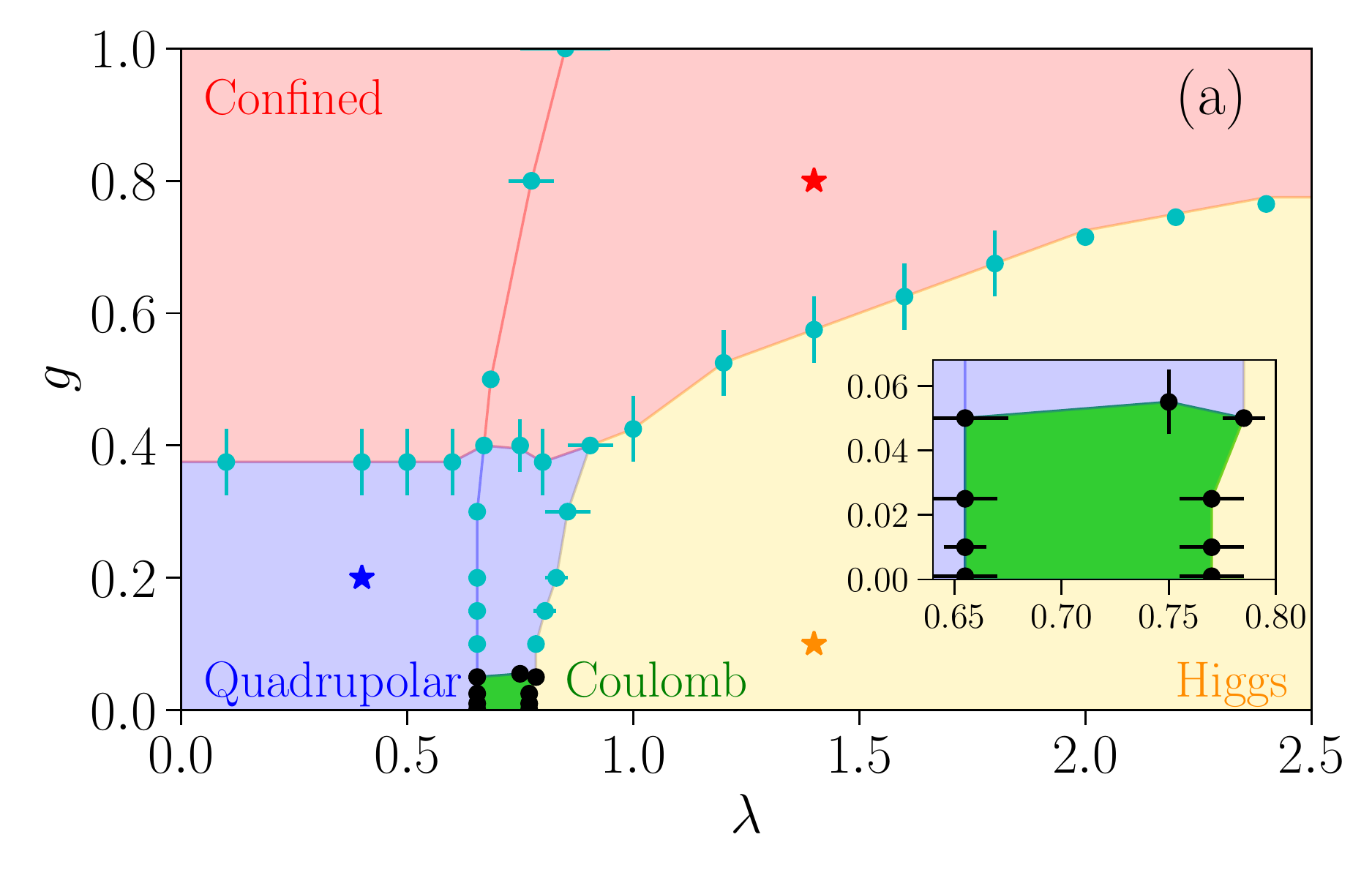}
\includegraphics[width=\columnwidth]{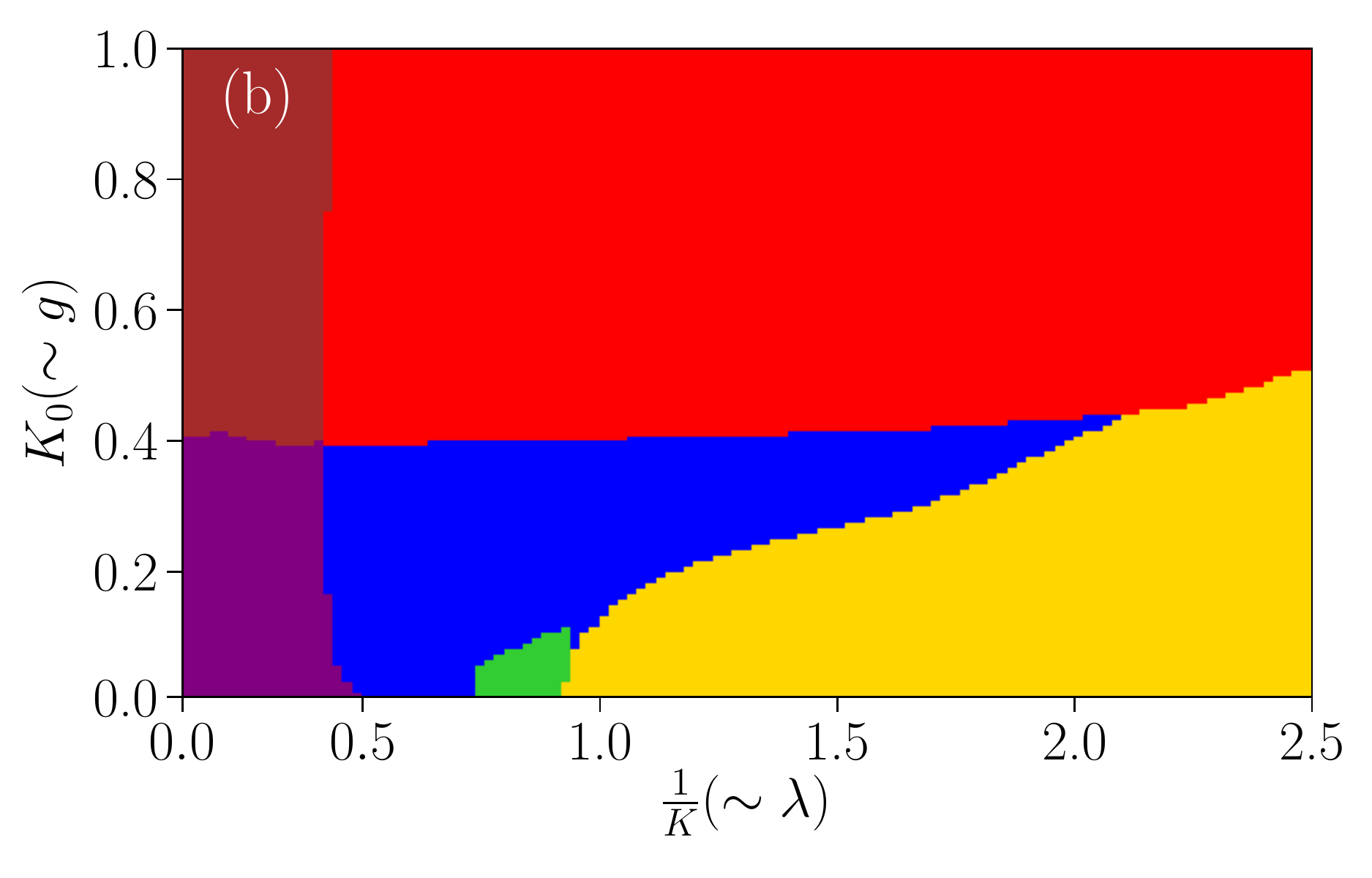}
\caption{
Phase diagram from DMRG (a) and second-order RG (b).
In panel (a), a gapless (green) and a gapped (other colors combined) phases are separated by a BKT phase transition (black dots).
The gapped phase is further distinguished into three ``phases'', the quadrupolar phase (blue), the Higgs phase (yellow), and the confined rung-dominated phase (red), separated by crossovers. The cyan dots indicate finite peaks in the fidelity susceptibility of the system.
The inset zooms in the Coulomb phase.
In panel (b), different phases are shown by different colors, including the deconfined phase (purple), the quadrupolar phase (blue), the Coulomb phase (green), the Higgs phase (yellow), the fully confined phase (brown), and the confined rung-dominated phase (red). See Table.~\ref{tab:phases} for their properties.
}
\label{fig:pd}
\end{figure}

\begin{table*}[t!]
\setlength{\tabcolsep}{0.25em}
\begin{tabular}{|c|c|>{\columncolor[RGB]{230, 230, 255}[\tabcolsep][\tabcolsep]}c|>{\columncolor[RGB]{220, 255, 220}[\tabcolsep][\tabcolsep]}c|>{\columncolor[RGB]{255, 255, 200}[\tabcolsep][\tabcolsep]}c|}
\hline
\textbf{Phases at $\bf g=0$} & Deconfined $(\lambda \to 0)$ & Quadrupolar / disordered  & Coulomb (gapless) & Higgs / ordered ($\lambda\gtrsim 1$)\\
\hline
Pinned fields & $\varphi_\sigma, \varphi_\rho$ & $\theta_\sigma, \varphi_\rho$ & $\theta_\sigma$ & $\theta_\sigma,\theta_\rho$\\
\hline
\makecell{Dominating\\ 
interactions} & $Q,C_\rho,C_\sigma$ & $T,C_\rho$ & $T$ & $P,T$\\
\hline
\makecell{Observables} & \makecell{$\left\langle \mathcal{O}_{r,s} \right\rangle = 0$ \\ Constant $\mathcal{G}_\rho$ \\ Exp. decay of $\mathcal{R}$ and mesons  } &  \tabularCenterstack{c}{$\left\langle \mathcal{O}_{r,s} \right\rangle = 0$\\ Exp. decay of $M_\rho$\\ Constant $\mathcal{G}_\rho$, $M_\sigma$ and $\mathcal{R}$} & \makecell{Alg. decay of $\mathcal{O}_s$ \\ Alg. dec. of $\mathcal{G}_\rho$ and $M_\rho$\\  Constant $M_\sigma$ and $\mathcal{R}$} & \makecell{$\left\langle \mathcal{O}_{r,s} \right\rangle \neq 0$ \\ Exp. decay of $\mathcal{G}_\rho$ \\ Constant $M_\rho,\, M_\sigma$ and $\mathcal{R}$ } \\
\hline
\end{tabular}

\vspace{1em}

\begin{tabular}{|c|c|>{\columncolor[RGB]{255, 230, 230}[\tabcolsep][\tabcolsep]}c|}
\hline
\textbf{Additional phases at $\bf g>0$} & Fully confined $(\lambda \to 0)$ & Confined rung-dominated \\
\hline
Pinned fields & $\varphi_\sigma, \varphi_\rho, \varphi_0$ & $\varphi_\rho, \, \sqrt{2}\theta_\sigma - \theta_0,\, \varphi_\sigma +\sqrt{2}\varphi_0 $ \\
\hline
\makecell{Dominating interactions} & $Q,Q_0,G,C_\rho,C_\rho',C_\sigma,C_\sigma'$ & $T,G,C_\rho,C_\rho',C_\sigma'$ \\
\hline
Observables & \makecell{Constant $\mathcal{G}_\rho $ \\ Exp. decay of $\mathcal{R}$ \\ Exp. decay of $M_\rho$ and $M_\sigma$} & \makecell{Constant $\mathcal{G}_\rho $ \\ Constant $\mathcal{R}$\\ Exp. decay of $M_\rho$ and $M_\sigma$ } \\
\hline
\end{tabular}

\caption{Summary of the regimes appearing in the phase diagram as predicted by the second-order RG flow. The first table displays the phases of the clock limit $g=0$ (in which $\theta_0$ is always pinned). The gapless Coulomb phase extends for finite $g$. The gapped phases, instead, cross over into the additional confined phases listed in the second table. Columns with white or colored background distinguish the phases of the pure LGT ($\lambda \to 0$) and the ones appearing at finite $\lambda$ respectively [the background colors match the regions in Fig. \ref{fig:pd} (a)]. 
The listed dominating interactions refer to the coupling constants in Eqs. \eqref{boson}, \eqref{boson2} and \eqref{SI}.}

\label{tab:phases}

\end{table*}

In the following, we analyze the phase diagram for $N=5$ based on density matrix renormalization group (DMRG)~\cite{PhysRevLett.69.2863,PhysRevB.48.10345,SCHOLLWOCK201196} and Wilsonian second-order renormalization group (RG).
Our results are displayed in Fig. \ref{fig:pd}.
Our bosonization description, however, does not rely on specific assumptions about $N$ and qualitatively similar results hold for $N>5$ as well, as shown in Appendix \ref{app:2nd}.

The ladder model presents a single extended gapped phase that surrounds a gapless phase appearing around $\lambda \sim 0.75$ for $g\lesssim 0.05$ (depicted in green in Fig. \ref{fig:pd}). These phases are separated by a Berezinskii-Kosterlitz-Thouless (BKT) phase transition, which can be detected by evaluating the fidelity susceptibility (FS) of the system \cite{sun2019,vekua2015}.

The gapless phase displays the properties of a Coulomb phase, characterized by an emergent $U(1)$ symmetry, in which electric fields may propagate without mass gaps along the legs of the ladder and they appear to be only weakly screened by the dynamical matter.

The gapped phase displays instead several crossovers connecting different regimes which include a confined and a Higgs regime, as typical for Abelian LGTs \cite{fradkin1979}. These crossovers are signaled by extrema in both the susceptibilities associated with the Hamiltonian terms and fidelity susceptibility. In both cases these susceptibilities do not diverge with the system size, consistently with crossovers rather than phase transitions. For simplicity, we will refer to the different gapped regimes as ``\textit{phases}'', although they are not separated phases in the thermodynamic meaning, but rather adiabatically connected regimes, as in the case of Fradkin-Shenker LGTs (in the fundamental representation for the matter degrees of freedom) \cite{fradkin1979}. 

In the following, we will discuss the origin of the different phases based on their bosonized description and we will identify the phase transitions and crossovers based on numerical simulations. Then, we will focus on the properties and identification of the thermodynamic regimes in the phase diagram which can be obtained by the analysis of the behavior of several observables over the ground states of the model, and by the screening properties that characterize the system upon the introduction of static charges. Table \ref{tab:phases} offers a summary of the thermodynamic regimes appearing in the model.

\subsection{The analysis of the $\mathbb{Z}_5$ gauge model in the clock limit}

We begin our analysis of the $\mathbb{Z}_5$ LGT from the clock limit $(g\to 0)$, which is described by the Hamiltonian \eqref{boson2}. In the previous section, we emphasized that a simple scaling analysis predicts that the competition of the background interactions $P$ and $Q$ yields the onset of two fully gapped phases, separated by an additional gapless phase for $N\ge 5$. In this critical phase, a gap is opened in the spin sector only, as effect of the rung tunneling $T$; therefore its central charge is $c=1$, and its charge sector is characterized by an emerging $U(1)$ symmetry $\left(\theta_\rho \to \theta_\rho + \alpha\right)$ in a way analogous to the one-dimensional quantum clock models \cite{ortiz2012}. We call this phase \textit{Coulomb phase}, in analogy with its higher dimensional counterparts.

According to the first-order scaling analysis of the Hamiltonian \eqref{boson2}, the three phases alternate in the following way: for $K<4/N$, $P$ dominates and the model is fully gapped; for $4/N < K< N/\sqrt{2}$, the tunneling term dominates and only the spin sector is gapped; for $K>N/\sqrt{2}$, the background $Q$ term dominates and the gap in the charge sector is restored. In these estimates we consider $K=K_\sigma=K_\rho\sim 1/\lambda$.

The $P$-dominated phase (large $\lambda$) corresponds to both the Higgs phase of the LGT and the ordered phase in the corresponding clock model. In this phase the $\theta$ fields are semiclassically pinned.

In the gapless Coulomb phase, only the field $\theta_\sigma$ is semiclassically pinned. Finally, the $Q$-dominated phase would correspond to a symmetric and disordered phase in the clock model, with the $\varphi$ fields pinned, which can be mapped into the deconfined phase of the LGT.

These simple predictions obtained with a first-order renormalization approach, however, are not sufficient to completely describe the behavior of the system. Due to the presence of non-commuting interactions, the renormalization group flow yields in general the appearance of novel effective interactions, which can be more relevant than the original terms in Eq. \eqref{boson2} and must be taken into account for a more rigorous study of the phase diagram. These emergent interactions appear naturally when considering higher orders of the sine-Gordon interactions. In particular, among the set of operators appearing in its second-order analysis, we focus on the following terms:
\begin{equation} \label{ham2g0}
H^{(2)}_{\rm int}(g=0) = \sum_{q=\rho,\sigma}\int \rd x\, C_q \cos\left( \sqrt{2} N \varphi_q\right) .
\end{equation}
In Appendix \ref{app:2nd}, their derivation is presented in detail. A key feature of the second-order interactions \eqref{ham2g0} is that the term $C_\rho$ commutes with the rung tunneling. Therefore, it allows for the appearance of a new gapped phase in which both $T$ and $C_\rho$ flow towards strong coupling, such that both $\varphi_\rho$ and $\theta_\sigma$ can be qualitatively considered pinned to a semiclassical minimum.
Therefore, for values of $\lambda$ intermediate between the deconfined (small $\lambda$) and the Coulomb phase ($\lambda \sim 0.75$), a new phase appears, which corresponds to a clock model ordered along the rungs but disordered along the two legs. We call this regime quadrupolar, since it is characterized by the condensation of pairs of mesons with opposite dipoles along the two legs (see Sec. \ref{sec:obs2}).

We additionally observe that the $C_\rho$ interaction is responsible for completely gapping the first-order gapless phase for $N=3$ and $N=4$ (see Appendix \ref{app:mpd}). For $N=5$, instead, the gapless phase is reduced at second order but it survives over an extended region (see Fig. \ref{fig:pd}). The four phases appearing in the $\mathbb{Z}_5$ clock limit (and, in general for $N\ge5$) are summarized in the first part of Table \ref{tab:phases} and their main properties are discussed in Sec. \ref{sec:obs2}.

To understand the phase diagram of the model, it is useful to compare the two panels of Fig.~\ref{fig:pd}, which depict the results of the numerical DMRG simulations [panel (a)] and the numerical solution of the second order RG equations in Appendix \ref{app:2nd} [panel (b)]. The DMRG results clearly provide a more rigorous scenario, although our numerical analysis is based on finite system sizes and an accurate scaling analysis of the phase boundaries is beyond the scope of this work. The RG results, instead, give an insight on the thermodynamical behavior of the system and provide a useful reference to compare its different regimes. They can be easily extended to $N>5$ (see Appendix \ref{app:2nd}), but they suffer from several approximations adopted in the bosonization procedures and their results are not reliable in the extreme regions $\lambda \ll 1$ and $\lambda \gg 1$. 

The numerical solution of the RG flow equations derived from the Hamiltonian \eqref{boson2} determine the $g=0$ axes of the phase diagram in Fig.~\ref{fig:pd}. The so-obtained phase diagram displays all four phases (deconfined, quadrupolar, Coulomb, Higgs) in its lowest part. 
We observe that both the DMRG simulations and the second-order RG equations confirm the appearance of the gapless Coulomb phase between the disordered and Higgs phases. The main difference between the two approaches in the limit $g\to 0$ is about the deconfined phase. The DMRG results present a smooth behavior of the low $\lambda$ region compatible with the quadrupolar phase extending everywhere but in the limit $\lambda=g=0$. The second-order RG calculation would instead indicate the onset of an extended deconfined phase for small $g$ and $\lambda$. 

This difference is due to a limitation of our second-order RG approach, which relies on the assumption that the bare values of the Luttinger parameters are given by $1/\lambda$. This assumption is realistic only for $\lambda$ in a neighborhood of 1, but it is likely that the bare values of $K_s$ do not diverge for $\lambda \to 0$. Taking into account this limitation, the region on the extreme left of the phase diagram \ref{fig:pd} (b) must be considered non-physical and, similarly to what we discussed for the pure LGT case, the deconfined phase shrinks to the single point $g=\lambda=0$.

In order to understand the extension of the phase diagram for $g>0$, in the following we present numerical tensor network simulations of the model and we compare them with the bosonization predictions provided by the Hamiltonian \eqref{boson}.

\subsection{DMRG phase diagram}

In Fig.~\ref{fig:pd}(a) we show the phase diagram obtained from DMRG simulations with bond dimension up to 300. Our results confirm the presence of two extended phases: the gapless Coulomb phase appearing for intermediate values of $\lambda$ and small $g$ [green area in Fig.~\ref{fig:pd}(a)], and the surrounding gapped phase, which is depicted in blue, red and yellow in Fig. ~\ref{fig:pd}(a), to distinguish the quadrupolar, confined rung-dominated and Higgs regimes respectively.
The Coulomb phase is separated from the gapped phase by BKT phase transitions, whereas the gapped regimes are adiabatically connected by smooth crossovers. 

The typical DMRG truncation errors are about $10^{-9}$ around the BKT transition points, and are smaller than $10^{-10}$ in other region including the gapped and gapless phases.

\begin{figure}
\includegraphics[width=1\columnwidth]{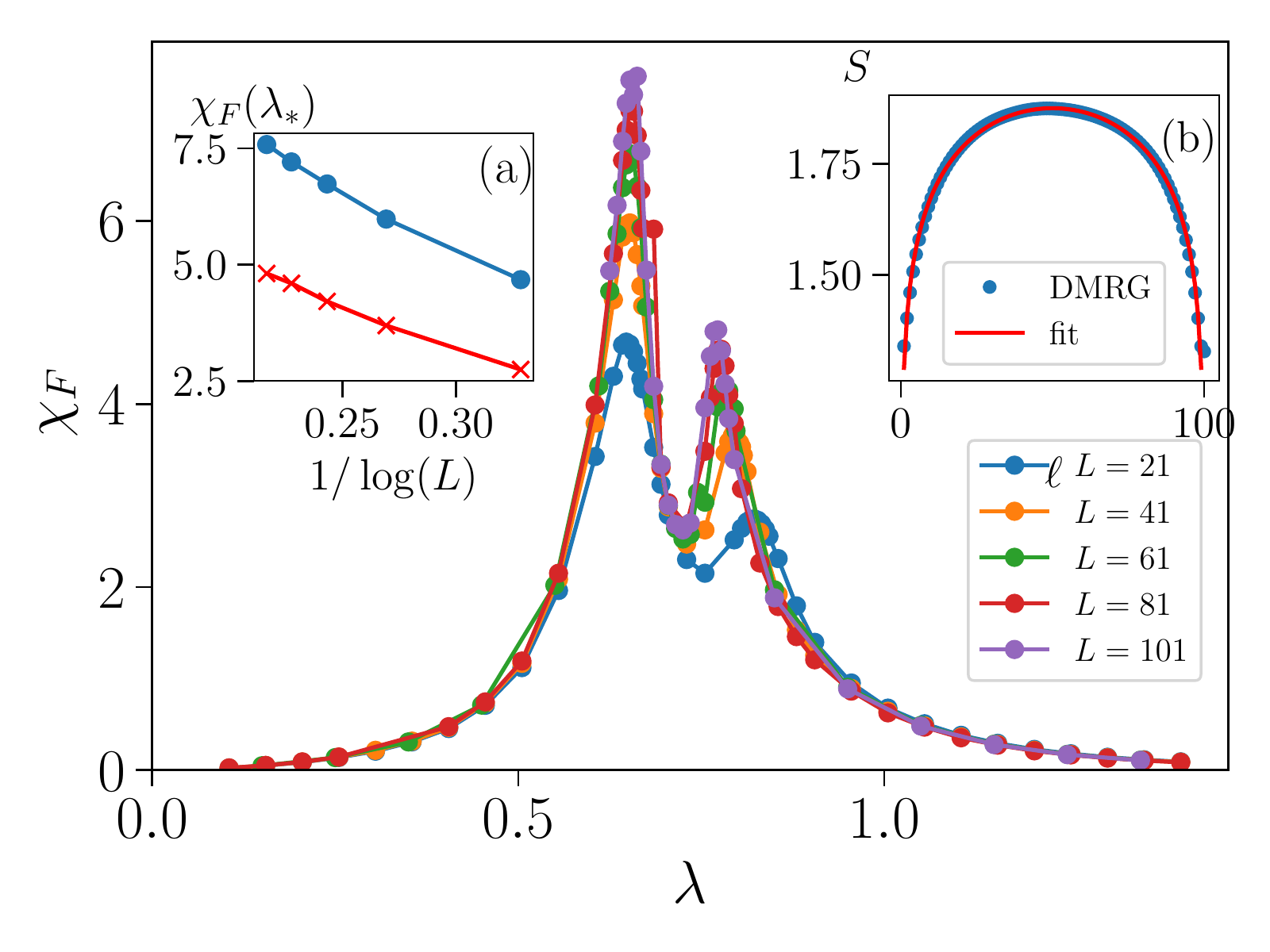}
\caption{
Fidelity susceptibility per link $\chi_F$ as a function of $\lambda$ for systems in the clock limit ($g=0$) with smooth-smooth boundary conditions.
(a) Finite size scaling of the peak values of $\chi$ for the left (blue) peak and the right (red) peak.
(b) Entanglement entropy as a function of $\ell$ for $\lambda=0.75$ (in the gapless phase), for $L=101$. The blue dots are from DMRG and the red curve is the fitting by the Calabrese and Cardy formula.
}
\label{fig:FS}
\end{figure}

The gapless Coulomb phase is the only one characterized by a logarithmic growth of the entanglement entropy as a function of the subsystem size $\ell$; therefore, it can be easily distinguished from the others. The entanglement entropy of the system in this phase follows indeed the Calabrese and Cardy formula~\cite{Calabrese2004}
\begin{equation}
S_\ell = \frac{c}{6}\log\left(\frac{2L}{\pi} \sin \frac{\pi\ell}{L}\right)+\frac{c_\alpha}{2},
\end{equation}
where $c$ is central charge of the underlying conformal field theory and $c_\alpha$ is a non-universal constant.
In Fig.~\ref{fig:FS}(b) we show the entanglement entropy as a function of $\ell$ for $\lambda=0.75$ in the gapless phase.
The central charge from the fitting is $1.02$ which is consistent with the bosonization prediction $c=1$, confirming that only one of the Luttinger liquid sectors remains gapless.

To verify that the phase transitions between the gapless and gapped phases are of the BKT kind, we analyze the behavior of fidelity susceptibility (FS).
The FS per link is defined by
\begin{equation}
    \chi_F = \frac{1}{3L} \lim_{\delta\Lambda\to 0} \frac{-2\log\left(F(\Lambda,\Lambda+\delta\Lambda)\right)}{(\delta\Lambda)^2},
\end{equation}
where the fidelity $F(\Lambda,\Lambda+\delta\Lambda)=|\langle\psi_0(\Lambda)|\psi_0(\Lambda+\delta\Lambda)\rangle|$ is defined as the overlap between two ground states of Hamiltonians $H(\Lambda)$ and $H(\Lambda+\delta\Lambda)$, with $\Lambda$ any parameter in the Hamiltonian.
The susceptibility $\chi_F$ remains finite across a BKT phase transition and approaches its thermodynamic value with a characteristic $1/\log(L)$ dependence~\cite{sun2019,vekua2015}, which distinguishes BKT from other phase transitions.
In Fig.~\ref{fig:FS} we show $\chi_F$ as a function of $\lambda$ for different system sizes. The two peaks correspond to the transitions between the gapless Coulomb and the gapped quadrupolar and Higgs regimes. The BKT transition points are determined by these peaks and we verify in Fig.~\ref{fig:FS}(a) that their finite-size scaling approximately follows the predicted logarithmic behavior.

\begin{figure}
\includegraphics[width=1\columnwidth]{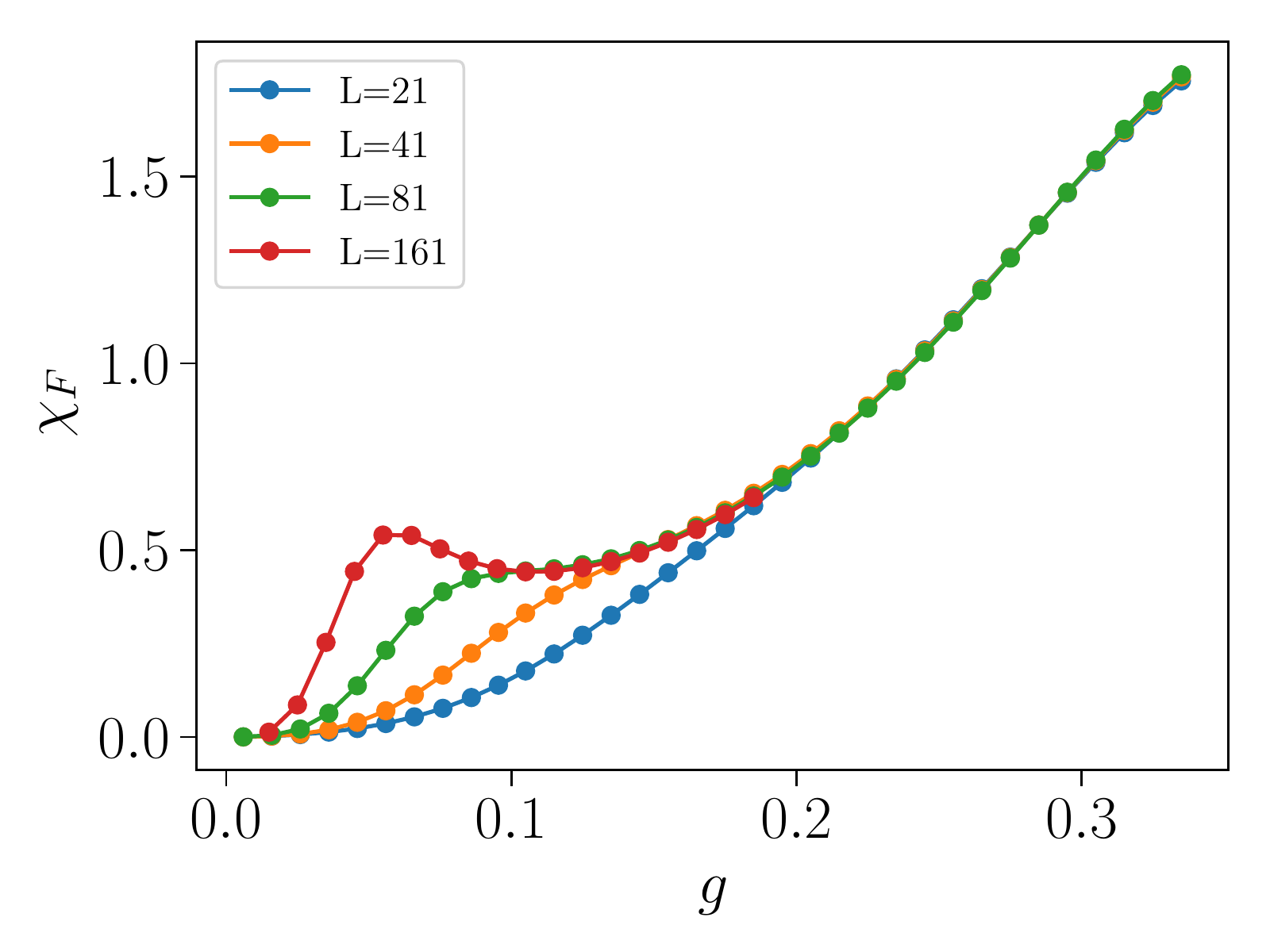}
\caption{
Fidelity susceptibility per link $\chi_F$ as a function of $g$ for $\lambda=0.75$ for systems with rough-smooth boundary conditions.
The peak in $\chi_F$ indicates a transition between the gapless and the gapped phases.
}
\label{fig:FS_g}
\end{figure}

We observe that the critical FS grows very weakly with the system size when varying the parameter $g$ (see Fig.~\ref{fig:FS_g}).
To characterize the transition driven by $g$, we consider a system with rough (smooth) boundary condition at the left (right) boundary,
where the critical FS grows faster than the case with smooth-smooth boundaries.
Fig.~\ref{fig:FS_g} shows the FS as a function of $g$ for $\lambda=0.75$.
Based on the system sizes we can numerically access ($L\lesssim 161$), the data display the evolution of the FS towards the formation of a peak for $g\gtrsim 0.05$, without the possibility of a well-defined system size scaling. Given this difficulty, in the definition of the phase diagram in Fig. \ref{fig:pd} (a), we set the critical value of $g$ by the position of the maximum for $L=161$, and the upper edge of the Coulomb phase must be considered a tentative line.

\begin{figure}
\includegraphics[width=1\columnwidth]{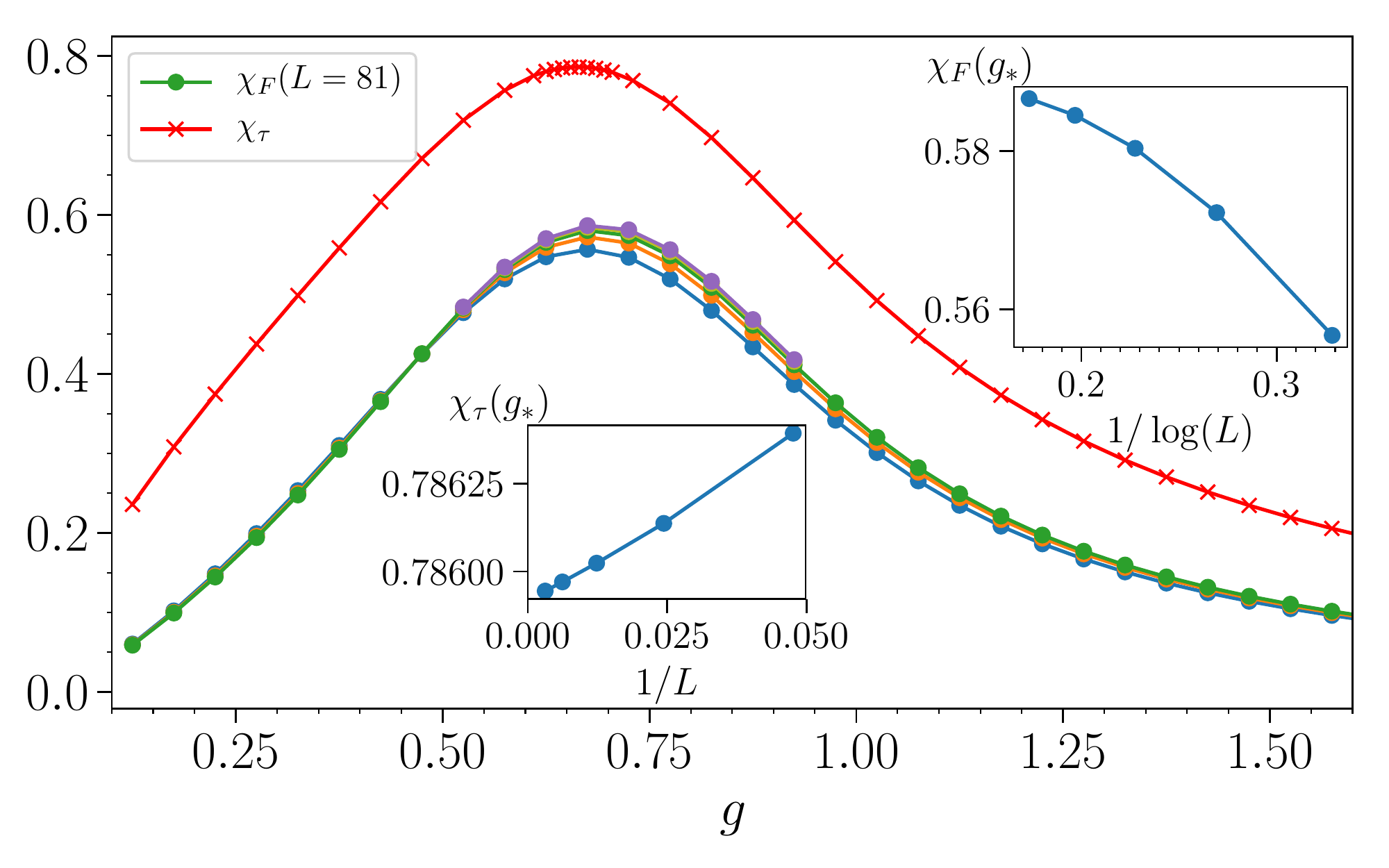}
\caption{
Fidelity susceptibility (dots), $\chi_F$, and the susceptibility of $\tau$ (crosses), $\chi_\tau$, as functions of $g$ for $L=21$, $41$, $81$, $161$ and $321$, for $\lambda=1.8$.
$\chi_\tau$ of different $L$ are indistinguishable in the scale of the figure.
The insets shows the finite-size analysis of the peak values of $\chi_F$ and $\chi_\tau$.
}
\label{fig:suscept}
\end{figure}

The fidelity susceptibility across the gapped phase displays additional maxima, characterized by curves smoother than the BKT behavior (compare Fig. \ref{fig:suscept} with Fig. \ref{fig:FS}). These maxima typically appear in correspondence with the maxima of other susceptibilities of the system and, in particular, we consider the susceptibilities $\chi_\tau$ and $\chi_\sigma$, defined by:
\begin{equation}
\chi_\tau=\frac{1}{3L}\frac{\partial \left\langle H_\tau\right\rangle}{\partial g}\,,\qquad \chi_\sigma=\frac{1}{3L}\frac{\partial \left\langle H_\mathrm{tunnel}\right\rangle}{\partial \lambda}\,,
\end{equation}
 where 
\begin{align}
H_\tau =& \sum_{s=\Up,\Dn,0}\sum_{r=1}^{L} \left(\tau_{r,s} +\tau^\dag_{r,s}\right)\,, \\
H_\mathrm{tunnel} =& \sum_{s=\Up,\Dn}\sum_{r=1}^{L-1}\zeta^\dag_{r,s}\sigma^\dag_{r+1,s}\zeta_{r+1,s}\\
&\;+ \sum_{r=1}^{L}\zeta^\dag_{r,\Up}\sigma_{r,0}\zeta_{r,\Dn} + {\rm H.c.}\,,
\end{align}
correspond to the second and the fourth terms (proportional to $g$ and $\lambda$) in the Hamiltonian (Eq.~\ref{Hinv}).
The maxima of the susceptibilities $\chi_F$, $\chi_\tau$ ($\chi_\sigma$) appear at the same value of $g$ ($\lambda$) when scanning $\lambda$ ($g$).
We present an example in Fig.~\ref{fig:suscept}, where $\chi_F$ and $\chi_\tau$ are shown as functions of $g$ for $\lambda=1.8$, thus across the crossover between the Higgs and confined (rung-dominated) regimes.
The maxima of $\chi_F$ and $\chi_\tau$ indicate a crossover at $g\approx 0.67$.
These susceptibilities, however, do not diverge when increasing the system size.
The insets of Fig.~\ref{fig:suscept} show the system size dependence of $\chi_F$ and $\chi_\tau$: both susceptibilities clearly converge to a finite value for larger and larger system sizes (see the insets in Fig. \ref{fig:suscept}).

As already mentioned, $\chi_F$ does not diverge even at a BKT phase transitions. However, by comparing its behavior in the insets of Fig. \ref{fig:FS} and Fig. \ref{fig:suscept}, we observe that in the crossovers within the gapped phase, $\chi_F$ grows even slower with the system size. 

Within the gapped phase, we can distinguish three main crossovers that correspond with the boundaries between the Higgs (yellow), quadrupolar (blue) and confined rung-dominated (red) regimes in Fig. \ref{fig:pd} (a). In these cases we observe clear maxima in all the relevant susceptibilities and, based on the observables we will introduce in the next subsections, we can distinguish these three phases consistently with the properties listed in Table \ref{tab:phases}.

In particular, the crossover between the Higgs and confined regime is qualitatively the same with respect to the analogous crossover in the Fradkin and Shenker LGTs in higher dimensions \cite{fradkin1979}. As a function of $g$ we find not only a maximum of $\chi_F$ and $\chi_g$, but also a maximum in the susceptibilities associated with the plaquette energy and other observables. Analogously to other LGTs with Higgs matter \cite{bertle2004,wenzel2005,nussinov2005}, these features suggest that such crossover can be a Kert\'esz line, namely a percolation phase transition in a corresponding two-dimensional classical model at finite temperature, which is not accompanied by any singularity in the thermodynamic properties of the system.

Finally, let us mention that inside the confined and quadrupolar phases, there appears an additional line of maxima of $\chi_F$ (associated with either very weak local maxima or inflection points of $\chi_\lambda$). We did not consider this line to be an additional crossover because of the fundamentally equal behavior of the system for $\lambda$ larger and smaller than these maxima. These maxima of $\chi_F$, in particular, seem to be associated with a prolongation within the gapped phase of the BKT phase transition separating the quadrupolar and Coulomb phases [see Fig. \ref{fig:pd} (a)].

\subsection{The observables of the system} \label{sec:obs1}

The results from the RG analysis of the low-energy bosonized description of the system \eqref{boson} do not allow us to clearly distinguish between crossovers and phase transitions within the gapped phases. However they confirm that the gapless Coulomb phase survives at finite values of $g$. In the following we summarize the main results from the complete study of second-order RG equations derived from the Hamiltonian \eqref{boson}. The detail of their derivation is presented in Appendix \ref{app:2nd}.

When considering $g>0$ the two fundamental interactions which determine the properties of the system are the rung tunneling and the electric field energy. We can express them in the following form:
\begin{multline} \label{TG}
-T \int \rd x\, \cos \left(\sqrt{2} \theta_\sigma -\theta_0\right) \\
-G \int \rd x\, \left[\cos \left(\frac{\varphi_\rho}{\sqrt{2}} + \frac{\varphi_\sigma}{\sqrt{2}} +\varphi_0 \right) \right.+\\ \left. \cos \left(\frac{\varphi_\rho}{\sqrt{2}} - \frac{\varphi_\sigma}{\sqrt{2}} -\varphi_0 \right)\right]\,.
\end{multline}
These interactions commute and can be simultaneously minimized. When relevant, they tend to pin the combinations of fields $\sqrt{2} \theta_\sigma -\theta_0$, $\frac{\varphi_\sigma}{\sqrt{2}} +\varphi_0$ and $\varphi_\rho$.

The solution of the second-order RG equations shows the onset of the two additional confined phases listed in the second part of Table \ref{tab:phases} for $g \gtrsim 0.4$ [red and brown phases in Fig. \ref{fig:pd} (b)]. The fully confined phase appears only when the electric field and background $Q$ terms dominate, thus pinning all the $\varphi$ fields. This implies, in particular, that the rung clock degrees of freedom are in a disordered state and, based on our DMRG results, this happens only in the $\lambda=0$ limit (for any value of $g>0$). As soon as $\lambda >0$, indeed, the numerical results display ordered rung operators ($\sigma_0$ in the unitary gauge).

The different regimes appearing in the phase diagram can be characterized and distinguished based on suitable observables calculated over the ground state of the system. Therefore, in the following, we introduce several gauge-invariant correlations and string-order parameters that constitute a diagnostic toolbox to characterize the thermodynamic phases of the model, in order to be able to compare the field-theory prediction with the numerical results.

To investigate the clock limit of the system, we introduced the order parameter $\mathcal{O}_{r,s}$ which allows us to distinguish the phases for $g\to 0$. This string-order parameter can be extended to define the creation operators of the mesons in the system:
\begin{equation} \label{meson}
M_{s}(x,y) \equiv \zeta_{x,s} \left(\prod_{j=x}^{y-1} \sigma_{j,s}\right) \zeta^\dag_{y,s} \rightarrow \e^{i\left[\theta_s(y) - \theta_s(x)\right]}\,.
\end{equation}
The operator $M_s$ introduces a meson lying on the $s$ leg of the system by introducing opposite dynamical charges on the sites $x$ and $y$ linked by an electric flux line. By considering the axial gauge, it is straightforward to derive the bosonized description of these string operators in the right hand side of Eq. \eqref{meson} through the mapping \eqref{map1}.

Our former analysis of the low-energy sector of the model displayed a separation between spin and charge degrees of freedom. To this purpose it is useful to introduce also the following combination of the meson operators:
\begin{align}
 M_\sigma(x,y) &\equiv M_{\Up}(x,y) M_{\Dn}^\dag(x,y) \rightarrow  \e^{i\sqrt{2}\left[\theta_\sigma(y) - \theta_\sigma(x)\right]}\,, \label{msigma}\\
 M_\rho(x,y) &\equiv M_{\Up}(x,y) M_{\Dn}(x,y) \rightarrow \e^{i\sqrt{2}\left[\theta_\rho(y) - \theta_\rho(x)\right]}\,,
\end{align}
where we specified their explicit form in terms of the bosonic fields. $M_\sigma$ creates a pair of opposite dipoles on the rungs $x$ and $y$, which are connected by two opposite electric flux lines lying on the two legs of the ladder. Due to this configuration, the resulting doubled meson presents a vanishing total electric dipole, but a non-vanishing electric quadrupole. In the case of $M_\rho$, instead, two parallel mesons are created with two negative charges in the rung $x$ and two positive charges in the rung $y$, connected by parallel electric fluxes along the legs of the ladder.

The meson strings are mostly useful to investigate the properties of the matter in the system: depending on their behavior at large space separation $x-y$ one can distinguish phases in which the mesons condense (for example the Higgs phase), and phases in which the dynamical matter is screened or confined. In this respect, the quadrupolar phase is characterized by a condensation of the spin meson $M_\sigma$ accompanied by an exponential decay of the the charge meson $M_\rho$.

We introduce next another family of observables which describes on one side the charge fluctuations of the system, and, on the other, the behavior of the electric field. We call these string operators t'Hooft operators since they are a generalization of the t'Hooft string $\mathcal{G}$. In particular, we define them for a system displaying smooth boundary conditions on the right edge and rough boundary conditions on the left as in Fig. \ref{fig:ladder} (the extension to smooth boundaries on both sides is straightforward):
\begin{align}
\mathcal{G}_{\Up}(r) &\equiv \prod_{j = r}^L \eta_{\Up,r}^\dag = \tau_{r,\Up} \prod_{j = r}^L \tau_{j,0} \rightarrow \e^{i\varphi_\Up(r)}\, ,\\ 
\mathcal{G}_{\Dn}(r) &\equiv \prod_{j = r}^L \eta_{\Dn,r}^\dag = \tau_{r,\Dn} \prod_{j = r}^L \tau_{j,0}^\dag \rightarrow \e^{i\varphi_\Dn(r)}\,.
\end{align} 
The string operator $\mathcal{G}_s$ corresponds to the exponential of the total charge laying on the leg $s$ between the $r^{\rm th}$ site and the end of the ladder. Through the Gauss law, it also corresponds to the exponential of the total electric field generated generated by these charges. It can also be considered an operator moving a magnetic flux from the right edge, along all the plaquettes of the ladder until the $r^{\rm th}$ rung and then pushing it out of the ladder  through the operator $\tau_{r,s}$. Having smooth boundary conditions on the right edge (see the boundary conditions in Appendix \ref{app:bosonization}), the t'Hooft operators $\mathcal{G}_s(r)$ additionally exemplify the physical meaning of the fields $\varphi_s$.

In analogy with the meson strings (which can be considered the dual operators of the t'Hooft strings), it is convenient to define the following t'Hooft operators addressing the spin and charge sectors:
\begin{align}
\mathcal{G}_{\sigma}(r) &\equiv \mathcal{G}_{\Up}(r)\mathcal{G}_{\Dn}^\dag(r) = \tau_{r,\Up}\tau^\dag_{r,\Dn} \prod_{j=r}^L \tau^2_{j,0} \rightarrow \e^{i\sqrt{2} \varphi_\sigma(r)}\,, \label{Gsigma}\\
\mathcal{G}_{\rho}(r) &\equiv \mathcal{G}_{\Up}(r)\mathcal{G}_{\Dn}(r) = \tau_{r,\Up} \tau_{r,\Dn} \rightarrow  \e^{i\sqrt{2} \varphi_\rho(r)}\,.\label{Grho} 
\end{align} 
We observe, in particular, that the operator $\mathcal{G}_{\rho}(r) = \e^{-i\frac{2\pi}{N}\mathcal{Q}_r}$ is local (due to the smooth right boundary conditions we have chosen) and measures the charge $\mathcal{Q}_r$ of all the matter sites on the right of the $r^{\rm th}$ links. $\mathcal{G}_{\sigma}$ is instead related to the charge differences between the two legs.

Finally we introduce the two-point correlation function of the rung tunneling operators:
\begin{multline} \label{eq:R}
R(x,y) = \zeta^\dag_{x,\Up}\sigma_{x,0}\zeta_{x,\Dn} \zeta_{y,\Up}\sigma^\dag_{y,0}\zeta^\dag_{y,\Dn} \\
\rightarrow \e^{i\left(\sqrt{2}\theta_\sigma -\theta_0\right)(x) - i\left(\sqrt{2}\theta_\sigma -\theta_0\right)(y)}\,.
\end{multline}
Its expectation value gives us information about the gap opened by the rung tunneling term, corresponding to the $T$ interaction in Eq.\eqref{boson} and it provides a direct evidence of the fact that both the fully confined and deconfined phases do not extend for finite values of $\lambda$.

\subsection{Features of the thermodynamic phases} \label{sec:obs2}

After defining the gauge-invariant observables of the system, we can proceed with the study of the phases and regimes we observe in the numerical simulation, summarized in the phase diagram Fig. \ref{fig:pd} (a). In particular, our main findings are that for small values of $g$, the Coulomb phase is stable for $g\lesssim 0.05$, whereas both the Higgs and quadrupolar phase smoothly cross over towards the confined-rung dominated regime.  

\begin{figure}
\includegraphics[width=1\columnwidth]{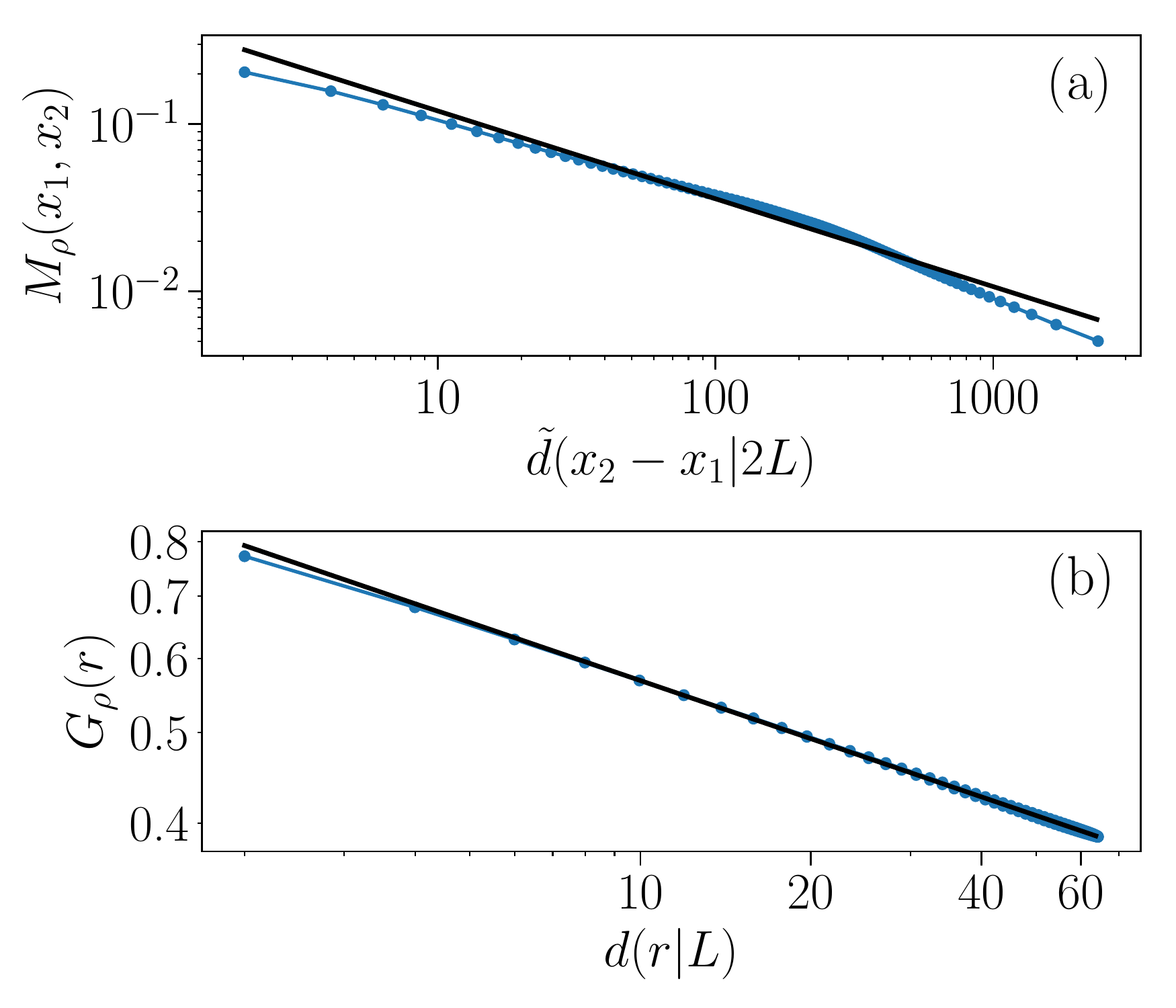}
\caption{
$M_\rho(x_1,x_2)$ and $G_\rho(r)$ in the gapless Coulomb phase as functions of $\tilde{d}(x_1-x_2|2L)$ (with $x_2=2$) and $d(r|L)$ in log-log scale.
The system is a ladder of length $L=101$ with smooth boundary conditions for $g=0.001$ and $\lambda=0.75$.
The black lines are the results of fits with a power-law decay.
}
\label{fig:gapless}
\end{figure}

From the field theoretical description, we can predict the behavior of the two-point and string correlation functions introduced above. In particular, when a bosonic field, for example $\theta_\sigma$, is semiclassically pinned by a relevant interaction, its fluctuations are suppressed and its correlation are approximately constant across the system;  the fluctuations of its dual field, $\varphi_\sigma$, are instead maximal, yielding an exponential decay of the two-point correlation functions of the related vertex operators, thus of $\mathcal{G}_\sigma$  in the example [see Eq. \eqref{Gsigma}].

Our bosonization and RG analysis predicts that the gapless Coulomb phase is characterized by a gap in the spin $\sigma$ and rung $0$ sectors, whereas the charge $\rho$ sector is gapless. This implies that the expectation value of both the meson $ M_\rho $ and the t'Hooft operator $\mathcal{G}_\rho$ decay as a power law. In Fig. \ref{fig:gapless} we display the typical behavior of these two observables in the ground state of the system within the Coulomb phase. In particular, we considered a ladder with smooth boundary conditions on both sides of the system. This implies Dirichlet boundary conditions for $\varphi_\sigma$ and Neumann boundary conditions for $\theta_\sigma$ at the two edges. Consequently one obtains the following leading behaviors:
\begin{align}
\left\langle \mathcal{G}_\rho(r) \right\rangle &\propto \left[d\left(r|L+a\right)\right]^{-\frac{1}{NK_\rho}}\,, \label{grho2}\\
\left\langle M_\rho(x,y)\right\rangle & \propto \left[\tilde{d}\left(x-y|2L\right)\right]^{-\frac{2K_\rho}{N}}\,; \label{mrho2}
\end{align}
where we introduced the chord distance $d$ and a modified chord distance $\tilde{d}$ (see \cite{cazalilla2004} for detailed calculations):
\begin{align}
d\left(r|L\right) & = \frac{L}{\pi} \left|\sin \frac{\pi r}{L}\right|\,,\\
\tilde{d}\left(x-y|2L\right) & = \frac{d\left(x+y|2L\right)d\left(x-y|2L\right)}{\sqrt{d\left(2x|2L\right)d\left(2y|2L\right)}}\,.
\end{align}
These analytical predictions are roughly compatible with the numerical results in Fig.~\ref{fig:gapless}: the t'Hooft operator $\mathcal{G}_\rho$ decays as a power law with its distance from the edges of the system and, for smooth boundary conditions, it is symmetric under space inversion; the meson $M_\rho$ approximately decays as a power law of the modified chord distance between its charges, although it presents a bent shape in the logarithmic plot \ref{fig:gapless} (a), most probably caused by subleading terms in its bosonized description (in analogy with bosonic systems \cite{cazalilla2004}). Strong subleading deviations from the predictions in Eqs. (\ref{grho2},\ref{mrho2}) are observed approaching the BKT transitions.

\begin{figure}[t]
\includegraphics[width=1\columnwidth]{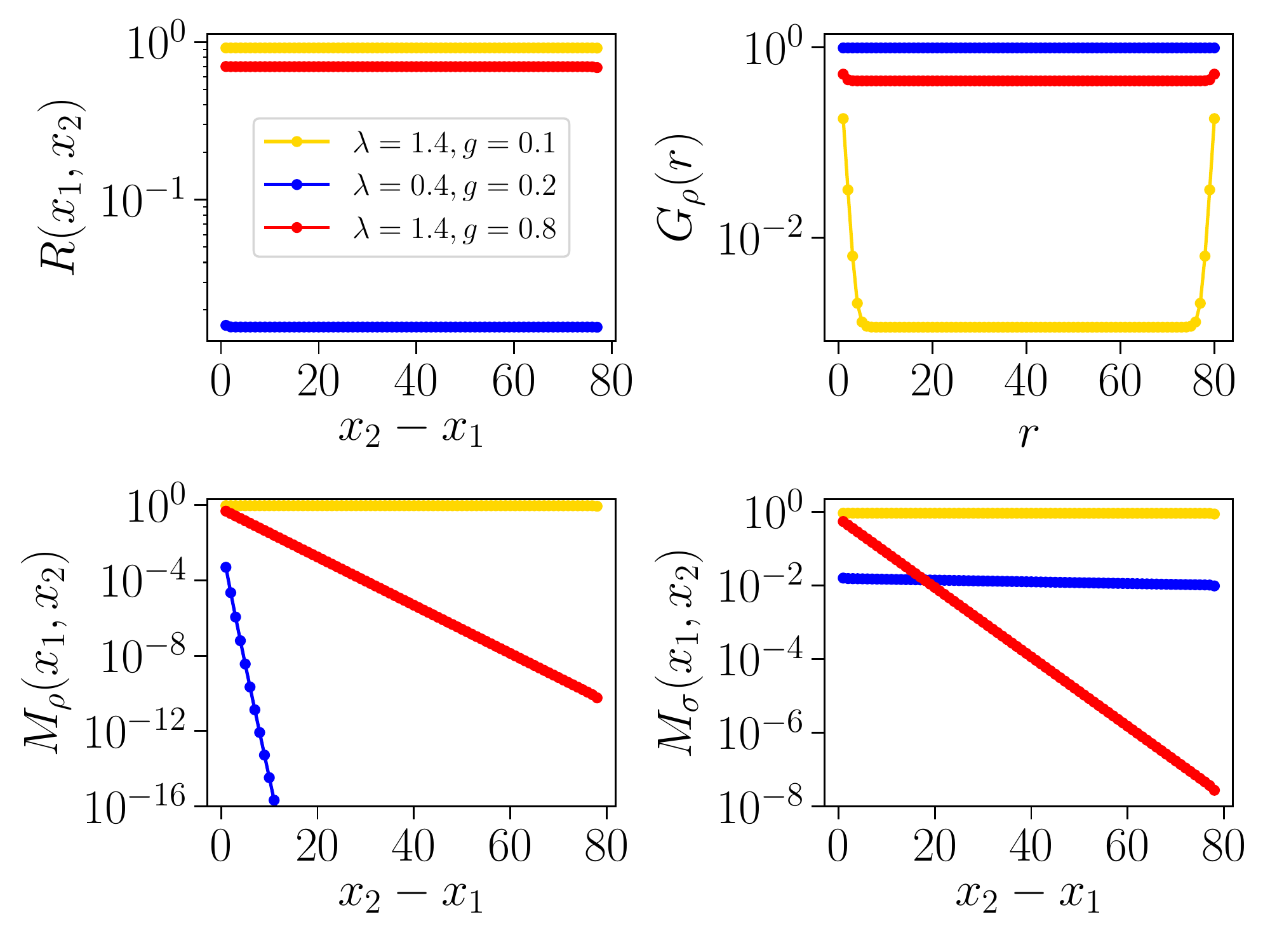}
\caption{
Examples of $R(x_1,x_2)$, $G_\rho(r)$, $M_\rho(x_1,x_2)$ and $M_\sigma(x_1,x_2)$ in semi-log scale for each regime of the gapped phase. The related values of the coupling constants are indicated by the stars with the corresponding colors in Fig.~\ref{fig:pd}.
The system is a $L=81$ ladder with smooth boundaries.
$x_1$ is chosen to be $2$.}
\label{fig:gapped}
\end{figure}

Concerning the gapped thermodynamical phase we begin our analysis by observing that the numerical results are consistent with having an approximately constant expectation value of the rung two-point correlation function $\left\langle R(x,y)\right\rangle$ for any $\lambda >0$ (see Fig.~\ref{fig:gapped}). In the limit $\lambda \to 0$, our findings suggest that $R$ maintains its distance-independent behavior, although its value considerably decreases and is smaller than $10^{-8}$ when $\lambda\lesssim 0.01$. As previously mentioned, the constant behavior of $R$ indicates that the fully confined and deconfined regimes exist only in the pure lattice gauge theory limit $\lambda=0$, whereas the gapped phase for any finite $\lambda > 0$ falls in one of the following three regimes: quadrupolar, confined rung-dominated and Higgs [see Table \ref{tab:phases} and  Fig. \ref{fig:pd}(a)]. This is consistent with the fact that the Luttinger parameter approximation $K_{\sigma/\rho} \sim 1/\lambda$ breaks for small values of $\lambda$ and the bare values of the Luttinger parameter must be considered bounded. In particular, our findings suggest that the rung tunneling term $T$ is relevant for all values $\lambda>0$.

To distinguish the quadrupolar, confined and Higgs gapped regimes, we can compare the behavior of the meson strings $M_\rho$ and $M_\sigma$ and the t'Hooft parameter $\mathcal{G}_\rho$. As summarized in Table \ref{tab:phases} these three regimes are identified, from the RG analysis, by different behaviors of the gauge-invariant observables, which we discuss in the following.

The confined (rung-dominated) regime appears for $g\gtrsim 0.4$ (red region in Fig. \ref{fig:pd}). In this phase both the electric field energy and the interleg tunneling flow to strong coupling, thus pinning the field combinations $\varphi_\rho, \, \sqrt{2}\theta_\sigma - \theta_0,\, \varphi_\sigma +\sqrt{2}\varphi_0 $, as can be derived by the Hamiltonian \eqref{boson} [see also the action \eqref{SI}]. Consequently $\mathcal{G}_\rho$ is constant, whereas all the mesons display an exponential decay (see the red curves in Fig. \ref{fig:gapped} for a typical scenario) and, in this region, we have a good agreement between the RG and DMRG predictions.

\begin{figure}[t]
\includegraphics[width=0.85\columnwidth]{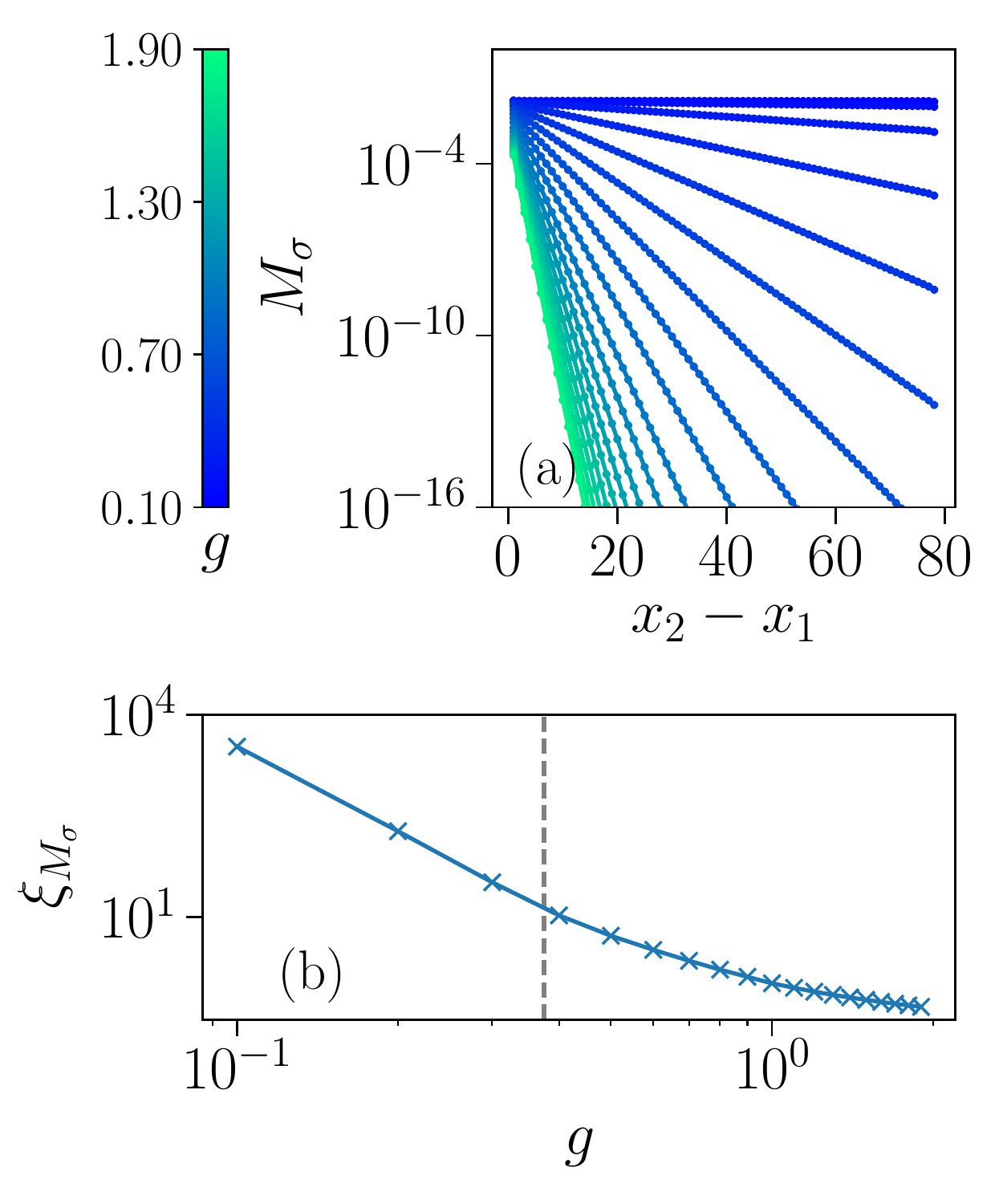}
\caption{(a) Meson $M_\sigma$ for $\lambda=0.4$ as a function of its length. Different colors represent different values of $g$. $M_\sigma$ is a constant in the limit $g\to 0$ and decays exponentially for $g>0$. (b) Decay length $\xi_{M_\sigma}$ of $M_\sigma$ as a function of $g$. For small $g$, $\xi_{M_\sigma}$ is approximately proportional to $g^{-4}$. In the crossover between the quadrupolar and confined phases, the behavior of $\xi_{M_\sigma}$ changes. The vertical dashed line in panel (b) indicates the value of $g$ corresponding to the peak of the fidelity susceptibility at $g\approx 0.375$.} \label{fig:mes_quadr}
\end{figure}

The picture is more complicated in the gapped phase at low values of $g$. In the quadrupolar phase the bosonization analysis yields that the interactions $T$ and $C_\rho$ pin the fields $\theta_\sigma$ and $\varphi_\rho$, whereas $\theta_0$ is pinned by the background $P_0$ interaction. Therefore, based on RG, the meson string $M_\sigma$ should display a constant behavior in this regime; this implies that pairs of mesonic strings with opposite dipoles at each end condense. This kind of mesons are indeed compatible with having ordered rung operators.

However, what is observed by the DMRG results is that this picture captures the behavior of the system only for $g=0$ [blue limit in Fig. \ref{fig:mes_quadr} (a)]. For small but finite values of $g$, instead, $M_\sigma$ always displays a weak exponential decay. Nevertheless, for $g\lesssim 0.2$ the corresponding decay length $\xi_{M_\sigma}$ is larger than the typical system sizes we can probe ($L\sim 100$) and it diverges for $g \to 0$. This smooth evolution from a constant value at $g=0$ to a weak exponential decay is the first signature of the crossover between the quadrupolar and confined phase. We additionally observe that $\xi_{M_\sigma}$ transitions from a behavior proportional to $g^{-4}$ to a slower decay for $g$ increasing towards the confined phase [see Fig. \ref{fig:mes_quadr} (b)]. 

\begin{figure*}
\includegraphics[width=2\columnwidth]{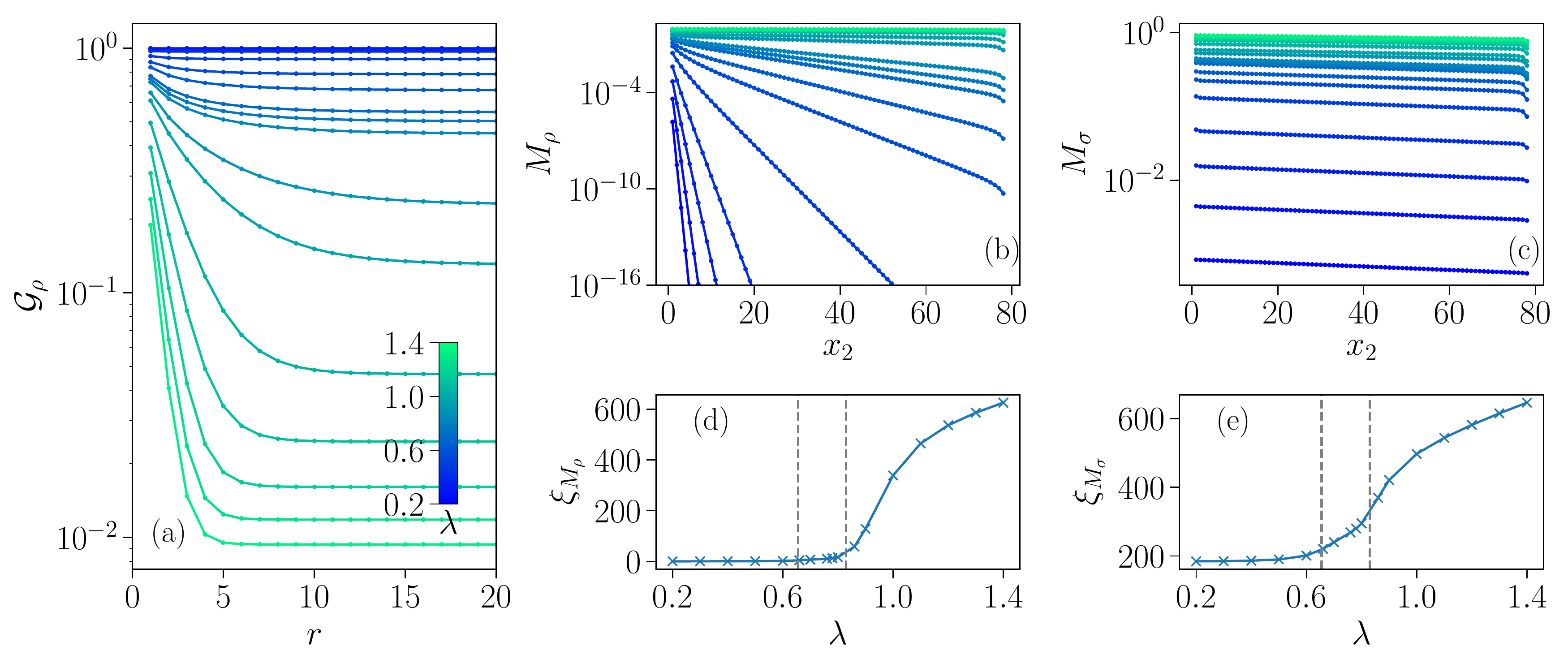}
\caption{
Upper panels: $\mathcal{G}_\rho(r)$ (a), $M_\rho(x_1,x_2)$ (b) and $M_\sigma(x_1,x_2)$ (c) for $\lambda$ from $0.2$ to $1.4$ and $g=0.2$ in semi-log scale.
The system is a $L=81$ ladder with smooth boundaries.
$x_1$ is chosen to be $2$.
Lower panels: (d) and (e) are the correlation lengths of $M_\rho$ and $M_\sigma$. The dashed lines indicate the positions of the peaks of $\chi_F$ at $\lambda\approx 0.655$ and $0.83$ [see Fig. \ref{fig:pd} (a)].
}
\label{fig:gapped_g}
\end{figure*}

\begin{figure*}
\includegraphics[width=2\columnwidth]{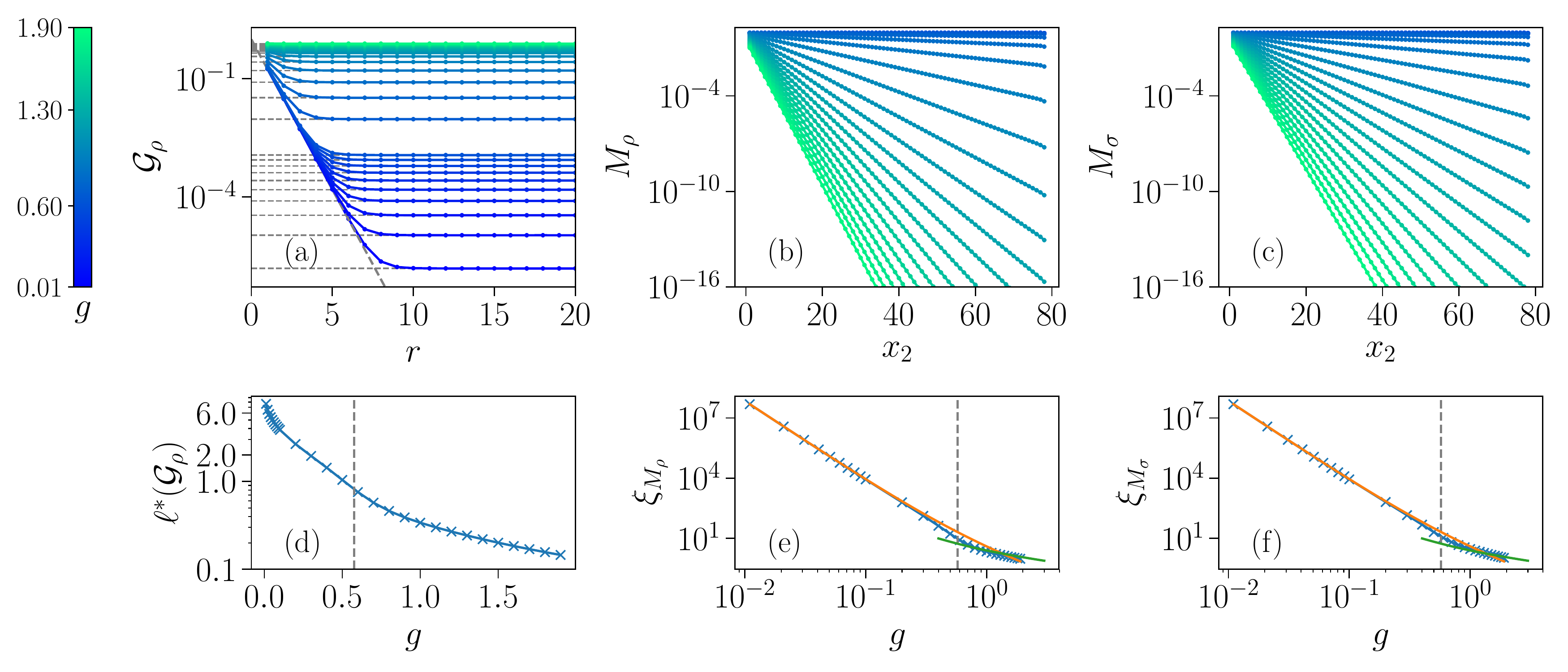}
\caption{
Upper panels: $\mathcal{G}_\rho(r)$ (a), $M_\rho(x_1,x_2)$ (b) and $M_\sigma(x_1,x_2)$ (c) for $g$ from $0.02$ to $1.9$ and $\lambda=1.4$ in semi-log scale.
The system is a $L=81$ ladder with smooth boundaries.
$x_1$ is chosen to be $2$.
Lower panels: (d) Estimate of the distance from the edge $\ell^*$ at which $\mathcal{G}_\rho$ transitions from an exponential decay to a constant as a function of $g$.
(e,f): Decay lengths of $M_\rho$ and $M_\sigma$ in log-log scale. The dashed lines indicate the crossover at $g\approx 0.575$.
The orange (green) curves are the analytic predictions for $g\ll 1$ ($g\gg 1$ and $\lambda \gg 1$); see Eq.~\ref{xiqa}.
}
\label{fig:gapped_lambda}
\end{figure*}

Concerning the meson $M_\rho$, its expectation value clearly decays exponentially with a very short decay length for any value of $g$ in the quadrupolar phase. This can be seen by considering  the data for $\lambda\lesssim 0.7$ (blue curves) in Fig. \ref{fig:gapped_g}, which depicts the observables of the system for $g=0.2$ as a function of $\lambda$. This is consistent with the fact that this meson string does not commute with the relevant and ordered $T$ interaction. Finally, $\mathcal{G}_\rho$ is approximately constant, showing that also in this phase the fluctuations of the dynamical charge are suppressed [blue curves in Fig.~\ref{fig:gapped_g} (a)].

Concerning the Higgs regime, the features of the crossover into the confined phase for growing $g$ are even stronger.
The bosonization prediction is that the tunneling and background $P$ terms pin all the $\theta$ fields to their semiclassical minima. This implies that all the mesons should condense in this regime and all the $M$ strings should display an approximately constant expectation value. On the contrary, the fluctuations of the charge are maximal and $G_\rho$ must decay exponentially from the boundary of the system. What we observe in the DMRG is that, once again, the bosonization predictions are accurate in the clock limit. For any $g>0$, instead, all the meson strings acquire a weak exponential decay, with a decay length diverging as $g^{-4}$ for $g\to 0$ (see Fig.~\ref{fig:gapped_lambda}), compatibly with results from a quasiadiabatic continuation estimating the following:
\begin{equation} \label{xiqa}
\xi_{M_\rho} \approx \xi_{M_\sigma} \approx \frac{\left(\lambda g + 1\right)^2\left(1-\cos \frac{2\pi}{5}\right)}{g^4} \,.
\end{equation}
See Appendix \ref{app:qa} for detail. Eq. \eqref{xiqa} relies on $g\ll 1$ in general, and we display its result for small values of $g$ in Figs.~\ref{fig:gapped_lambda} (e) and (f) (orange lines). For large values of $g$ and $\lambda$, instead, the approximate behavior of the mesons can be deduced by approximating the ground state with a product state that minimizes the electric field and tunneling energies only. The results of this approximation are depicted as the green lines in Figs.~\ref{fig:gapped_lambda} (e) and (f) and provide a reasonable estimate of the behavior of the meson even for the chosen value $\lambda=1.4$ and $g \gtrsim 1$. The smooth crossover between these behaviors of the decay length is a further signature of the Higgs / confined crossover.

The Higgs / confined crossover has an even more interesting effect over $\mathcal{G}_\rho$. In systems with smooth boundary conditions, $\mathcal{G}_\rho$ decays exponentially away from the edge as predicted for the clock limit. However, for finite $g$, this exponential decay stops at a certain distance $\ell^*$ from the edge, and $\mathcal{G_\rho}$ stabilizes to a bulk constant (see Fig. \ref{fig:gapped_lambda} (a), showing the results for $\lambda=1.4$). In the limit $g \to 0$, $\ell^*$ increases and the bulk region shrinks with smaller and smaller bulk values of $\mathcal{G_\rho}$. For $g$ approaching the crossover line, instead, $\mathcal{G_\rho}$ becomes essentially flat, corresponding to $\ell^* \lesssim 1$.

We finally comment on the regime at very low $\lambda$.  As we previously mentioned, the two-point correlation function $R$ displays approximately a constant behavior for any $\lambda>0$, and this is the main reason for which we consider the regions at small $\lambda$ in the phase diagram \ref{fig:pd} (a) to belong to the quadrupolar and confined rung-dominated phase, rather than the deconfined and fully confined phases. However, based on the DMRG data, we observe a very weak exponential decay of $\left\langle R(x,y) \right\rangle$ for the regions at very small $\lambda$: the corresponding decay length is typically $\xi_{R}>10^4$ for $\lambda = 0.02$. This tiny decay may be interpreted as an effective crossover from the fully confined phase of the pure LGT to the quadrupolar and confined rung-dominated phases. At the level of the field theoretical description, this may also be due to the weak mixing between the spin $\sigma$ and gauge $0$ sectors of the theory which appears at second order in the perturbative RG analysis and we neglected in our flow equations (see App. \ref{app:2nd}).

\subsection{Static charges and screening} \label{sec:screen} 

\begin{figure}[t]
\includegraphics[width=1\columnwidth]{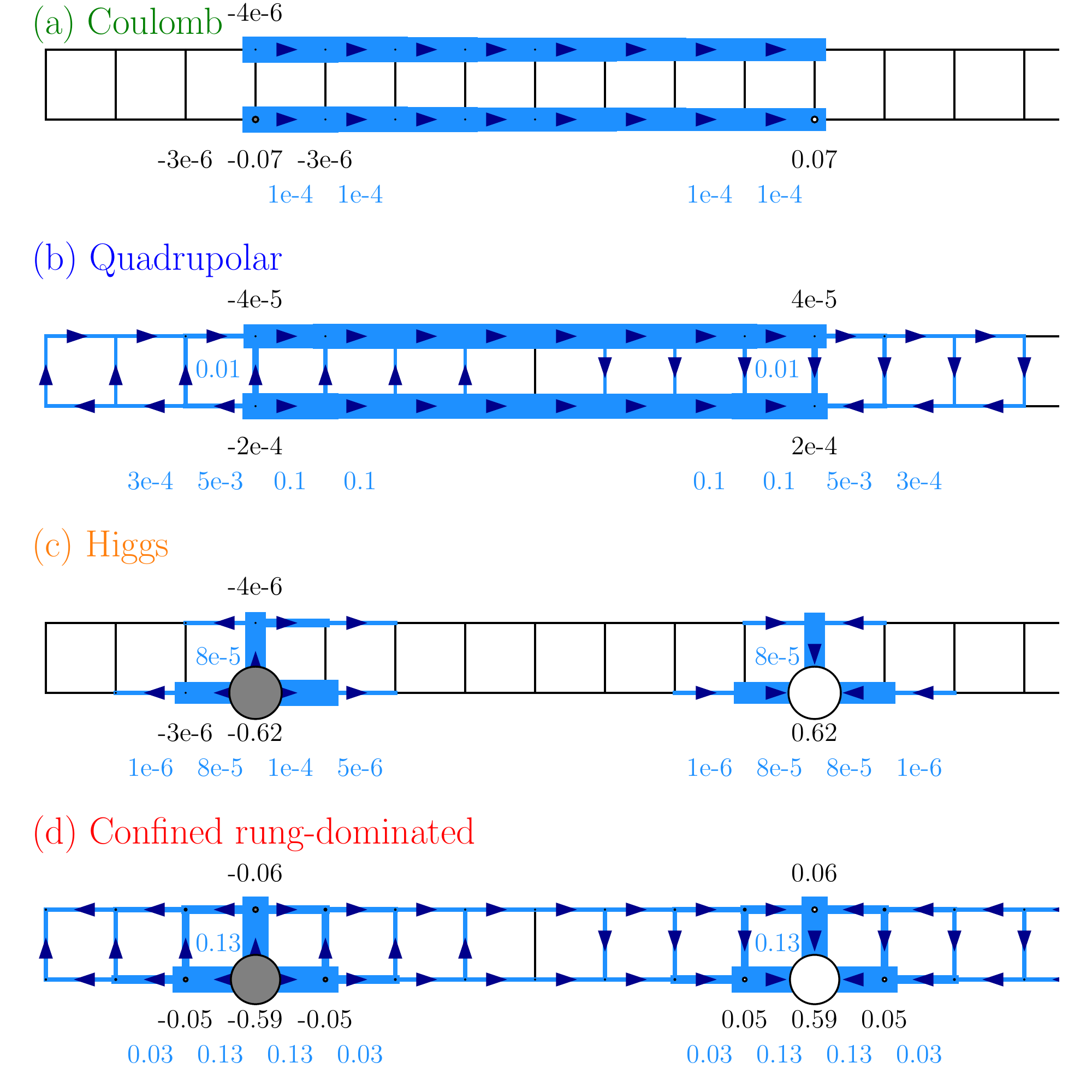}
\caption{Examples of the dynamical charge and electric field distributions in the intermediate region between two opposite static charges. The size of the dots represent the expectation value of the dynamical charge $q$; the thickness of the links represents the expectation value of the electric field $E$ in arbitrary units. The electric field and charge smaller than $10^{-6}$ are not shown in the figures. Black and blue numbers label examples of the $q$ and $E$ expectation values.  Panel (a) represents a typical example in the Coulomb phase ($g=0.01$, $\lambda=0.75$); panels (b), (c) and (d) correspond to the values of the coupling constants in Fig. \ref{fig:gapped} (the stars in Fig.~\ref{fig:pd}) and they represent typical examples taken within the quadrupolar, confined rung-dominated and Higgs regimes respectively. } \label{fig:staticc}
\end{figure}

\begin{figure}[th]
\includegraphics[width=1\columnwidth]{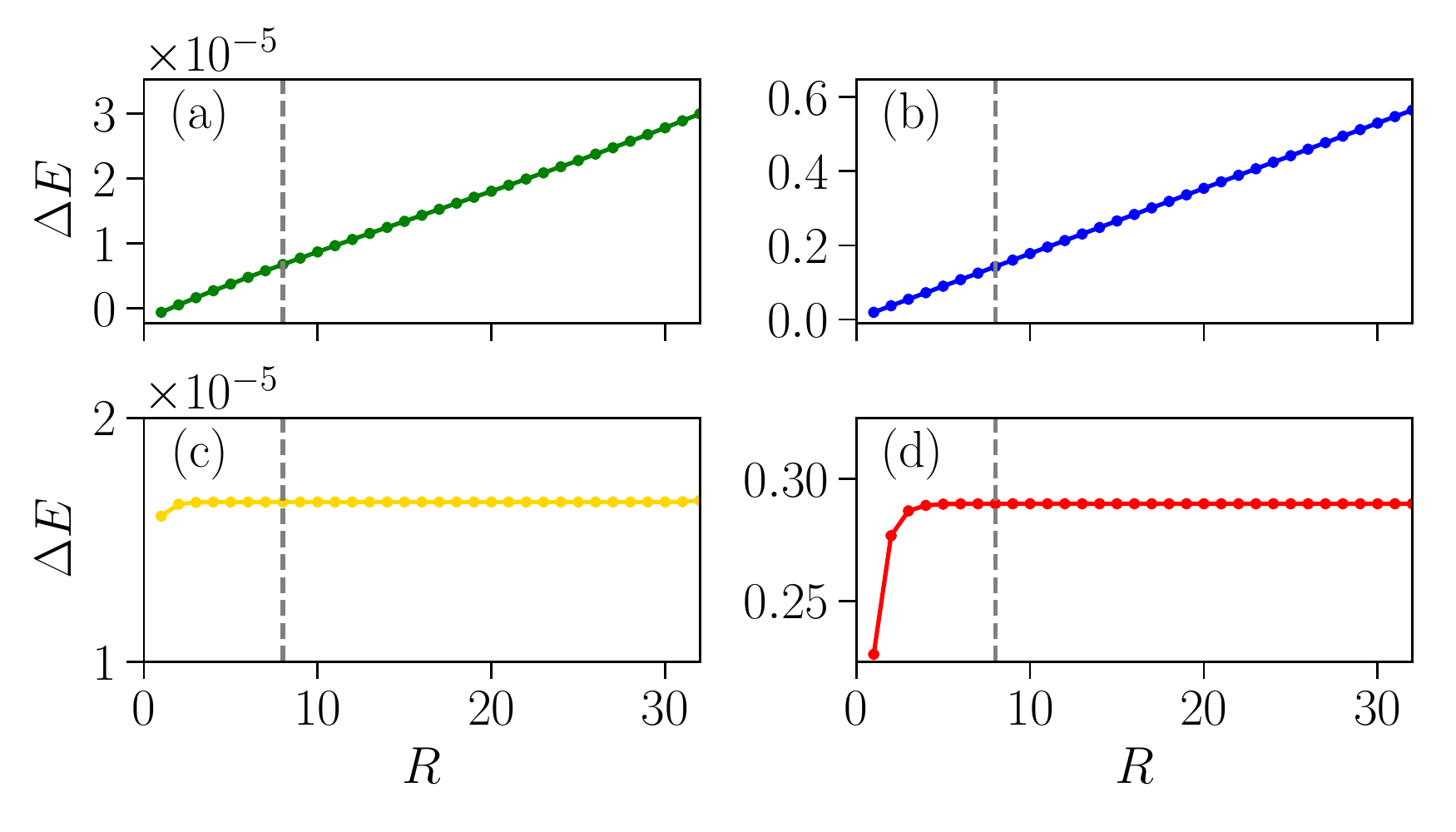}
\caption{Energy cost $\Delta E$ for the introduction of two static charges in the same leg of the ladder as a function of their distance $R$. The four panels correspond to the same coupling constants chosen in Fig. \ref{fig:staticc}. The gray dashed lines indicate the static charge distance $R=8$ used in Fig. \ref{fig:staticc}. } \label{fig:deltae}
\end{figure}

When introducing dynamical charges in the system by considering $\lambda > 0$, the confinement of the pure lattice gauge theory is disrupted in general by screening. In particular, when we insert two opposite static charges through the violation of the Gauss law in two specific sites at a distance $R$, as done in Sec. \ref{sec:pure}, dynamical charges can accumulate in a screening cloud around them and, in general, they will suppress the electric field propagation in the intermediate region. 

To roughly estimate the extent of this screening mechanism, we can compare the electric field string energy $\mathcal{T}R$ (see Fig. \ref{fig:tension}) and the mass of a pair of dynamical charges $2m=4(1-\cos{2\pi/N})/\lambda$. We define the length scale $R^*$ such that $\mathcal{T}(g)R^* = 2m(\lambda)$; $R^*$ provides a rough approximation of the size of the screening clouds of the dynamical charges around the static ones. When $\mathcal{T}R < 2m$, thus $R<R^*$, the energy cost of an electric field string connecting the static charges is smaller than the mass cost required to screen them. Therefore $\Delta E$ increases linearly with $R$ as in the pure lattice gauge theory, and the corresponding states do not display a complete screening of the static charges such that electric lines clearly propagate between them [see, for example,  Fig. \ref{fig:staticc}(b)]. When $R>R^*$, instead, screening dominates, and $\Delta E$ stabilizes towards an asymptotic value (typically smaller than $2m$). In this case, localized clouds of dynamical particles completely screen the static charges and the electric field rapidly decays away from them [see, for example, the behavior in Fig. \ref{fig:staticc}(c)].

\begin{figure}[th]
\includegraphics[width=1\columnwidth]{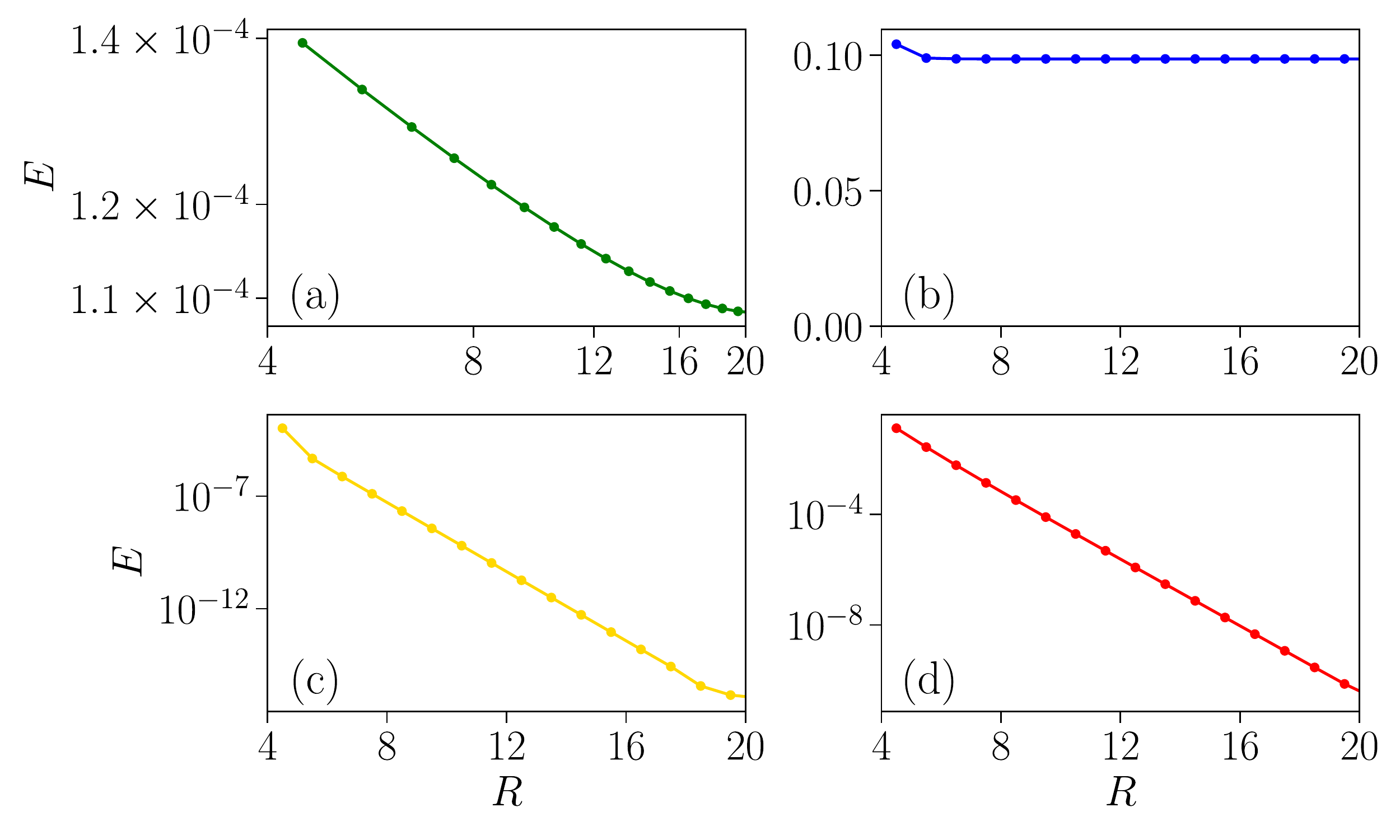}
\caption{Decay of the electric field $\left\langle E\right\rangle$ between two static charges as a function of the distance $r$ from the negative charge. The distance between the static charges is $R=34$. (a) Coulomb phase (log-log scale): $\left\langle E\right\rangle$ decays approximately as a power law for $r<R/2$. (b) Quadrupolar phase: $\left\langle E\right\rangle$ is approximately constant (no screening). (c) and (d) Screened states in the Higgs and confined regimes (log-normal scale). $\left\langle E\right\rangle$ decays approximately exponentially for $r<R/2$.  The four panels correspond to the same coupling constants chosen in Fig. \ref{fig:staticc}. } \label{fig:electr}
\end{figure}

In Fig. \ref{fig:deltae} we display the behavior of $\Delta E$ as a function of the static charge distance $R$ within the different regions of the phase diagram. In the quadrupolar phase (blue) the string tension is small whereas the mass of the Higgs charges is large, therefore screening does not occur on the length scales here presented. $R^*$ is indeed large, and, as expected, for $R<R^*$ the electric field propagates along both legs in the intermediate regime [see Fig. \ref{fig:staticc}(b)]. In this regime, in particular, the expectation value of the electric field $E$ decreases only weakly away from the static charges and it is approximately constant along the leg connecting them [see Fig. \ref{fig:electr} (b)]. The electric field along the rungs is typically much smaller than the one along the legs, consistently with the rung tunneling being a relevant interaction.

By increasing $g$, the system smoothly evolves into the confined rung-dominated phase [see Fig. \ref{fig:deltae}(d)]; in this regime the string tension is stronger, thus $R^*$ decreases. For $R<R^*$ the state is again analogous to the confined limit and, depending on $g$, the electric field either propagates on both legs (intermediate values of $g$) or only in the leg of the static charges (large $g$). In the former case, we observe that the electric field predominantly flows from one leg to the other along the same rungs of the static charges. The case $R>R^*$ is exemplified instead by Fig. \ref{fig:staticc}(d) and  \ref{fig:electr}(d): the electric field is exponentially suppressed away from the static charges by a screening cloud of dynamical charges.

In the Higgs phase, screening is extremely strong such that $R^*$ is typically around one [see Fig. \ref{fig:deltae}(c)]. Also in this case the electric field is exponentially suppressed with the distance from the static charges [Figures \ref{fig:staticc}(c) and  \ref{fig:electr}(c)]. In this regime the amplitude $|\left\langle \tau\right\rangle|$ is in general very small due to the strong plaquette and tunneling interactions, thus concurring in suppressing further the expectation values $\left\langle E\right\rangle$.

Finally, in the Coulomb phase, we observe that $R^*$ is very large due to the weak string tension; thus, the static charges display confinement over long distances  [see Fig. \ref{fig:deltae}(a)]. In this gapless phase, the electric field $\left\langle E\right\rangle$ decays weakly as a power law away from the static charges [Figures \ref{fig:staticc}(a) and  \ref{fig:electr}(a)] and it propagates on both legs, consistently with the predicted algebraic decay of the correlation functions in the charge $\rho$ sector.

\section{Extension to multiple legs} \label{sec:2D}

The field theoretical approach we adopted for the analysis of the system can be extended to investigate wider ladders with a finite number of legs $L_y$. To this purpose, we can apply a so-called coupled-wire construction (see, for example, the review \cite{meng2020} and references therein): we decompose the system (in the axial gauge) into a set of Luttinger liquids that describe each horizontal stripe of the lattice and interact with each other. In particular, we consider the axial gauge and we generalize the matter bosonic fields $\varphi_s$ and $\theta_s$ in Eq. \eqref{map1} into pairs of dual fields $\varphi_y$ and $\theta_y$ labeled by the coordinate $y=1,\ldots,L_y$ which specifies the row they refer to. In a similar way, the fields $\varphi_0$ and $\theta_0$ in Eq. \eqref{map2} are extended to $\varphi_{0,y}$ and $\theta_{0,y}$ (with $y=1,\ldots, L_y-1$), in order to represent the gauge bosons on the vertical links between the matter rows $y$ and $y+1$. As before, each pair of dual bosonic fields requires the addition of a pair of background interactions of the $P$ and $Q$ kinds. 

The major difference between two and multiple legs relies in the form of the electric field and tunneling interactions; Eq. \eqref{TG} takes a more symmetric form:
\begin{multline} \label{TG2}
-T \sum_y \int \rd x\, \cos \left(\theta_{y+1} -\theta_{y} - \theta_{0,y}\right) \\
-G \sum_y \int \rd x\, \cos \left( \varphi_{0,y+1} - \varphi_{0,y} - \varphi_{y+1} \right) \,.
\end{multline}
The electric field energy on each horizontal link can indeed be described by a combination of the difference of electric fluxes ingoing and outgoing from the vertical links and the total Higgs charge on the same row. These interactions must be supported by suitable boundary conditions. 

This form of the coupling between subsequent rows allows for the existence of a Coulomb phase also for multiple legs (with smooth boundaries at $y=1$ and $y=L_y$). Such a phase can be understood by the emergence of a gapless sector characterized by the dual fields:
\begin{equation}
\theta_{\rm bulk} = \frac{1}{\sqrt{L_y}}\sum_y \theta_y \,,\quad \varphi_{\rm bulk} = \frac{1}{\sqrt{L_y}}\sum_y \varphi_y\,.
\end{equation}
These fields are a linear superposition of all the matter fields and they generalize the charge sector into a bulk mode of the system; the operator $\e^{i\varphi_{\rm bulk}}$, for example, is linked to the motion of a magnetic flux along the vertical direction of the lattice. Analogously to its two-leg counterpart, this sector remains gapless for intermediate values of $\lambda$ (such that the related background terms are irrelevant) and small $g$ (such that the $G$ term in \eqref{TG2} is irrelevant and does not open a gap in this sector).

A key aspect of the interactions \eqref{TG2} is that the electric field energy involves three bosonic fields in the multileg case (differently from the corresponding equation \eqref{e3} for the ladder). This implies that its scaling dimension increases and its growth in the RG flow is slightly suppressed.
This has the important implication of increasing the extension of the Coulomb phase to larger values of $g$. The extension of the gapless phase is further increased due to the second-order terms generated by the $G$ interactions being suppressed as well (the generalization of the $C'_\rho$ terms in \eqref{SI} in particular).

Concerning the further extension to a fully two-dimensional system, the analysis becomes qualitatively more complex: an extended deconfined and topological phase appears for small values of $g$ and $\lambda$, whereas the gapless phase is supposed to evolve into a $U(1)$ symmetric weakly gapped phase \cite{Borisenko2014}. 
The mechanism gapping the critical phase is non-trivial and can be understood based on the mapping, for $g=0$, into the 2D $\mathbb{Z}_N$ quantum clock model and its classical 3D analog. In three dimensions, it is indeed known that the classical clock model does not display an extended gapless phase due to the $P$ background terms being ``dangerously irrelevant'' perturbations and the system displays only a single phase transition in the 3D XY model universality class. This is due to the fact that the irrelevant $P$ operators become relevant when approaching the $U(1)$ symmetric gapless Nambu-Goldstone fixed point \cite{Oshikawa2000}, thus causing a second step in the RG flow towards the ordered ferromagnetic phase of the clock model (see the phase diagrams evaluated in \cite{Shao2020,Patil2020}). As a consequence, the system is characterized by two different length scales $\xi <\xi'$, both diverging at the critical point, which are associated with the onset of $U(1)$ and $\mathbb{Z}_N$ symmetric features respectively. This causes indeed the appearance of a crossover between a U(1) symmetric and a $\mathbb{Z}_N$ symmetric regime in the ordered phase. This analysis has been numerically well-verified for the classical 3D model (see, for example, \cite{Borisenko2014,Shao2020}), and it has been recently confirmed also for the quantum 2D case \cite{Patil2020}, thus leading to the conclusion that no gapless phase exists for the clock limit $g\to 0$ and the dual pure LGT $\lambda \to 0$ \cite{Borisenko2014}. 

We conclude our comparison between the quasi-one-dimensional ladder geometry and the fully two-dimensional systems by observing that, in general, the Hamiltonian \eqref{Hinv} presents neither true topological order nor symmetry-protected topological phases.

Concerning topological order, even in the limit $\lambda \to 0$ and $g\to 0$, in the ladder system local operators such as $\mathcal{G}_\rho (r)$ in Eq. \eqref{Grho} mix the degenerate ground states appearing with periodic (or rough) boundary conditions. Despite that, we observe that the Higgs matter excitations and the magnetic flux excitations obey a mutual anyonic statistics, as in the analogous 2D $\mathbb{Z}_N$ models with topological order \cite{brennen2007,orus2012,burrello13,Zarei2020}.

This suggests to investigate under which conditions it is possible to obtain symmetry-protected topological phases in the system, for example, by extending the $\mathbb{Z}_2$ surface code on the ladder \cite{Vaezi2018} to its $\mathbb{Z}_N$ generalizations. We must observe that, in the Hamiltonian \eqref{Hinv}, the electric field energy term along the legs of the ladder and the tunneling term along the rungs break the  $\mathbb{Z}_N \times \mathbb{Z}_N$ symmetry required to design symmetry-protected phases analogous to \cite{Vaezi2018} (see also \cite{burrello2018} for similar non-Abelian constructions). By excluding these terms from our system, it is possible to design error-resilient model to store quantum information. With this purpose, the symmetry-protected $\mathbb{Z}_2$ model in \cite{Vaezi2018} was used to study the self-localization of anyons for highly excited states. The study of similar dynamical properties of the $\mathbb{Z}_N$ anyons is an interesting open problem which we leave for future studies.

\section{Conclusions} \label{sec:concl}

Ladder setups offer the simplest realization of a lattice gauge theory whose dynamics crucially relies on the plaquette interactions. In the path towards the experimental analog quantum simulations of gauge theories, therefore, the realization of gauge models in the ladder geometry would constitute an important milestone bridging one-dimensional chains and higher dimensional setups. The first steps in this direction have already been accomplished in ultracold atom systems trapped in optical lattices: a recent experiment \cite{Aidelsburger2019} has proved that a tunneling term mediated by an effective $\mathbb{Z}_2$ gauge degrees of freedom can be realized based on density-dependent laser-assisted tunneling techniques. This is indeed the rung tunneling interaction required for the realization of a gauge theory in the ladder based on the axial gauge \cite{Barbiero2019}. With respect to the double-well systems proposed in Ref. \cite{Barbiero2019}, our model additionally includes the plaquette interaction, which, in the axial gauge, must be engineered as an operator acting on two neighboring rung double-wells.

Previous works focused on the analysis of several aspects of dynamical gauge theories in ladders \cite{burrello2018,Barbiero2019,klco2020,santos2020,Vaezi2018,Gonzalez2020}. In this article, we explored the general features of the full Kogut-Susskind model with a $\mathbb{Z}_N$ LGT and Higgs matter degrees of freedom. We analyzed its phase diagram based on both a low-energy field theoretical description inspired by bosonization, and DMRG numerical simulations. The model displays different features for $N\le 4$, where a single critical point is observed in the limit of strong plaquette interactions ($g\to 0$), and $N\ge 5$, where instead an extended critical Coulomb phase, with emergent U(1) symmetry appears. 

Our numerical analysis focused mostly on the $N=5$ case, but, based on the renormalization group study of the field theory description of the model, we conclude that the existence of this extended gapless phase is stable for larger (but finite) values of $N$, in which the gapless phase covers broader and broader domains in the $\lambda$ coupling constant.

On the technical side, our bosonization description of this gauge theory can be applied, in general, to describe a broad family of quantum clock models and our description can be extended to the study of more complicated quasi-one-dimensional ribbon geometries.

\acknowledgments

M.B. warmly thanks A. Negretti, E. Rico Ortega and D. Rossini for discussing in depth the physics of LGTs in the ladder geometry and sharing their numerical result for the U(1) truncated gauge symmetry. M.B. is also indebted with E. Cobanera, A. Milsted and G. Ortiz for a useful collaboration about 1D systems with $\mathbb{Z}_N$ symmetry, and with P. Silvi and H.-H. Tu for insightful correspondence. This work was supported by the Villum Foundation (Research Grant No. 25310) and by the EU’s Horizon 2020 programme under the Marie Sk\l odowska-Curie grant agreement No. 847523.
The DMRG calculations were performed using the ITensor C++ library~\cite{itensor}, https://itensor.org/.

\appendix

\section{Details on the pure lattice gauge theory} \label{app:pure}

In this appendix we provide further information about the pure lattice gauge theory and the calculation of its electric string tension. 

Eq. \eqref{pureLGT} shows that the pure lattice gauge theory in the axial gauge is equivalent to a $\mathbb{Z}_N$ quantum clock model with the addition of non-local disorder operators. This Hamiltonian can be reduced in a completely local form by use of a unitary bond-algebraic duality transformation, as presented in \cite{Fradkin,ortiz2012}:
\begin{align}
    &\sigma_{r}^{\dagger}\sigma_{r-1} \rightarrow \tilde{\tau}_{r}^{\dagger} \,  \quad \text{for}\; r  \ge 2\,;  \quad \sigma_{1} \rightarrow \tilde{\tau}_{1} \,;\\
		& \tau_{r} \rightarrow \tilde{\sigma}_{r+1}^{\dagger} \tilde{\sigma}_{r} \,  \quad \text{for}\; r  \neq L\,; \quad \tau_{L} \rightarrow \tilde{\sigma}_{L}\,;\\
		&\prod_{j=r}^{L}\tau_{j} \rightarrow \tilde{\sigma}_{r}  \,.
    \label{app:mapping}
\end{align}
This mapping preserves indeed all the commutation relations and maps the disorder operators into a local term.  The dual Hamiltonian reads:
\begin{multline}
    \tilde{H}= -\frac{1}{g}\sum_{i=2}^{L}(\tilde{ \tau}_{i}^{\dagger} + \tilde{\tau}_{i}) - g\sum_{i=1}^{L-1}( \tilde{\sigma}_{i}^{\dagger} \tilde{\sigma}_{i+1} + \text{H.c.}) \\
   -2g\sum_{i=1}^{L}(\tilde{\sigma}_{i} + \tilde{\sigma}_{i}^{\dagger}) - g(\tilde{\sigma}_{L} + \tilde{\sigma}_{L}^{\dagger}). 
    \label{app:hamil}
\end{multline}
This dual Hamiltonian describes the $N$-clock model in a longitudinal field, with the important feature that $\tilde{\tau}_{1}$ does not appear in the dual model. As a consequence, the global  symmetry $\mathcal{G}$ is mapped into the holographic symmetry defined by $ \tilde{\mathcal{G}} =  \tilde{\sigma}_{1}$, such that  $[\tilde{\sigma}_{1},\tilde{H}]=0$ (see \cite{cobanera2012} for more detail on holographic symmetries in $\mathbb{Z}_N$ models with Higgs matter).

 From Eq. \eqref{app:hamil} we see that the ground state is degenerate in the limit $g\to 0$, because the first clock degree of freedom may assume any of its $N$ states. This corresponds to the ordered ferromagnetic limit for the Hamiltonian \eqref{pureLGT}. For any value $g>0$, however, the degeneracy is split by the term proportional to $\tilde{\mathcal{G}}$ entering the Hamiltonian. The ground state is then symmetric and non-degenerate for any $g>0$ and the gap separating it from the first excited states grows linearly with $g$.
  For $N=2$ this corresponds indeed to the Ising model in a longitudinal and transverse field which presents only one gapped phase \cite{coldea2010,banuls2011}. 

This feature distinguishes our ladder model from its two-dimensional counterpart, since, for pure $\mathbb{Z}_N$ lattice gauge theories in 2D square lattices, there exists a phase transition between a confined $\left(g>g_c\right)$ and a deconfined $\left(g<g_c\right)$ phase for a finite $g_c>0$ \cite{Fradkin,fradkin1979}. 

To verify that the extended gapped phase characterizing the pure lattice gauge model \eqref{pureLGT} at $g>0$ corresponds to a confined phase, we consider the behavior of the system in the presence of external static charges and we estimate their string tension. 

In order to simplify our analysis, we modify the rough boundary on the left of the ladder by including an additional 3-site boundary plaquette term which does not violate any of the gauge constraints:
\begin{equation} \label{leftplaq}
	H_{\text{left plaq.} } = -\frac{1}{g_b}\left(   \sigma_{1,\uparrow}\sigma_{1,0}^{\dagger}    \sigma_{1,\downarrow}^{\dagger}  + \text{H.c.}  \right)\,.
\end{equation} 
This magnetic field term must be added to the boundary term in Eq. \eqref{hambound} and its inclusion explicitly breaks the $\mathcal{G}$ symmetry, favouring a symmetry-broken ferromagnetic ground state with all the operators $\sigma_0$ aligned along their eigenvalue 1 in the axial gauge choice of Eq. \eqref{pureLGT}.

A confined phase is characterized by the presence of a linear string tension for the electric flux lines. When introducing two opposite static charges, the energy of the ground state displays a linear dependence on the distance $R$ between them. 
The static charges are introduced by modifying the gauge constraints \eqref{gconstr} associated to two arbitrary ladder sites. In particular, we impose that the physical Hilbert space fulfills $G_{x,\Dn}\ket{\psi}=e^{i\frac{2\pi}{N}}\ket{\psi}$ and $G_{y,\Dn}\ket{\psi}=e^{-i\frac{2\pi}{N}}\ket{\psi}$. For simplicity, we introduce the charges in the same leg and we consider $y>x$. In this physical subspace, the Hamiltonian in the axial gauge takes the form 
\begin{multline} \label{pureLGT2}
H_{\rm gauge} = -\frac{1}{g} \sum_{r=1}^{L-1} \left(\sigma_{r}\sigma_{r+1}^\dag + {\rm H.c.}\right) -g \sum_{r=1}^{L} \left(\tau_{r} +\tau^\dag_{r}\right) \\
-g\sum_{ y \geq r > x} \left[(e^{-i\frac{2\pi}{N}} + 1) \prod_{j = r}^{L}\tau_{j} + (e^{\frac{i2\pi}{N}}+ 1) \prod_{j = r}^{L}\tau_{j}^{\dagger}  \right]   \\ - 2g\sum_{\substack{r \leq x \, {\text{or}} \, r>y }} \left[\prod_{j = r}^{L}\tau_{j} +  \prod_{j = r}^{L}\tau_{j}^{\dagger} \right]-\frac{1}{g}\left( \sigma_{1,0} + \sigma_{1,0}^{\dagger}\right) .
\end{multline}
To verify that this Hamiltonian supports only a confined phase for any $g>0$, we consider the behavior in the two limits $g\rightarrow 0,  \infty$. For $g\rightarrow \infty$ the ground state is a (paramagnetic) product state with all sites obeying $\tau_{r}\ket{\psi_{\rm axial}}=\ket{\psi_{\rm axial}}$. The electric energy of this product state with the two static charges is $E(g \rightarrow \infty) = -6gL + 2gR(1-\cos(2\pi/N)) $ such that:  
\begin{multline} \label{deltaEg1}
	\Delta E(g \rightarrow \infty) = 2g(1-\cos(2\pi/N))R = \mathcal{T} R \,, 
\end{multline}
where $\Delta E$ describes the energy difference between the ground states with and without the static charges separated by the distance $R=|x-y|$ and $\mathcal{T}$ denotes the string tension (see Fig. \ref{fig:tension}). The phase in this limit is therefore confined. Including also the plaquette interaction, for $g \gg 1$ the system still supports a confined phase, and from perturbation theory one finds
\begin{multline} \label{deltaEg2}
	\Delta E (g \gg 1) = \left[2g\left(1-\cos{(\frac{2\pi}{N})}\right)  \right.	\\ \left.
	- \frac{1}{g^{3}}\left( \frac{1}{2(3-2\cos{(2\pi/N)}-\cos{(4\pi/N)})}\right)\right] R + \mathcal{O}\left(\frac{1}{g^{7}}\right).
\end{multline}
In the other limit, $g\rightarrow 0$, the ground state is instead a product state with all sites obeying $\sigma_{r}\ket{\psi_{axial}}=\ket{\psi_{\rm axial}}$. Exactly at the $g=0$ limit, the ground state is deconfined since $\Delta E(g \rightarrow 0)=0$. For small values of $g>0$ we apply a standard non-degenerate perturbation theory and the lowest order correction to the energy yields
\begin{equation}
	\Delta E(g \ll 1 )= 2g^{3}R  + \mathcal{O}(g^{7}).
\end{equation}
Also in this case, the static charge energy presents a linear dependence with respect to their distance $R$, thus showing that that the deconfined phase is unstable under any small $g$ perturbation (see Fig. \ref{fig:tension}). In this respect, the ladder model behaves like a fully 1D system and the ground state of the pure lattice gauge theory displays confinement of the static charges for any $g>0$.

\section{The case $N=4$ in the limit $g=0$}\label{app:4}

In the limit $g=0$, the lattice gauge theory model is equivalent to the quantum clock model \eqref{Potts} in the ladder geometry. In the specific case $N=4$, we can apply a unitary mapping from the $\mathbb{Z}_4$ clock operators into two pairs of $\mathbb{Z}_2$ Ising operators \cite{ortiz2012}. We introduce an additional index $j=1,2$ to label these two sets of Pauli matrices:
\begin{align}
&\zeta_{r,s} = \frac{\e^{\frac{i\pi}{4}}}{\sqrt{2}}\left(\sigma^{z}_{r,s,1}-i\sigma^{z}_{r,s,2}\right)\,,\\
&\eta_{r,s} = \frac{1}{2}\left(\sigma_{r,s,1}^x + \sigma_{r,s,2}^x\right) + \frac{i}{2}\left(\sigma^z_{r,s,1}\sigma^{y}_{r,s,2}- \sigma^{y}_{r,s,1}\sigma^{z}_{r,s,2}\right)\,;
\end{align}
where $\sigma^{a}$ label the Pauli matrices.
Based on this unitary mapping, the model of Eq. \eqref{Potts} for $N=4$ becomes:
\begin{multline}
H_{N=4}(g=0) =  \\ 
-\lambda \left[\sum_{r,s,j} \sigma^z_{r,s,j}\sigma^z_{r+1,s,j} + \sum_{r,j} \sigma^z_{r,\Up,j}\sigma^z_{r,\Dn,j} \right]
-\frac{1}{\lambda} \sum_{r,s,j} \sigma^x_{r,s,j}\,.
\end{multline}
The resulting Hamiltonian corresponds to two copies $j=1,2$ of the Hamiltonian \eqref{Potts} for $N=2$ (up to an overall rescaling of the energy by a factor 1/2). Therefore the critical value of $\lambda$ at $g=0$ is the same for $N=2$ and $N=4$.

\section{Details about bosonization} \label{app:bosonization}

In this appendix, we present the detail about the construction of the effective Hamiltonian \eqref{boson}.

The fundamental criterion to built a low-energy description in continuum space of the $\mathbb{Z}_N$ LGT on the ladder is to create a mapping from the clock operators to the bosonic fields $\theta_s$ and $\varphi_s$ which preserves their algebraic properties. To this purpose, we verify first that Eqs. (\ref{comm},\ref{map1},\ref{map2}) fulfill the commutation relation in Eq. \eqref{clockpr2}:
\begin{multline}
\zeta_{r,s} \eta_{r',s'} \to \e^{-i\theta_s(r)}e^{-i\varphi_{s'}(r')+i\varphi_{s'}(r'+a)} =\\
 e^{-i\varphi_{s'}(r')+i\varphi_{s'}(r'+a)}\e^{-i\theta_s(r)}\e^{-\left[\theta_s(r),\varphi_{s'}(r')-\varphi_{s'}(r'+a)\right] }=\\
e^{-i\varphi_{s'}(r')+i\varphi_{s'}(r'+a)}\e^{-i\theta_s(r)}\e^{i\frac{2\pi}{N}\left[\Theta\left(r-r'\right)-\Theta\left(r-r'-a\right)\right]\delta_{s,s'}} =\\
e^{-i\varphi_{s'}(r')+i\varphi_{s'}(r'+a)}\e^{-i\theta_s(r)}\e^{i\frac{2\pi}{N}\delta_{r,r'}\delta_{s,s'}} \to\\
\eta_{r',{s'}}\zeta_{r,s} \e^{i\frac{2\pi}{N}\delta_{r,r'}\delta_{s,s'}}\,,
\end{multline}
which verifies Eq. \eqref{clockpr2}. In the previous calculation we used the Campbell-Baker-Haussdorf formula and we adopted the convention that $\Theta(x)=1$ for $x\geq 0$. The analogous property in Eq. \eqref{clockpr} can be verified in the same way.

Concerning finite systems, we observe that for the right edge, characterized by smooth boundary conditions (see Fig. \ref{fig:ladder}), the definition of $\eta_{L,s}$ and $\tau_{L,0}$ must be taken with Dirichlet boundary conditions $\varphi_s(L+a)=0$ such that:
\begin{equation} \label{mapbound}
\eta_{L,s} \to \e^{-i\varphi_s(L)}\,,\quad \tau_{L,0} \to \e^{-i\varphi_0(L)}\,.
\end{equation}
By introducing different boundary conditions for $\varphi_{\Up/\Dn}$ at the two edges it is possible to add a background electric field thus modifying the $\uptheta$ vacuum of the theory.

Based on the mapping (\ref{map1},\ref{map2}), we are now ready to derive the effective Hamiltonian \eqref{boson}. We list in the following all the Hamiltonian terms in the axial gauge and their continuum approximations. The intraleg tunneling terms read:
\begin{multline}
-\lambda\sum_{r}\zeta^\dag_{r,s} \zeta_{r+1,s} + {\rm H.c.} \to \\
-\frac{\lambda}{a}\int \rd x \, \e^{i\left(\theta_s(x)-\theta_s(x+a)\right)} + {\rm H.c.} =\\
-\frac{2\lambda}{a}\int \rd x \, \cos{\left(\theta_s(x)-\theta_s(x+a)\right)} \approx \\
\int \rd x \,a\lambda \left(\partial_x\theta_s(x)\right)^2\,,
\end{multline}
where we considered that the bosonic fields vary slowly over the length scale $a$, and we neglected constant contributions. This term clearly contributes to the first line of Eq. \eqref{boson} for $s=\Up,\Dn$.
An analogous contribution, for $s=0$, is obtained from the plaquette term:
\begin{equation}
-\frac{1}{g}\sum_{r}\sigma^\dag_{r,0} \sigma_{r+1,} + {\rm H.c.}  \approx \\
\int \rd x \, \frac{a}{g} \left(\partial_x\theta_0(x)\right)^2\,.
\end{equation}
The additional quadratic terms of the fields $\varphi_s$ are derived from the mass and rung electric field contributions. The former reads:
\begin{multline}
-\frac{1}{\lambda}\sum_r \left(\eta_{r,s} + \eta^\dag_{r,s}\right) \to \\
-\frac{2}{a\lambda} \int \rd x \, \cos\left(\varphi_s(x)-\varphi_s(x+a)\right) \\
\approx
\int \rd x \, \frac{a}{\lambda} \left(\partial_x\varphi_s(x)\right)^2\,,
\end{multline}
for $s=\Up,\Dn$. The rung electric energy has an analogous form:
\begin{equation}
-g\sum_r \left(\tau_{r,0} + \tau^\dag_{r,0}\right) \approx \int \rd x \, ag \left(\partial_x\varphi_0(x)\right)^2\,.
\end{equation}
The sum of these four quadratic terms determines the Luttinger liquid part of the Hamiltonian \eqref{boson} with the parameters:
\begin{align}
& K_{\Up,\Dn} = \frac{1}{\lambda}\,, \quad K_0 =g\,, \\
& v_\Up= v_\Dn = v_0 = \frac{4\pi a}{N} \equiv v\,.
\end{align} 
These values of the Luttinger parameters provide the bare values entering the renormalization group flow, whereas the velocities are equal in all the Luttinger sectors and are invariant through RG due to the emergent Lorentz symmetry of \eqref{boson}.

The interacting terms in \eqref{boson} are determined by the interleg tunneling and the leg electric field term. The former is straightforwardly obtained by the mapping (\ref{map1},\ref{map2}):
\begin{equation}
-\lambda \sum_{r=1}^{L}\zeta^\dag_{r,\Up}\sigma_{r,0}\zeta_{r,\Dn} + {\rm H.c.} \to -\frac{2\lambda}{a}\int \rd x\, \cos\left(\theta_\Up-\theta_\Dn-\theta_0\right)\,;
\end{equation}
The latter must be estimated by considering its string operator formulation in the axial gauge, which is derived from the form of the gauge constraints (in the case of a smooth right edge). In particular, on the physical states, we have:
\begin{equation}
\tau_{r,\Up} = \prod_{r'>r} \tau^\dag_{r',0}\eta^\dag_{r',\Up}\,.
\end{equation}
This expression is derived from Eqs. (\ref{gop},\ref{gop2},\ref{gconstr}) similarly to the pure LGT case in Eq. \eqref{pureLGT}. Based on Eqs. (\ref{map1},\ref{map2},\ref{mapbound}) we obtain:
\begin{equation} \label{e3}
-g \sum_r \left(\tau_{r,\Up} + \tau_{r,\Up}^\dag\right) \to -\frac{2g}{a}\int \rd x \, \cos\left(\varphi_\Up + \varphi_0\right)\,.
\end{equation}
The analogous expression for the lower leg completes the electric energy terms in \eqref{boson}.

Finally, we must consider the background terms. They are meant to restore the original $\mathbb{Z}_N$ symmetry of the system, and, indeed, when translated back to the clock operators, the background terms become proportional to the identity because they correspond to the $N^{\rm th}$ power of the clock operators. Despite this, they play a crucial role in determining the phase diagram of the system and their interplay is crucial to understand the transitions from gapped to gapless phases for $N>4$, in a way similar to the 1D quantum clock model.

The coupling constants $P_s$ and $Q_s$ must be determined by comparing the energy spectra of the field theory \eqref{boson} with the lattice model. This task is non-trivial but a rough estimate of their value can be obtained by considering the limiting cases $\lambda \to 0,\infty$ and $g \to 0,\infty$ and neglecting all the irrelevant terms.
Let us consider, in particular, the clock model limit with frozen gauge boson degrees of freedom. For simplicity, we also neglect the coupling between the two legs of the ladder, and we focus in the following on a single 1D clock chain. 

In the case $\lambda \to \infty$, thus $K_\Up,K_\Dn \to 0$, only the $P_{\Up,\Dn}$ terms are relevant and we neglect all the other interactions. In this case we expect that the corresponding clock model is deep in its ferromagnetic phase, where the elementary excitation is provided by the domain walls with mass $M_{\rm cl} = 2\lambda \left(1-\cos 2\pi/N\right)$. We compare this mass with the mass of a kink in the classical and static sine-Gordon model with quadratic part corresponding to the one in the Hamiltonian \eqref{boson}. This classical mass can be obtained by following standard techniques (see, for example, Chap. 16 in \cite{mussardo}):
\begin{equation}
M_{\rm kink} = 4\sqrt{\frac{2P_s v}{\pi N K_s}} \approx 8\sqrt{\frac{2P_s \lambda a}{N^2}}\,.
\end{equation}
By comparing the two masses we derive:
\begin{equation}
P_{\Up,\Dn} \approx \frac{\lambda N^2 \left(1-\cos 2\pi/N\right)^2}{32 a}\,,
\end{equation}
which provides Eq. \eqref{Peq} through $K=1/\lambda$ for $s=\Up,\Dn$. This expression sets the bare coupling constant of the $\theta_\Up$ and $\theta_\Dn$ background terms and it must be considered as an approximation valid for small $K_s$. The result for the small $\lambda$ limit, corresponding to the paramagnetic phase of the clock model, can be easily retrieved through the duality $\theta \leftrightarrow \varphi$ and $K \leftrightarrow K^{-1}$. It results into Eq. \eqref{Qeq}. The calculation of $P_0$ and $Q_0$ follows the same procedure. In this case, though, it is convenient to consider first the pure lattice gauge theory limit for $g\to 0$. In this way, the plaquette term plays the role of the ferromagnetic coupling of the clock model \eqref{pureLGT} and one obtains Eq. \eqref{Peq} for $K_0=1/g$. Finally, the duality $K_0 \leftrightarrow K_0^{-1}$ allows for the estimate of $Q_0$.

\section{The second-order renormalization group equations} \label{app:2nd}

Our renormalization group analysis is based on Wilson's approach and, in particular, on a second-order perturbative calculation in momentum space of the Euclidean action corresponding to the Hamiltonian \eqref{boson}.
In this Appendix we summarize the main steps for the derivation of the second-order RG equations we adopted in the study of the phase diagram, and, in particular, we focus on the onset of the main effective interactions that are generated through the flow of the sine-Gordon terms in \eqref{boson}.

Our perturbative approach relies on considering all the interaction terms in \eqref{boson} as perturbations of the Gaussian action $S_0$ corresponding to the quadratic terms in the bosonic fields. The quadratic action can be written as a function of either the $\varphi$ or the $\theta$ fields and, with the former choice, it reads:
\begin{equation}
S_0 =\frac{N}{4\pi} \int \rd^2 r \left[ \sum_{s=\sigma,\rho,0} \frac{K_s}{v} (\partial_\tau \varphi_s)^2 + K_s v (\partial_x \varphi_s)^2 \right],
\end{equation}
where we consider a two-dimensional Euclidean space-time. In the following, we will use both the charge and spin degrees of freedom, and the spin $\Up$ and $\Dn$ fields, depending on the most convenient notation.

Based on Wilson's prescription, we distinguish fast and slow oscillating modes for each of the bosonic fields. In particular, we introduce an effective cutoff $\tilde{\Lambda}$ in momentum space such that the slow modes are characterized by $k<\tilde{\Lambda}$, whereas the fast modes are defined by the choice $\tilde{\Lambda}<k<\Lambda$, with $\Lambda=2\pi/a$ being the ultraviolet cutoff of the system. To perform the Wilsonian RG, we will integrate out the fast modes of each field, and, in particular, we are interested in the limit $\Lambda/\tilde{\Lambda} = 1+\rd l$, with $\rd l$ infinitesimal. The decomposition of the bosonic fields in fast and slow modes reads:
\begin{align}
& \varphi_s(x,\tau)= \varphi_{{\sf s},s}(x,\tau)+\varphi_{{\sf f},s}(x,\tau)\,,\\
&  \theta_s(x,\tau)= \theta_{{\sf s},s}(x,\tau)+\theta_{{\sf f},s}(x,\tau)\,.
\end{align}

To derive the RG equations, we will define an effective action for the slow modes in the form:
\begin{multline} \label{seff}
 S_{\rm eff}(\tilde{\Lambda})=S_0(\varphi_{\sf s}) - \ln \left\langle e^{-S_I(\varphi_{\sf s}+\varphi_{\sf f})} \right\rangle_{\sf f}\\
 \approx S_0(\varphi_{\sf s}) + \underbrace{\left\langle S_I(\varphi_{\sf s}+\varphi_{\sf f})\right\rangle_{\sf f}}_{\mathcal{A}} \\
-\frac{1}{2} \left(\underbrace{\left\langle S_I^2(\varphi_{\sf s}+\varphi_{\sf f}) \right\rangle_{\sf f}}_{\mathcal{B}} - \underbrace{\left\langle S_I(\varphi_{\sf s}+\varphi_{\sf f}) \right\rangle^2_{\sf f}}_{\mathcal{A}^2} \right) + \ldots\,,
\end{multline}
where the average values are taken over the fast oscillating modes.
The interacting part $S_I$ of the action matches the interacting part of the Hamiltonian \eqref{boson}. $S_I$, however, collects also many effective interactions whose bare coupling constants vanish, but acquire non-trivial values during the RG flow. As discussed in the main text, we keep track only of the most relevant of these terms appearing at second order of perturbation, and we neglect simple powers of the terms appearing in the bare Hamiltonian which do not qualitatively affect the RG flow. We list here, for reference, the main interaction terms we consider:
\begin{multline} \label{SI}
S_I = -T\int \rd^2 r\, \cos\left(\sqrt{2}\theta_\sigma-\theta_0\right) \\
-G\int \rd^2 r\,\left[ \cos\left(\varphi_\Up+\varphi_0\right) + \cos\left(\varphi_\Dn-\varphi_0\right)\right] \\
-\sum_{s=\Up,\Dn} \int \rd^2 r \,\left[P \cos N \theta_s + Q \cos N\varphi_s  \right]\\
- \int \rd^2 r \,\left[P_0 \cos N \theta_0 + Q_0 \cos N\varphi_0  \right]\\
+\sum_{q=\rho,\sigma}\int \rd^2 r\, C_q \cos\left( \sqrt{2} N \varphi_q\right)\\
+\int \rd^2 r\, \left[C'_\rho \cos\sqrt{2}\varphi_\rho  + C'_\sigma\cos\left(\sqrt{2}\varphi_\sigma + 2 \varphi_0\right)\right]\,.
\end{multline}
The last two integrals, in particular, refer to interactions that appear only at second order in perturbation theory, but bear important implications for identifying the physical regimes of the system. The $C'$ interactions, in particular, appear only when $g>0$, differently from the $C$ interactions which influence the system in the clock limit as well.

We emphasize that these interactions  are only a small subset of all the terms appearing at second order. Let us consider, for example, the operators appearing in the clock limit. Beyond the interactions \eqref{ham2g0}, one should consider similar interactions in the $\theta$ fields. We point out, however, that such interactions would be in general less relevant than the $P$ and $T$ terms and commute with them, in such a way that their effect in determining the phase diagram is marginal, as we verified through a numerical solution of the second order RG equations including also these additional terms.

Before proceeding in the evaluation of the main second order terms, we summarize here some of the properties of the bosonic fields we will utilize. Concerning their duality relations in Euclidean time $\tau=it$, we have:
\begin{equation} \label{fielddual}
\partial_\tau \theta_j = ivK_j\partial_x\varphi_j \,,\quad \partial_\tau \varphi_j = i\frac{v}{K_j}\partial_x\theta_j\,.
\end{equation}
Concerning the correlation functions of the fast modes, we will adopt the following approximations:
\begin{align}
 &\left\langle \varphi_{{\sf f},j}^2(x) \right\rangle_{{\sf f},j} = \frac{1}{NK_j}\ln\frac{\Lambda}{\tilde{\Lambda}}\,, \label{cRG1} \\
 &\left\langle \varphi_{{\sf f},j}(x_1,\tau_1) \varphi_{{\sf f},j}(x_2,\tau_2) \right\rangle_{\sf f}   \approx \frac{\mathcal{C}(r)}{NK_j} \ln\frac{\Lambda}{\tilde{\Lambda}}\,,\label{cRG2}\\
 &\left\langle \theta_{{\sf f},j}^2(x) \right\rangle_{{\sf f},j} = \frac{K_j}{N}\ln\frac{\Lambda}{\tilde{\Lambda}}\,, \\
 &\left\langle \theta_{{\sf f},j}(x_1,\tau_1) \theta_{{\sf f},j}(x_2,\tau_2) \right\rangle_{\sf f}   \approx \frac{K_j\mathcal{C}(r)}{N} \ln\frac{\Lambda}{\tilde{\Lambda}}\,.
\end{align}
Here the logarithm captures the dominant scaling behavior, whereas $\mathcal{C}(r)$ is a function of $r=\sqrt{v^2(\tau_1-\tau_2)^2+(x_1-x_2)^2}$. In the following we will consider $\mathcal{C}(r)$ to be suitably short-ranged; for a sharp momentum cutoff, $\mathcal{C}(r) \approx J_0(\Lambda r)$ and this assumption is not satisfactorily fulfilled; however, $\mathcal{C}(r)$ can be made sufficiently short-ranged with more refined cutoffs \cite{Kogut1979,Haller2020}.

The first-order contribution $\mathcal{A}$ of the interacting action \eqref{seff} provides the standard dependence from the scaling dimensions of the RG equations. We focus in the following in the second-order contributions and, in particular, on the non-trivial terms appearing in $\mathcal{B}$.

We begin our analysis by studying a part of $\mathcal{B}$ which appears already in the clock limit $(g=0)$ and we consider, in particular, the following terms:
\bwt
\begin{multline} \label{Btermgen}
\mathcal{B} \supset \int d^2r_1 d^2r_2\, Q^2 \sum_{q,q'=\Up,\Dn}\left\langle \cos N\left(\varphi_{q,{\sf s}}(r_1) +\varphi_{q,{\sf f}}(r_1) \right)  \cos N\left(\varphi_{q',{\sf s}}(r_2) +\varphi_{q',{\sf f}}(r_2) \right) \right\rangle_{\sf f}\\
=\int \rd^2r_1 \rd^2r_2 \, \frac{Q^2}{4}\sum_{\substack{q,q'=\Up,\Dn \\\mu,\mu'=\pm1}}\left\langle \e^{i N\mu \left(\varphi_{q,{\sf s}}(r_1) +\varphi_{q,{\sf f}}(r_1) \right)}  \e^{i N\mu' \left(\varphi_{q',{\sf s}}(r_2) +\varphi_{q',{\sf f}}(r_2) \right)} \right\rangle_{\sf f}\\
=\int \rd^2r_1 \rd^2r_2 \, \frac{Q^2}{4}\sum_{\substack{q,q'=\Up,\Dn \\\mu,\mu'=\pm1}}e^{i N \left(\mu\varphi_{q,{\sf s}}(r_1) +\mu' \varphi_{q',{\sf s}}(r_2) \right)}\left\langle \e^{i N\left(\mu \varphi_{q,{\sf f}}(r_1) +\mu'\varphi_{q',{\sf f}}(r_2) \right)} \right\rangle_{\sf f}
\end{multline}
In this expression: (i) the terms with $q\neq q'$ and $\mu=\mu'$ return the $C_\rho$ interaction; (ii) the terms with $q\neq q'$ and $\mu=-\mu'$ return the $C_\sigma$ term; (iii) the terms with $q=q'$ and $\mu = -\mu'$ provide a correction to the quadratic part of the action which we must evaluate to obtain the RG equations for the Luttinger parameters. The last terms, $q=q'$ and $\mu=\mu'$, result instead in highly irrelevant interactions which we neglect.

We consider the contributions (i) first. We obtain:
\begin{multline}
\mathcal{B} \supset \int \rd^2r_1 \rd^2r_2 \frac{Q^2}{2}e^{i N \left(\varphi_{\Up,{\sf s}}(r_1) + \varphi_{\Dn,{\sf s}}(r_2) \right)}\left\langle \e^{i  N\left( \varphi_{\Up,{\sf f}}(r_1) +\varphi_{\Dn,{\sf f}}(r_2) \right)} \right\rangle_{\sf f} +{\rm H.c.}=\\
=\int \rd^2r_1 \rd^2r_2 \frac{Q^2}{2}e^{i \frac{N}{\sqrt{2}} \left(\varphi_{\rho,{\sf s}}(r_1) + \varphi_{\sigma,{\sf s}}(r_1)+ \varphi_{\rho,{\sf s}}(r_2)-\varphi_{\sigma,{\sf s}}(r_2) \right)}  
\left(\frac{\tilde{\Lambda}}{\Lambda}\right)^{\frac{N}{2K_\rho} + \frac{N}{2K_\sigma} + \frac{N\mathcal{C}}{2K_\rho} - \frac{N\mathcal{C}}{2K_\sigma}} +{\rm H.c.}\\
=\int \rd^2r_1' \rd^2r_2' \left(1+4\rd l\right) \frac{Q^2}{2}e^{i \frac{N}{\sqrt{2}} \left(\varphi_{\rho}(r_1') + \varphi_{\sigma}(r_1')+ \varphi_{\rho}(r_2')-\varphi_{\sigma}(r_2') \right)} \times \\ 
\left[1-\rd l\left(\frac{N}{2K_\rho} + \frac{N}{2K_\sigma} + \frac{N\mathcal{C}}{2K_\rho} - \frac{N\mathcal{C}}{2K_\sigma}\right)\right]+{\rm H.c.}\,,
\end{multline}
\ewt
where we used Eqs. (\ref{cRG1},\ref{cRG2}), we decomposed the fields in the charge and spin sectors, we imposed $\tilde{\Lambda} = \Lambda (1-\rd l)$, and we applied a general rescaling of the coordinates $\rd^2r = (1+2\rd l)\rd^2r'$. The coordinate dependence of the correlation $\mathcal{C}$ has been suppressed for ease of notation. Among the terms in the previous expression, only the ones proportional to $\mathcal{C}$ contribute to the second-order correction of the action. The others are simplified by the analogous terms in $\mathcal{A}^2$ in Eq. \eqref{seff}. In this expression, we approximate  $\mathcal{C}(r) \approx \left(a^2/v\right) \delta(r)$. This results in the following contribution to the renormalized action:
\begin{equation} \label{Crhoterm}
 \frac{Q^2a^2}{v}\rd l\left(\frac{N}{4K_\rho} - \frac{N}{4K_\sigma}\right)\int \rd^2r\cos\left(\sqrt{2}N \varphi_\rho(r) \right)\,.
\end{equation} 
By following the same approach, in the case (ii), we obtain instead:
\begin{equation} \label{Csigmaterm}
 -\frac{Q^2a^2}{v}\rd l\left(\frac{N}{4K_\rho} - \frac{N}{4K_\sigma}\right)\int \rd^2r\cos\left(\sqrt{2}N \varphi_\sigma(r) \right)\,.
\end{equation}
From these results we can immediately derive the second order RG equations for the coupling constants $C_\rho$ and $C_\sigma$ [see Eqs. \eqref{CcRG} and \eqref{CsRG}].

The terms (iii) in Eq. \eqref{Btermgen} provide a paradigmatic example of the feedback of the interactions in the definition of the Luttinger parameters. For $q=q'$ and $\mu = -\mu'$ we obtain the following contribution in $\mathcal{B}$:
\begin{multline}
\int \rd^2r_1 \rd^2r_2 \frac{Q^2}{4} \times \\
\sum_{q=\pm 1} e^{i \frac{N}{\sqrt{2}} \left(\varphi_{\rho,{\sf s}}(r_1) + q\varphi_{\sigma,{\sf s}}(r_1)- \varphi_{\rho,{\sf s}}(r_2)-q\varphi_{\sigma,{\sf s}}(r_2) \right)} \times  \\ 
\left(\frac{\tilde{\Lambda}}{\Lambda}\right)^{\frac{N}{2K_\rho} + \frac{N}{2K_\sigma} - \frac{N\mathcal{C}}{2K_\rho} - \frac{N\mathcal{C}}{2K_\sigma}} +{\rm H.c.}\,.
\end{multline}
Also in this case, we must consider that $\tilde{\Lambda}/\Lambda = 1-\rd l$, and the effective action will include only the terms that are proportional to the correlation function $\mathcal{C}(r_1-r_2)$. To manipulate this expression, we take into account that $\mathcal{C}$ is localized and peaked around $r_1 - r_2 =0$. In particular, we approximate it with a function different from zero only in a range of width $a$ (see, for example, \cite{giamarchi}). It is thus convenient to rewrite the former expression in terms of the center of mass and relative coordinates. The integral over the relative coordinate $r_1 - r_2$ gives a non-negligible contribution only around zero. After rescaling the center of mass coordinate we obtain an effective contribution to the action of the kind:
\begin{multline}
-\frac{NQ^2a^2\rd l }{8v} \int \rd^2 r  \left(\frac{1}{K_\rho} + \frac{1}{K_\sigma}\right) \times \\
\left[\cos\frac{Na}{\sqrt{2}}\left(\nabla\varphi_\rho +\nabla\varphi_\sigma\right) + \cos\frac{Na}{\sqrt{2}}\left(\nabla\varphi_\rho -\nabla\varphi_\sigma\right)\right] \approx \\
\frac{N^3Q^2a^4\rd l }{16v}  \left(\frac{1}{K_\rho} + \frac{1}{K_\sigma}\right) \times\\
\int \rd^2 r \sum_{q=\rho,\sigma} \left[\left(\partial_x\varphi_q\right)^2 +\frac{\left(\partial_\tau\varphi_q\right)^2}{v^2}\right]\,.
\end{multline}
By considering the relations \eqref{fielddual}, one can add these second order terms to the Gaussian action $S_0$ and derive the corresponding corrections to the Luttinger parameters, which contribute to Eqs. (\ref{KcRG},\ref{KsRG}).

All the second-order terms we considered so far stemmed from the background $Q$ interaction. The analogous results for the $P$ interaction can be easily derived by applying the field duality $\varphi \leftrightarrow \theta$ and $K \leftrightarrow 1/K$. This allows us to determine the dependence of $\rd K_q/\rd l$ on $P^2$, and it yields to the appearance of additional sine-Gordon interactions analogous to the ones in (\ref{Crhoterm},\ref{Csigmaterm}) for the $\theta$ fields. We numerically verified that their role in the solution of the RG equations and in the definition of the phase diagram of the system is negligible, therefore, for simplicity, we did not include them in our effective action \eqref{SI}.

To estimate the role of the fields $\varphi_0$ and $\theta_0$, we must consider instead the tunneling $T$, electric field $G$ and background $P_0$ and $Q_0$ interactions in the action \eqref{SI}. Their second order terms yield the additional interactions $C'$ and provide additional contributions to the Luttinger parameter flow equations. 

The analysis of the $P_0$ and $Q_0$ terms are completely analogous to the previous example. The tunneling $T$ term, at second order, presents instead a mixing of the matter $\sigma$ and gauge $0$ sectors. Such a mixing would require to modify our description of the field theory by introducing an $l$-dependent rotation of the field sectors. In the following we neglect the mixing terms (proportional to $\nabla \theta_\sigma \nabla \theta_0$) and we maintain the separation of the $\rho,\sigma$ and $0$ sectors: this simplification is justified only for small values of $l$ and may determine a tiny shift of the phase boundaries of the system resulting from the solution of the RG equations. Additionally, it can also yield a small systematic error in estimating the correlation functions and string parameters of the system.

A similar mixing appears also in the terms generated by the $G$ interaction. The contribution to $S_I$ obtained from the second-order estimate of the $G$ terms reads:
\begin{multline} \label{G2}
\frac{G^2dl}{16N} \int \rd^2r_1 \, \rd^2r_2 \sum_{\mu,\mu',\nu,\nu',=\pm 1} \left(\frac{\mu \mu'}{K_\rho} + \frac{\nu \nu'}{K_\sigma} +\frac{2\nu \nu'}{K_0}\right)\mathcal{C} \times \\
e^{i\left(\frac{\mu\varphi_\rho}{\sqrt{2}}+\frac{\nu\varphi_\sigma}{\sqrt{2}}+\nu\varphi_0\right)(r_1)} 
e^{i\left(\frac{\mu'\varphi_\rho}{\sqrt{2}}+\frac{\nu'\varphi_\sigma}{\sqrt{2}}+\nu'\varphi_0\right)(r_2)}\,.
\end{multline}
Analogously with the previous analysis, the terms with $\mu=-\mu'$ and $\nu=-\nu'$ generate a correction of the Gaussian action and yield the $G^2$ contribution in Eqs. (\ref{KcRG},\ref{KsRG},\ref{K0RG}) which are obtained neglecting the mixing between $\varphi_0$ and $\varphi_\sigma$.

The terms $\mu=\mu'$ and $\nu=-\nu'$, instead, generate the t'Hooft $\mathcal{G}_\rho$ interaction, corresponding to the $C'_\rho$ in the action \eqref{SI}:
\begin{equation}
\frac{G^2a^2 dl}{4Nv} \left[\frac{1}{K_\rho}-\frac{1}{K_\sigma} -\frac{2}{K_0}\right]\int \rd^2 r \, \cos \sqrt{2} \varphi_\rho\,.
\end{equation}
This is a highly relevant term which is crucial in determining the boundaries of the Coulomb phase. By increasing $g$, it is indeed the dominant operator that gaps the charge sector. The term $C'_\sigma$ originates in a similar way from the contribution $\mu=-\mu'$ and $\nu=\nu'$ in \eqref{G2}. Both the $C'$ terms, in turn, provide a second-order contribution to the flow of $G$ (see Eq. \eqref{GRG}).

We conclude by summarizing all the RG equations. The flow equations for the background coupling constants do not have second-order terms:
\begin{align}
 \frac{\rd P}{\rd l} &= \left[2 - \frac{N}{4}\left(K_\rho + K_\sigma\right)\right]P\\
 \frac{\rd Q}{\rd l} &= \left[2 - \frac{N}{4}\left(\frac{1}{K_\rho} + \frac{1}{K_\sigma}\right)\right]Q\\
 \frac{\rd P_0}{\rd l} &= \left[2 - \frac{NK_0}{2}\right]P_0\\
 \frac{\rd Q_0}{\rd l} &= \left[2 - \frac{N}{2K_0}\right]Q_0\,.
\end{align}
The second-order equations for the coupling constants read:
\bwt
\begin{align}
 \frac{\rd T}{\rd l} &= \left(2 - \frac{K_\sigma}{N} - \frac{K_0}{2N}\right)T\\
\frac{\rd G}{\rd l}&= \left(2-D_g\right)G -\frac{C'_\rho Ga^2}{4NK_\rho v}-\frac{C'_\sigma Ga^2}{4Nv}\left(\frac{1}{K_\sigma}+\frac{2}{K_0}\right)\label{GRG} \\ 
 \frac{\rd C_{\rho}}{\rd l} &= \left(2-2\frac{N}{K_\rho}\right)C_\rho +\frac{Q^2a^2}{v}\left(\frac{N}{4K_\rho} - \frac{N}{4K_\sigma}\right) \label{CcRG}\\
 \frac{\rd C_{\sigma}}{\rd l} &= \left(2-2\frac{N}{K_\sigma}\right)C_\sigma -\frac{Q^2a^2}{v}\left(\frac{N}{4K_\rho} - \frac{N}{4K_\sigma}\right) \label{CsRG}\\
\frac{\rd C'_\rho}{\rd l} &= \left(2-\frac{1}{K_\rho N}\right)C'_\rho+\frac{G^2a^2 }{4Nv} \left[\frac{1}{K_\rho}-\frac{1}{K_\sigma} -\frac{2}{K_0}\right]\\
\frac{\rd C'_\sigma}{\rd l} &= \left(2-\frac{1}{K_\sigma N}-\frac{2}{NK_0}\right)C'_\sigma -\frac{G^2a^2 }{4Nv} \left[\frac{1}{K_\rho}-\frac{1}{K_\sigma} -\frac{2}{K_0}\right] \,,
\end{align}
with $D_g= \frac{1}{4N}\left[\frac{1}{K_\rho} + \frac{1}{K_\sigma} +\frac{2}{K_0}\right]$. Finally, the RG equations for the Luttinger parameters read:
\begin{align}
\frac{\rd K_\rho}{\rd l} &= -\frac{\pi N^2 P^2 a^4 K^2_\rho\left(K_\sigma+K_\rho\right)}{4v^2} + \frac{\pi N^2 Q^2 a^4}{4v^2}\left(\frac{1}{K_\sigma}+\frac{1}{K_\rho}\right) + \frac{8\pi N^2 C_\rho^2 a^4 }{4v^2 K_\rho} + \frac{2\pi C^{\prime 2}_\rho a^4 }{N^2v^2 K_\rho}  + \frac{\pi G^2a^4}{4N^2v^2}\left[\frac{1}{K_\rho} + \frac{1}{K_\sigma} +\frac{2}{K_0} \right] \label{KcRG}\\
\frac{\rd K_\sigma}{\rd l} &= -\frac{\pi N^2 P^2 a^4 K^2_\sigma\left(K_\sigma + K_\rho\right)}{4v^2} + \frac{\pi N^2 Q^2 a^4}{4v^2}\left(\frac{1}{K_\sigma}+\frac{1}{ K_\rho}\right)  + \frac{2\pi N^2 C_\sigma^2 a^4 }{v^2 K_\sigma} \nonumber \\
& - \frac{2\pi T^2 a^4 K_\sigma^2}{N^2v^2} \left(K_\sigma+\frac{K_0}{2}\right) + \frac{2\pi C^{\prime 2}_\sigma a^4 }{N^2v^2} \left(\frac{1}{K_\sigma }+\frac{1}{K_0 }\right) + \frac{\pi G^2a^4}{4N^2v^2}\left[\frac{1}{K_\rho} + \frac{1}{K_\sigma} +\frac{2}{K_0} \right] \label{KsRG} \\
\frac{\rd K_0}{\rd l} &= \frac{\pi N^2 Q_0^2 a^4 }{v^2 K_0} -\frac{\pi N^2 P_0^2 a^4 K_0^3}{v^2 }- \frac{\pi T^2 a^4 K_0^2}{N^2v^2} \left(K_\sigma+\frac{K_0}{2}\right) + \frac{2\pi C_\sigma^{\prime 2} a^4 }{N^2v^2} \left(\frac{1}{K_\sigma}+\frac{2}{K_0}\right) + \frac{\pi G^2a^4}{2N^2v^2}\left[\frac{1}{K_\rho} + \frac{1}{K_\sigma} +\frac{2}{K_0} \right] \label{K0RG}
\end{align}
\ewt

To determine the phase diagram in Fig. \ref{fig:pd}(b) we adopted an implicit Runge-Kutta method to solve numerical these differential equations. We set a fixed lower threshold ($\sim 0.2$) below which the coupling constants were considered negligible, in order to determine the extension of the gapless phase. We set a variable upper cutoff whose level sets the scale above which the constants were considered in the strong coupling regime. This upper cutoff was taken to be larger than all the bare constants. We stopped the flow every time a set of interactions sufficient to gap all the sectors reached the strong-coupling upper cutoff, or when the $\sigma$ and $0$ interactions reached the upper thresholds whereas the $\rho$ interactions fell below the lower cutoff. In case of non-commuting operators reaching together the upper threshold, we considered the largest to classify the corresponding phase.

\begin{figure}[bt]
\includegraphics[width=\columnwidth]{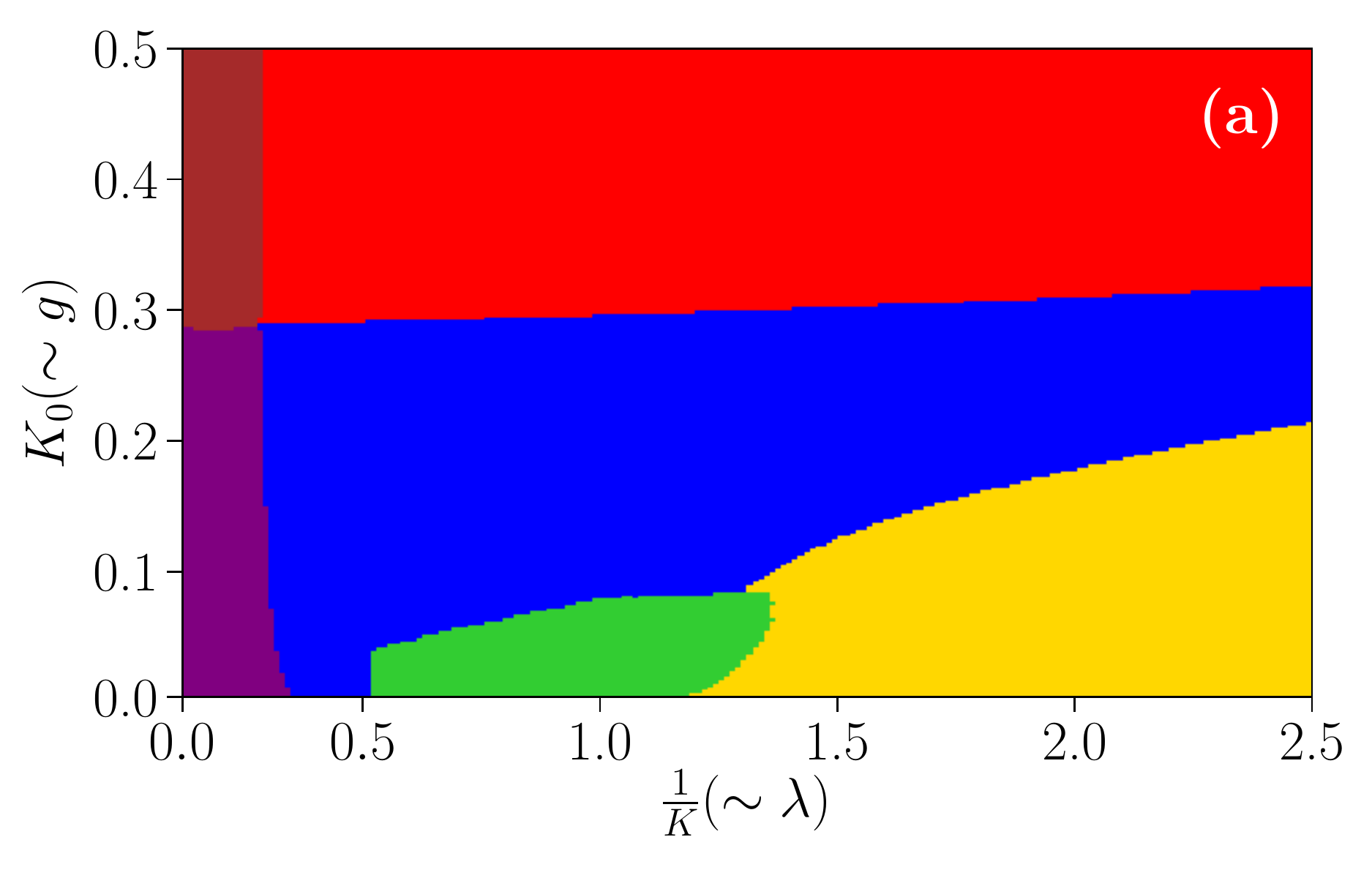}
\includegraphics[width=\columnwidth]{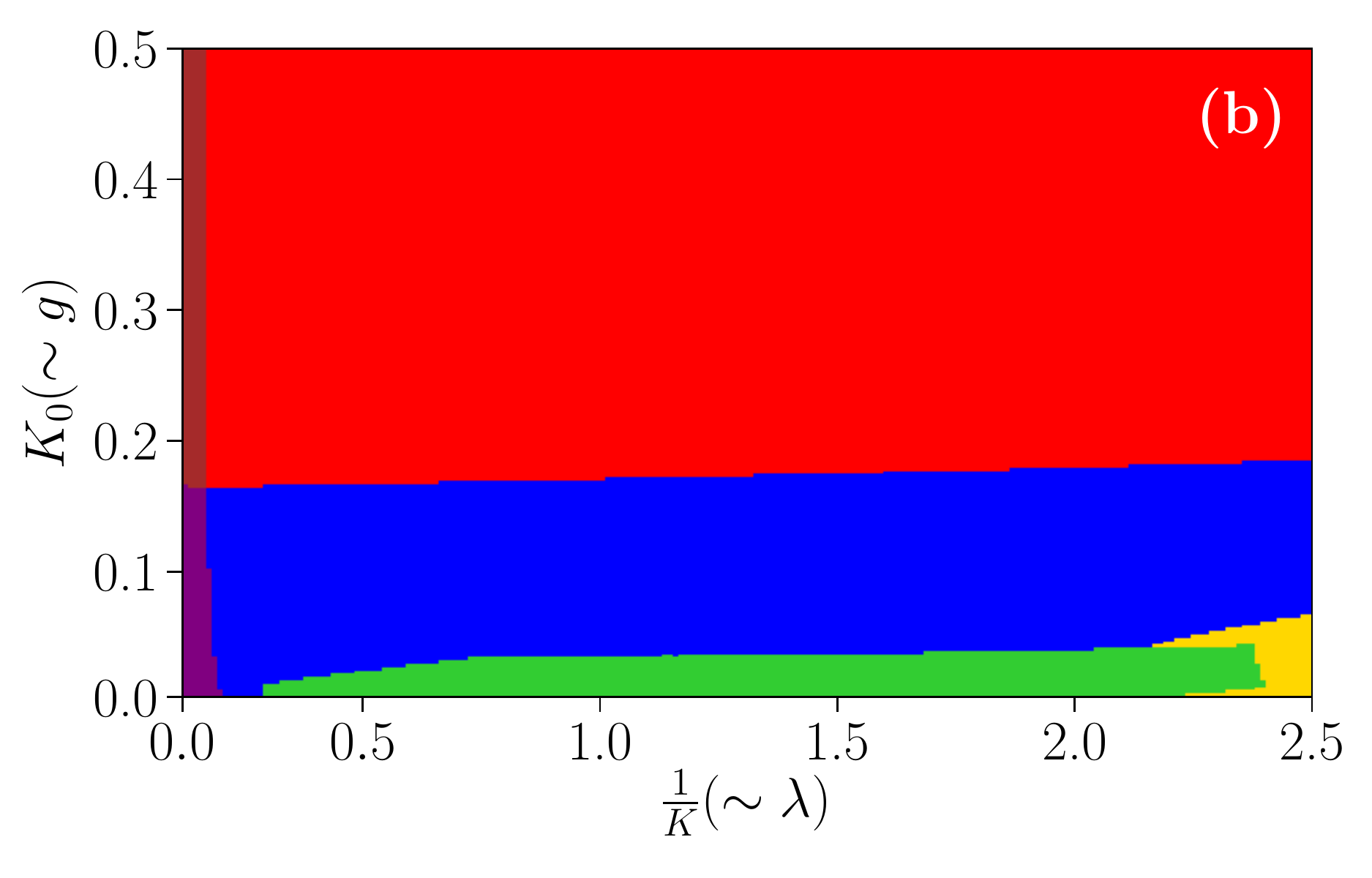}
\caption{
Phase diagram from second-order RG for $N=8$ (a) and $N=15$ (b). The different phases are shown by different colors, including the deconfined phase (purple), the quadrupolar phase (blue), the Coulomb phase (green), the Higgs phase (yellow), the fully confined phase (brown), and the confined rung-dominated phase (red).
}
\label{fig:app_pd}
\end{figure}

In Fig. \ref{fig:pd} (b) we presented the phase diagram obtained from the numerical solutions of the RG flow equations for $N=5$. The numerical solution of the flow equations can be straightforwardly extended to $N>5$: in Fig. \ref{fig:app_pd} we show the predicted phase diagrams for $N=8$ (a) and $N=15$ (b). These figures display how increasing $N$ leads the confined rung-dominated phase to grow and the gapless phase to be more extended in the $\lambda$ direction and less in the $g$ direction. For larger values of $N$ the gapless phase is seen to be spread out onto the $g=0$ line whereas the confined rung-dominated phase dominates all the gapped regions of the phase diagram.

\section{Modified first-order phase diagram for $g=0$} \label{app:mpd}

To simplify the full second-order renormalization group analysis, it is possible to consider a ``quasi''-second-order approach to illuminate the behavior of the higher order terms. This analysis shows that the phase diagram for $N\le 4$ does not display the extended gapless Coulomb phase and we present it here for the clock limit $g=0$ (Eq. \eqref{Potts}), based on the low-energy Hamiltonian \eqref{boson2}. 

In the first-order RG analysis both the rung tunneling term and the background $Q$ term are relevant for $K\in[N/4,2N]$. Since these terms do not commute,  the first order analysis demands the less relevant term of the two to be neglected. Therefore, the first-order analysis predicts a gapless phase to appear for $N>2$, which is ultimately shown to be wrong by the numerical simulations. A simple way of improving the Rg predictions is to adopt a two-step RG approach, where the flow is divided into separate parts, where each part is terminated when a coupling constant reaches a suitable upper threshold, which indicates when a given interaction semiclassically pins the related fields. After each separate flow, an effective Hamiltonian for the remaining unpinned sectors is considered. For $K<N/\sqrt{2}$, where the rung tunneling is the most relevant term, $\theta_{\sigma}$ is the first field being pinned to an energy minimum; hence, the effective Hamiltonian of the charge sector for the second RG step results
\begin{multline} 
H_{\rm{step} \, 2}(g=0)=\\
\frac{N}{4\pi} \int \rd x \, v\left[K \left(\partial_x \varphi_\rho\right)^2 +\frac{1}{K}\left(\partial_x \theta_\rho\right)^2 \right]\\
-2\int \rd x \, \tilde{P} \cos\frac{N\theta_\rho}{\sqrt{2}} ,
\end{multline}
where $\tilde{P}$ is a suitable renormalized parameter deriving from the background $P$ term. We observe that, after the initial flow pins $\theta_{\sigma}$, the scaling dimension of this background term decreases by a factor of 2, $D_{\tilde{P}}=KN/4$. Based on a scaling analysis of the second step Hamiltonian $H_{\rm{step} \, 2}$, the extended gapless phase disappears for $N=3$. 

In the above analysis the background $Q$ term was neglected, since we considered first-order contributions only. By following the two-step approach adopted in \cite{Kim1999}, we refine $H_{\rm{step} \, 2}$ by including the most relevant second-order term, which matches the $C_{\rho}$ interaction in Eq. \eqref{SI}:
\begin{multline} \label{boson2_rho}
H_{\rm{step} \, 2}(g=0)=\\
\frac{N}{4\pi} \int \rd x \, v\left[K \left(\partial_x \varphi_\rho\right)^2 +\frac{1}{K}\left(\partial_x \theta_\rho\right)^2 \right]\\
-2\int \rd x \, \left[ \tilde{P} \cos\frac{N\theta_\rho}{\sqrt{2}} + \tilde{C}_{\rho} \cos \sqrt{2}N\varphi_\rho \right].
\end{multline}
Here $\tilde{C}_{\rho}$ is a suitable renormalized coupling constant determined by the first step in the flow. Its scaling dimension is $D_{\tilde{C}_\rho} = N/K$, such that this term reduces the extension of the gapless phase for $N>4$, and it completely removes it for $N=4$. In conclusion, the two-step RG procedure indicates how the $\tilde{P}$ and $\tilde{C}_{\rho} $ terms are responsible for gapping the gapless phase for $N=3$ and $4$. In particular, the gapped phases for $N=2$ and $4$ coincide, consistently with Appendix \ref{app:4}.

\section{Estimate of the meson decay for large $\lambda$} \label{app:qa}

To obtain an estimate of the decay of the mesons in the Higgs phase, in the limit of large $\lambda$ and small $g$, we consider a quasiadiabatic continuation technique \cite{Hastings2005,Bernevig2016,burrello2018}. We begin our analysis from an unperturbed Hamiltonian (in the unitary gauge) given by:
\begin{multline}
H_0 =  -\frac{1}{g} \sum_{r} \left(\sigma_{r,0}\sigma_{r+1,\Up}\sigma_{r+1,0}^\dag\sigma_{r+1,\Dn}^\dag + {\rm H.c.}\right) \\ 
-\lambda\sum_{s,r}\left(\sigma^\dag_{r,s}+\sigma_{r,s}\right).
\end{multline}
The ground state $\ket{\Psi_0}$ of this Hamiltonian is a product state of clocks aligned on the state $\ket{\sigma=1}$. By applying either the operator $\tau$ or $\tau^\dag$ to any link along the legs of the ladder we create an excitation with energy $\delta E  = 2\left(g^{-1} +\lambda\right)\left(1-\cos 2\pi/N\right)$. We express the ground state of the Hamiltonian with small $g$ and large $\lambda$ by use of a unitary quasiadiabatic operator, such that $\ket{\Psi_g} = V(g) \ket{\Psi_0}$. A general definition of the quasiadiabatic operator $V$ can be found, for example, in Ref. \cite{hastings2010}.
In the following, we will approximate it by considering:
\begin{align}
&\ket{\Psi_g}=V(g) \ket{\Psi_0} \approx \e^{i g \mathcal{D}}\ket{\Psi_0}\,, \label{VPsi}\\
&\mathcal{D} =  \int_{-\infty}^{+\infty} \rd t \, \e^{i H_0 t} \cdot \nonumber\\
&\qquad\sum_{r,s=\Up,\Dn}\left( \tau_{r,s} + \tau^\dag_{r,s}\right) \e^{-i H_0 t} F\left[\delta Et\right]\,.
\end{align}
In the general case, the Hermitian operator $\mathcal{D}$ depends on $g$, and $V$ requires to be defined as an ordered exponential. Here, instead, we approximated $\mathcal{D}$ by considering a time evolution dictated by the Hamiltonian $H_0$ only and we neglected the electric field energy of the rungs which has only a minor effect on the mesons, in such a way that $\mathcal{D}$ is independent on $g$. In the previous equation, $F(t)$ is an odd and analytical filter function such that its Fourier transform results in
\begin{equation} \label{fFourier}
\tilde{F}(\omega) = \int_{-\infty}^{+\infty} {\rm d}t\, e^{i\omega t} F(t) = -\frac{1}{\omega} \quad \text{for} \quad |\omega| \ge 1\,,
\end{equation}
and $\tilde{F}(0)=0$ \cite{hastings2010}. 
Next, we consider that $V$ is applied over the product state $\ket{\Psi_0}$ and we observe that:
\begin{multline}
\mathcal{D} \ket{\Psi_0} = \\
 \int_{-\infty}^{+\infty} \rd t \, \e^{i\delta E t} \sum_{r,s=\Up,\Dn}\left( \tau_{r,s} + \tau^\dag_{r,s}\right)  F\left[\delta Et\right] \ket{\Psi_0} = \\
-\frac{1}{\delta E} \sum_{r,s=\Up,\Dn}\left( \tau_{r,s} + \tau^\dag_{r,s}\right)  \ket{\Psi_0}\,.
\end{multline}
Therefore, we approximate \eqref{VPsi} with:
\begin{equation}
V(g) \ket{\Psi_0} \approx \e^{-i\alpha  \sum_{r,s=\Up,\Dn}\left( \tau_{r,s} + \tau^\dag_{r,s}\right)}\ket{\Psi_0}\,,
\end{equation}
where:
\begin{equation}
\alpha=\frac{g^2}{2\left(1-\cos2\pi/N\right)\left(\lambda g + 1\right)}
\end{equation}
This approximation amounts to overestimate the energy of the higher excitations of the single links with multiples of $\delta E$, which, however, has a negliegible effect for long mesons.
Every meson string under these approximations factorizes into the product of expectation values of the kind:
\begin{equation}
\bra{\sigma=1}\e^{i\alpha  \left( \tau + \tau^\dag\right)} \sigma \e^{-i\alpha  \left( \tau + \tau^\dag\right)}\ket{\sigma=1} \approx \e^{-2\alpha^2\left(1-\cos2\pi/N\right)}\,,
\end{equation}
for small $\alpha$. From the previous relation one derives Eq. \eqref{xiqa} by considering the $M_\rho$ and $M_\sigma$ string operators.

In the regime with both $g$ and $\lambda$ much greater than $1$, the ground state of the system can be approximated by the product state which minimizes the Hamiltonian without the plaquette and mass terms (proportional to $1/g$ and $1/\lambda$ respectively). In this state each link is in the state $\ket{\chi(g/\lambda)}$. The estimate of the expectation value 
\begin{equation} \label{sigmaprod}
\bra{\chi(g/\lambda)} \sigma \ket{\chi(g/\lambda)} = \bra{\chi(g/\lambda)} \sigma^\dag \ket{\chi(g/\lambda)} 
\end{equation}
provides an approximation of $\e^{-2/\xi_{M}}$, for the $M_{\rho}$ and $M_{\sigma}$ mesons. For large value of $g/\lambda$ the expectation value \eqref{sigmaprod} can be approximated by $\sim 1.45\lambda/g$.

 \section{Detail of numerical calculations}

In this section we include further details about the numerical calculations by DMRG.

\begin{figure}[bt]
\includegraphics[width=\columnwidth]{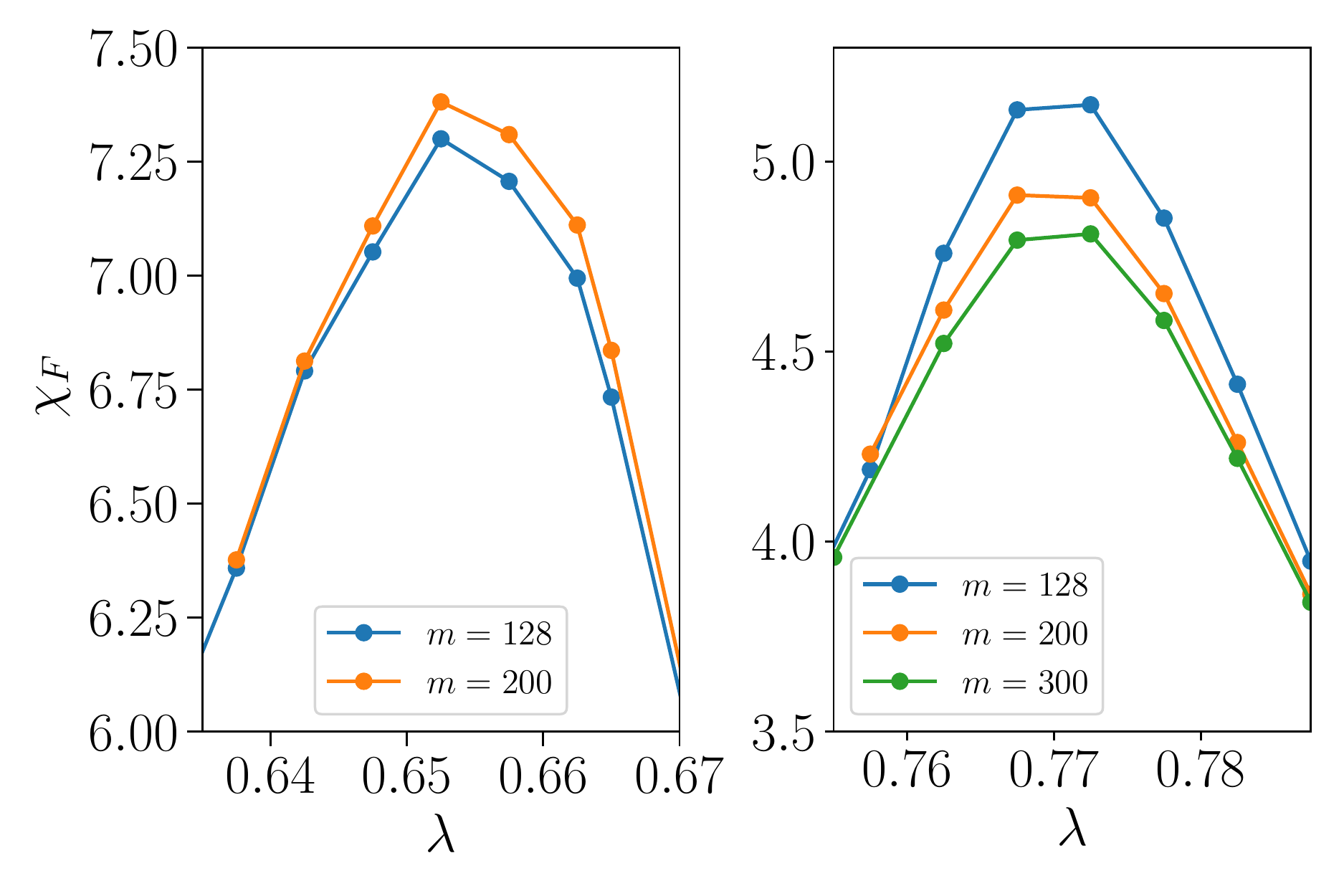}
\caption{
Fidelity susceptibility peaks at two BKT transitions at the boundaries of gapless phase for $g=0$
(the same as Fig.~\ref{fig:FS}).
The system length is $L=101$.
The results of different bound dimensions $m$ are shown.
}
\label{fig:app_fs}
\end{figure}

\subsection{Fidelity susceptibility}

The physical quantity that is most affected by the finite bond dimension $m$ of our simulations is the fidelity susceptibility (FS), which is based on the overlap between two ground states that are very close in parameter space.
For longer systems and smaller parameter differences, one typically needs larger bond dimension to obtain the same accuracy.
In Fig.~\ref{fig:app_fs} we illustrate the FS for several bond dimensions for the two peaks at BKT transitions shown in Fig.~\ref{fig:FS}.
Although the FS does has not completely converged at $m=300$, the peak positions do not change with the bond dimension in the parameter resolution we consider.
We note that the values of $\lambda$ we present here are in a quite small region (which means high resolution) compared to the phase diagram shown in Fig.~\ref{fig:pd}.

\begin{figure}[bt]
\includegraphics[width=\columnwidth]{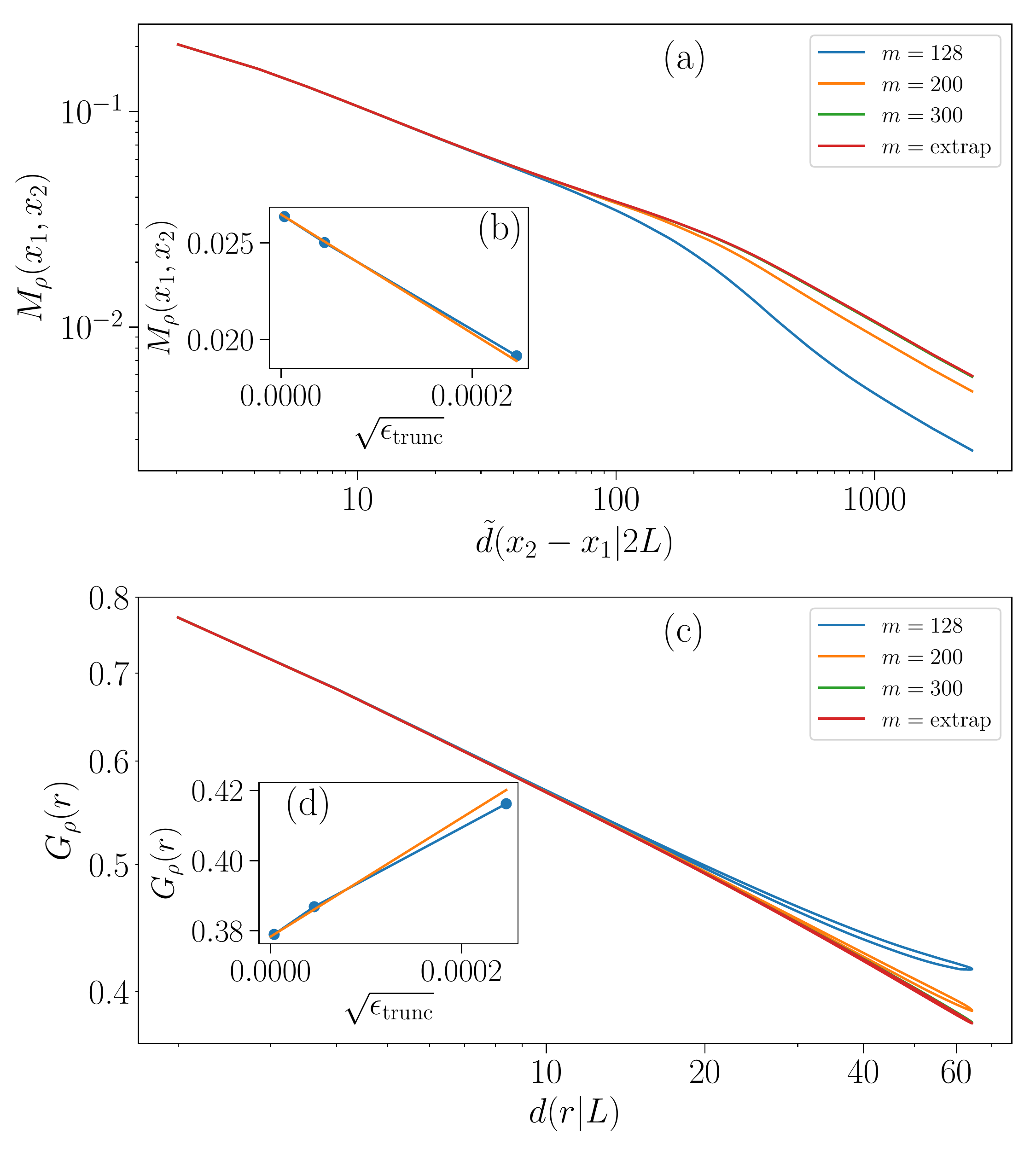}
\caption{
The correlations $M_{\rho}$ (a) and $G_{\rho}$ (c) in the gapless phase for different bond dimensions $m$ for $L=101$, $g=0.001$ and $\lambda=0.75$.
The results of the extrapolations based on the truncation error are also shown (red curves) and they overlap with the $m=300$ data.
(b), (d): The extrapolations of $M_{\rho}$ and $G_{\rho}$ with the square root of truncation error for $x_1=1,\,x_2=50$ and $r=50$ respectively.
The smallest truncation error is about $10^{-11}$ for $m=300$.
}
\label{fig:app_gapless}
\end{figure}

\subsection{Gapless phase}
The ground state in a gapless phase requires larger bond dimension in DMRG to converge.
In this work, we use bond dimensions up to $m=300$ to reach a truncation error $\epsilon_\mathrm{trunc}\approx 10^{-11}$ in the gapless phase.
In Figs.~\ref{fig:app_gapless}(a) and (c) we show $M_{\rho}$ and $G_{\rho}$ for different bond dimensions from $m=128$ to $m=300$.
It is useful to extrapolate the physical quantities based on the truncation error to estimate the values at infinite bond dimension.
We perform extrapolations of $M_{\rho}$ and $G_{\rho}$ with $\sqrt{\epsilon_\mathrm{trunc}}$, as shown in Figs.~\ref{fig:app_gapless} (b) and (d), and display the comparison with the finite bond dimension calculations in Fig.~\ref{fig:app_gapless} (a) and (c).
It can be seen that for $m=300$ the correlations are already very close to the extrapolated results and are indistinguishable in the figure. This demonstrates the high accuracy of our results.

\begin{figure}[bt]
\includegraphics[width=\columnwidth]{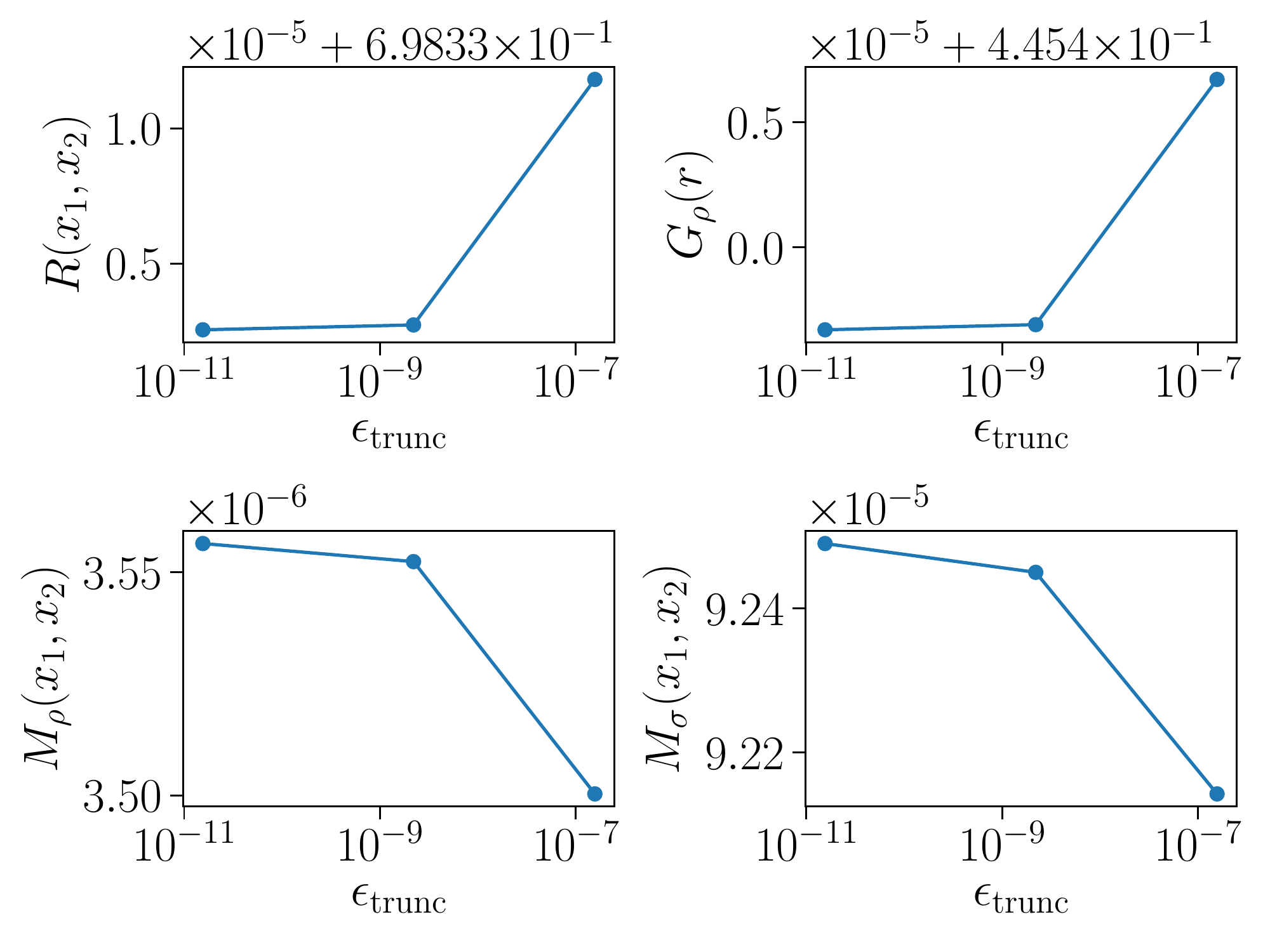}
\caption{
$R(x_1,x_2)$, $M_\rho(x_1,x_2)$, and $M_\sigma(x_1,x_2)$ for $x_1=2$ and $x_2=40$, and $G_\rho(r)$ for $r=40$, as functions of truncation errors, corresponding to bond dimension $m=32$, $64$ and $128$.
The system is a $L=81$ ladder with $\lambda=0.4$ and $g=0.2$.
(See red curves in Fig.~\ref{fig:gapped} for the correlations in full range.)
}
\label{fig:app_gapped}
\end{figure}

\subsection{Gapped phase}
In the gapped phase, the DMRG with bond dimension $m=128$ reaches a truncation error of about $10^{-11}$.
Fig.~\ref{fig:app_gapped} shows the convergences of several quantities in the gapped phase at $\lambda=0.4$ and $g=0.2$, which is a typical example in the gapped phase.
$R(x_1,x_2)$, $M_\rho(x_1,x_2)$, and $M_\sigma(x_1,x_2)$ for $x_1=2$ and $x_2=40$, and $G_\rho(r)$ for $r=40$, are shown as functions of truncation errors, corresponding to bond dimension $m=32$, $64$ and $128$.
It can be seen that the uncertainty between $m=64$ and $m=128$ is already very small ($\sim 0.1$\% for $M_\rho$, $\sim 0.01$\% for $M_\sigma$, and $\sim 0.00001$\% for $R$ and $G_\rho$), therefore no extrapolation is needed.

\end{document}